\documentstyle[12pt,epsfig,amssymb,amsbsy]{article}

\def\appendix#1{
  \addtocounter{section}{1}
  \setcounter{equation}{0}
  \renewcommand{\thesection}{\Alph{section}}
  \section*{Appendix \thesection\protect\indent \parbox[t]{11.715cm} {#1}}
  \addcontentsline{toc}{section}{Appendix \thesection\ \ \ #1}
  }

\def\eop{\vspace*{\fill}\pagebreak}
\newcommand{\rf}[1]{(\ref{#1})}
\newcommand{\eq}[1]{Eq.~(\ref{#1})}
\def\be{\begin{equation}}
\def\ee{\end{equation}}
\def\beq{\begin{equation}}
\def\eeq{\end{equation}}
\def\bea{\begin{eqnarray}}
\def\eea{\end{eqnarray}}

\newcommand{\fr}[2]{{\textstyle {#1 \over #2}}}

\def\fun#1#2{\lower3.6pt\vbox{\baselineskip0pt\lineskip.9pt
\ialign{$\mathsurround=0pt#1\hfil##\hfil$\crcr#2\crcr\sim\crcr}}}

\def\hybrid{\topmargin 0pt      \oddsidemargin 0pt
        \headheight 0pt \headsep 0pt
        \textwidth 17.5cm
        \textheight 25cm
        \voffset=-0.7cm
        \hoffset=-0.4cm
       \hoffset=-1.2cm
        \marginparwidth 0.0in
        \parskip 5pt plus 1pt   \jot = 1.5ex}
\catcode`\@=11
\def\marginnote#1{}

\newcount\hour
\newcount\minute
\newtoks\amorpm
\hour=\time\divide\hour by60
\minute=\time{\multiply\hour by60 \global\advance\minute by-\hour}
\edef\standardtime{{\ifnum\hour<12 \global\amorpm={am}%
        \else\global\amorpm={pm}\advance\hour by-12 \fi
        \ifnum\hour=0 \hour=12 \fi
        \number\hour:\ifnum\minute<10 0\fi\number\minute\the\amorpm}}
\edef\militarytime{\number\hour:\ifnum\minute<10 0\fi\number\minute}

\def\draftlabel#1{{\@bsphack\if@filesw {\let\thepage\relax
   \xdef\@gtempa{\write\@auxout{\string
      \newlabel{#1}{{\@currentlabel}{\thepage}}}}}\@gtempa
   \if@nobreak \ifvmode\nobreak\fi\fi\fi\@esphack}
        \gdef\@eqnlabel{#1}}
\def\@eqnlabel{}
\def\@vacuum{}
\def\draftmarginnote#1{\marginpar{\raggedright\scriptsize\tt#1}}

\def\draft{\oddsidemargin -0.1truein
        \def\@oddfoot{\sl preliminary draft \hfil
        \rm\thepage\hfil\sl\today\quad\militarytime}
        \let\@evenfoot\@oddfoot \overfullrule 3pt
        \let\label=\draftlabel
        \let\marginnote=\draftmarginnote
   \def\@eqnnum{{\rm (\theequation)}\rlap{\kern\marginparsep\tt\@eqnlabel}%
\global\let\@eqnlabel\@vacuum}  }

\newdimen\linethick  \linethick=0.4pt
\newdimen\hboxitspace    \hboxitspace=5pt
\newdimen\vboxitspace    \vboxitspace=5pt

\def\fr#1{%
\beq\new
\vcenter{
\hrule height\linethick
           \hbox{\vrule width\linethick
                 \kern\hboxitspace
                 \vbox{\kern\vboxitspace
                       \hbox{$\begin{array}{c}\displaystyle#1
          \end{array}$}%
                       \kern\vboxitspace}%
                 \kern\hboxitspace
                 \vrule width\linethick}%
           \hrule height\linethick}%
\eeq}

\newdimen\Squaresize \Squaresize=14pt
\newdimen\Thickness \Thickness=0.5pt

\def\Square#1{\hbox{\vrule width \Thickness
   \vbox to \Squaresize{\hrule height \Thickness\vss
      \hbox to \Squaresize{\hss#1\hss}
   \vss\hrule height\Thickness}
\unskip\vrule width \Thickness}
\kern-\Thickness}

\def\Vsquare#1{\vbox{\Square{$#1$}}\kern-\Thickness}

\def\numberbysection{\@addtoreset{equation}{section}
        \def\theequation{\thesection.\arabic{equation}}}
\numberbysection

\renewcommand{\theequation}{\thesection.\arabic{equation}}
\newcommand{\l@qq}[2]{\addvspace{2em}
 \hbox to\textwidth{\hspace{1em}\bf #1 \dotfill #2}}

\newcounter{app}

\def\app{\setcounter{equation}{0}
\def\theequation{\Alph{app}.\arabic{equation}}\par
   \addvspace{4ex}
   \@afterindentfalse
  \secdef\@app\@dapp}
\newcommand\@app{\@startsection {app}{1}{0ex}%
                                   {-3.5ex \@plus -1ex \@minus -.2ex}%
                                   {2.3ex \@plus.2ex}%
                                   {\normalfont\Large\bf}}

\def\@dapp#1{%
{\parindent \z@ \raggedright  \bf #1}\par\nobreak}
\def\l@app#1#2{\ifnum \c@tocdepth >\z@
    \addpenalty\@secpenalty
    \addvspace{1.0em \@plus\p@}%
    \setlength\@tempdima{2.5em}%
    \begingroup
      \parindent \z@ \rightskip \@pnumwidth
      \parfillskip -\@pnumwidth
      \leavevmode \bfseries
      \advance\leftskip\@tempdima
      \hskip -\leftskip
      #1\nobreak\hfil \nobreak\hb@xt@\@pnumwidth{\hss #2}\par
    \endgroup\fi}
\newcounter{sapp}[app]

\def\sapp{\def\theequation{\Alph{app}.\arabic{equation}}\par
   \@afterindentfalse
  \secdef\@sapp\@dsapp}
\newcommand\@sapp{\@startsection{sapp}{2}{\z@}%
                                     {-3.25ex\@plus -1ex \@minus -.2ex}%
                                     {1.5ex \@plus .2ex}%
                                     {\normalfont\large\bfseries}}

\def\@dsapp#1{%
{\parindent \z@ \raggedright  \bf #1}\par\nobreak}
\newcommand{\l@sapp}{\@dottedtocline{2}{1.5em}{3em}}

\newcounter{ssapp}[sapp]

\def\ssapp{\def\theequation{\Alph{app}.\arabic{equation}}\par
   \@afterindentfalse
  \secdef\@ssapp\@dssapp}
\newcommand\@ssapp{\@startsection{ssapp}{2}{\z@}%
                                     {-3.25ex\@plus -1ex \@minus -.2ex}%
                                     {1.5ex \@plus .2ex}%
                                     {\normalfont\normalsize\bfseries}}

\def\@dssapp#1{%
{\parindent \z@ \raggedright  \bf #1}\par\nobreak}
\newcommand{\l@ssapp}{\@dottedtocline{2}{1.5em}{3em}}

\def\titlepage{\@restonecolfalse\if@twocolumn\@restonecoltrue\onecolumn
     \else \newpage \fi \thispagestyle{empty}\c@page\z@
        \def\thefootnote{\fnsymbol{footnote}} }

\def\endtitlepage{\if@restonecol\twocolumn \else  \fi
        \def\thefootnote{\arabic{footnote}}
        \setcounter{footnote}{0}}  
\relax

\hybrid

\newtoks\@stequation

\def\subequations{\refstepcounter{equation}%
  \edef\@savedequation{\the\c@equation}%
  \@stequation=\expandafter{\theequation}
  \edef\@savedtheequation{\the\@stequation}
  \edef\oldtheequation{\theequation}%
  \setcounter{equation}{0}%
  \def\theequation{\oldtheequation\alph{equation}}}

\def\endsubequations{%
  \setcounter{equation}{\@savedequation}%
  \@stequation=\expandafter{\@savedtheequation}%
  \edef\theequation{\the\@stequation}%
  \global\@ignoretrue}

\makeatletter
\newdimen\normalarrayskip              
\newdimen\minarrayskip                 
\normalarrayskip\baselineskip
\minarrayskip\jot
\newif\ifold             \oldtrue            \def\new{\oldfalse}
\def\arraymode{\ifold\relax\else\displaystyle\fi} 
\def\@arrayskip{\ifold\baselineskip\z@\lineskip\z@
     \else
     \baselineskip\minarrayskip\lineskip1\baselineskip\fi}

\def\@arrayclassz{\ifcase \@lastchclass \@acolampacol \or
\@ampacol \or \or \or \@addamp \or
   \@acolampacol \or \@firstampfalse \@acol \fi
\edef\@preamble{\@preamble
  \ifcase \@chnum
     \hfil$\relax\arraymode\@sharp$\hfil
     \or $\relax\arraymode\@sharp$\hfil
     \or \hfil$\relax\arraymode\@sharp$\fi}}

\makeatother

\def\bse{\begin{subequations}}                
\def\ese{\end{subequations}}                 %

\begin{document}
\begin{titlepage}
\begin{flushright}
ITEP--TH--02/07\\
hep-th/0703123\\
February, 2007
\end{flushright}
\vspace{1.3cm}

\begin{center}
{\LARGE Wilson Loops in 2D Noncommutative  \\
Euclidean Gauge Theory: \\[.4cm] 
2. $1/\theta$ Expansion}\\
\vspace{1.4cm}
{\large Jan Ambj{\o}rn$^{a),\;c)}$, Andrei Dubin$^{b)}$ 
and Yuri Makeenko$^{a),\; b)}$}
\\[.8cm]
{$^{a)}$\it The Niels Bohr Institute,} \\
{\it Blegdamsvej 17, 2100 Copenhagen {\O}, Denmark}\\[.4cm]
{$^{b)}$\it Institute of Theoretical and Experimental Physics,}
\\ {\it B. Cheremushkinskaya 25, 117259 Moscow, Russia}\\[.4cm]
{$^{c)}$\it Institute for Theoretical Physics, 
Utrecht University,} \\
{\it Leuvenlaan 4, NL-3584 CE Utrecht, The Netherlands. 
}
\end{center}
\vskip 1.5 cm

\begin{abstract}

We analyze the $1/\theta$ and $1/N$ expansions of the Wilson loop
averages $<W(C)>_{U_{\theta}(N)}$ in the two-dimensional noncommutative
$U_{\theta}(N)$ gauge theory with the parameter of noncommutativity $\theta$.
For a generic rectangular contour $C$, a concise integral representation is
derived (non-perturbatively both in the coupling constant $g^{2}$ and in
$\theta$)
for the next-to-leading term of the $1/\theta$ expansion. In turn, in the limit
when ${\theta}$ is much larger than the area $A(C)$ of the surface bounded by
$C$, the large $\theta$ asymptote of this representation is argued to yield
the next-to-leading term of the $1/\theta$ series. For both of the
expansions, the next-to-leading contribution exhibits only a {\it power-like}\/
decay for areas $A(C)>>\sigma^{-1}$
(but $A(C)<<{\theta}$) much larger than the inverse of the string
tension $\sigma$ defining the range of the exponential decay of 
the leading term.
Consequently, for large $\theta$, it hinders a direct stringy interpretation
of the subleading terms of the $1/N$ expansion in the spirit of Gross-Taylor
proposal for the $\theta=0$ commutative $D=2$ gauge theory.

\end{abstract}

\end{titlepage}

\newpage

\section{Introduction}

 In short, given a commutative field theory defined in the Euclidean
space ${\bf R}^{D}$ by the action $S=\int d^{D}x~{\cal L}(\phi(x))$, the
corresponding noncommutative theory is implemented replacing the products of
the fields $\phi({\bf x})$ by the so-called star-products
introduced according to the rule%
\footnote{For a review see \cite{DN02,Sza03} and references therein.}
\be
\left( f_{1}\star{f_{2}}\right)({\bf x})\equiv
exp\left(-\frac{i}{2}\theta_{\mu\nu}\partial^{y}_{\mu}
\partial^{z}_{\nu}\right)~
f_{1}({\bf y})~f_{2}({\bf z})\Bigg|_{y=z=x}~,
\label{1.8}
\ee
where the parameter of noncommutativity $\theta_{\mu\nu}$, entering the
commutation relation $[x_{\mu},x_{\nu}]=-i\theta_{\mu\nu}$ satisfied by $D$
noncommuting coordinates, is real and
antisymmetric. In particular, the action of the standard $D-$dimensional
$U(N)$ Yang-Mills theory is superseded by 
\be
S=\frac{1}{4g^{2}}\int d^{D}x~
tr\left({\mathcal{F}}^{2}_{\mu\nu}({\bf x})\right)~~~~;~~~~~
{\mathcal{F}}_{\mu\nu}=\partial_{\mu}{\mathcal{A}}_{\nu}+
\partial_{\nu}{\mathcal{A}}_{\mu}
-i({\mathcal{A}}_{\mu}\star{\mathcal{A}}_{\nu}-{\mathcal{A}}_{\nu}\star
{\mathcal{A}}_{\mu})~,
\label{1.7}
\ee
where ${\mathcal{A}}_{\mu}\equiv{{\mathcal{A}}^{a}_{\mu}t^{a}}$ with
$tr(t^{a}t^{b})=\delta^{ab}$, and $\theta_{21}=-\theta_{12}=\theta$ in the
$D=2$ case in question.

The noncommutative two-dimensional $U_{\theta}(N)$ system (\ref{1.7}) provides 
the simplest example of the noncommutative gauge theory. As well as in the
$\theta=0$ case, investigation of non-perturbative effects in a
low-dimensional model is expected to prepare us for the analysis of a more
complicated four-dimensional quantum dynamics. An incomplete list of papers,
devoted to this direction of research, is presented in references 
\cite{BNT01}-\cite{RS07}.

The aim of the present work is to extend the perturbative analysis of our
previous publication \cite{ADM04} and examine, non-perturbatively in the
coupling constant $g^{2}$, the two alternative expansions
of the Wilson loop-average $<W(C)>_{U_{\theta}(N)}$ in the $D=2$
$U_{\theta}(N)$ theory on a plain. The first one is the $1/\theta$ series
\be
<W(C)>_{U_{\theta}(N)}=\sum_{k=0}^{\infty}~\theta^{-k}
<{\cal W}(C)>^{(k)}_{N}
\label{CO.02}
\ee
that is to be compared with the more familiar 't Hooft $1/N$
topological expansion
\be
<W(C)>_{U_{\theta}(N)}=\sum_{G=0}^{\infty}~
N^{-2G}<W(C)>^{(G)}_{U_{\theta}(1)}~,
\label{CO.01}
\ee
where $G$ can be identified with the genus of the auxiliary surface
canonically associated to any given diagram of the
weak-coupling series of the $N-$independent quantity
$<W(C)>^{(G)}_{U_{\theta}(N)}$. Also, the contour $C$ is always restricted
to be closed.

The $\theta\rightarrow{\infty}$ limit of
the $U_{\theta}(N)$ theory is known \cite{Filk} to retain the same set of the
planar diagrams (described by the same amplitudes) as the
$N\rightarrow{\infty}$ limit does so that
the leading terms of both of the above expansions coincide,
\be
<W(C)>^{(0)}_{U_{\theta}(N)}=<{\cal W}(C)>^{(0)}_{N}~,
\label{1.41a}
\ee
provided the appropriate identification of the coupling constants.
As the $G=0$ term of Eq. (\ref{CO.01}) is
$\theta-${\it independent}, it therefore reduces to
the corresponding average in the commutative variant of the gauge theory.
In consequence, the leading term of the series (\ref{CO.02}) reduces,
\be
<{\cal W}(C)>^{(0)}_{N}=<W(C)>^{(0)}_{U(N)}
~~~~~,~~~~~<{\cal W}(\Box)>^{(0)}_{N}=exp[-\sigma A(\Box)]~,
\label{1.41b}
\ee
to $G=0$ term of the $\theta=0$ expansion (\ref{CO.01}) of the average
$<W(C)>_{U(N)}$ in the ordinary commutative $U(N)$
gauge theory. In particular, it fits in the simple Nambu-Goto pattern for an
arbitrary non-self-intersecting contour $C$.

In this paper, for an arbitrary rectangular contour $C=\Box$, we evaluate the
next-to-leading term $<W(\Box)>^{(1)}_{U_{\theta}(1)}$ of the topological
expansion (\ref{CO.01}) and argue that its large $\theta$ asymptote
exactly reproduces,
\be
<{\cal W}(C)>_{N}^{(2)}=
\frac{1}{N^{2}}~\lim_{\theta\rightarrow{\infty}}~
\theta^{2}<W(C)>_{U_{\theta}(1)}^{(1)}~,
\label{CO.03}
\ee
the $k=2$ term $<{\cal W}(\Box)>^{(2)}_{N}$ of the $1/\theta$
series (\ref{CO.02}) (while $<{\cal W}(C)>^{(1)}_{N}=0$). The proof
of the relation (\ref{CO.03}) will be presented in a separate publication
\cite{ADM05b}. As for the computation of $<{\cal W}(C)>_{N}^{(2)}$, for this
purpose we perform a resummation of the genus-one diagrams
for a generic $C=\Box$, that is facilitated by the choice 
of the axial gauge where,
at the level of the $D=2$ action (\ref{1.7}), only tree-graphs (without
self-interaction vertices) are left. Nevertheless, the problem remains to be
nontrivial: due to the noncommutative implementation \cite{Ish}-\cite{DK01}
of the Wilson loop, an infinite number of different connected $G=1$
diagrams contributes to the average $<W(C)>_{U_{\theta}(N)}$ even in the case
of a non-self-intersecting contour $C$ that is in
contradistinction with the commutative case, where
$<W(\Box)>_{U(N)}=<W(\Box)>^{(0)}_{U(N)}$.
To deal with this problem, we propose a specific method of resummation. 

Application of the method allows to unambiguously split the whole set of the
relevant perturbative $G=1$ diagrams into the three subsets. Being
parameterized by the two integer numbers $r$ and $v$ with
$0\leq r\leq v\leq 1$, each
subset can be obtained starting with the corresponding protograph (with
$2+r-v$ lines) and then dressing it through the addition of extra lines in
compliance with certain algorithm. For a rectangle $C=\Box$, it yields an
integral representation of the $G=1$ term of the $1/N$ expansion in the form
\be
<W(\Box)>_{U_{\theta}(1)}^{(1)}~=~
\frac{1}{(2\pi\sigma\theta)^{2}}\sum_{0\leq r\leq v\leq 1}
h_{rv}{\cal Z}_{rv}(\bar{A},\bar{\theta}^{-1})~,
\label{SU.01z}
\ee
where ${\cal Z}_{rv}(\bar{A},\bar{\theta}^{-1})$ denotes the effective
amplitude which, after multiplication by the factor $h_{rv}=2+r-v$
separated for a later convenience, accumulates the entire $rv-$subset of the
perturbative amplitudes. Besides a dependence on $\bar{\theta}=\sigma\theta$,
${\cal Z}_{rv}(\cdot)$ depends only on the dimensionless area
$\bar{A}=\sigma RT$ of $C=\Box$ rather than separately on the lengths $T$ and
$R$ of the temporal and spatial sides of $\Box$.

Correspondingly, in the large $\theta$ limit,
\be
\theta>>A(C)~,
\label{LI.01}
\ee
the $N=1$ relation (\ref{CO.03}) can be rewritten as
\be
<{\cal W}(\Box)>_{1}^{(2)}=\frac{1}{(2\pi\sigma)^{2}}
\sum_{0\leq r\leq v\leq 1}h_{rv}{\cal Z}_{rv}(\bar{A},0)~,
\label{CO.03f}
\ee
where ${\cal Z}_{rv}(\bar{A},0)$ is obtained
from ${\cal Z}_{rv}(\bar{A},\bar{\theta}^{-1})$ (which is {\it continuous}\/ in
$\bar{\theta}^{-1}$ in a vicinity of $\bar{\theta}^{-1}=0$) simply
replacing\footnote{The peculiarity of this replacement 
is that it can {\it not}\/
be applied directly to the perturbative amplitudes describing individual
Feynman diagrams. It matches the observation \cite{ADM04} that the large
$\theta$ asymptote of the leading perturbative contribution to
$<W(C)>_{U_{\theta}(1)}^{(1)}$ scales as $\theta^{0}$ rather than as
$\theta^{-2}$. In turn, it implies a nontriviality of the relation
(\ref{CO.03}).} $\bar{\theta}^{-1}$ by zero. Then, performing the Laplace
transformation with respect to $\bar{A}$, the image
$\tilde{\cal Z}_{rv}(\beta,0)$ of the large $\theta$ asymptote
${\cal Z}_{rv}(\bar{A},0)$ assumes the concise form
\be
\tilde{\cal Z}_{rv}(\beta,0)=\frac{1}{(\beta+1)^{2}}
\int\limits_{-\infty}^{+\infty}
d\bar\zeta d\bar\eta~ \frac{{\cal K}_{rv}(\bar\zeta,\bar\eta)}
{(\beta+|1-\bar\zeta|)^{h_{rv}-1}~(\beta+|1+\bar\eta|)~
(\beta+|1+\bar\eta-\bar\zeta|)}~,
\label{LA.03}
\ee
where 
\be
{\cal K}_{rv}(\bar\zeta,\bar\eta)=
\sum_{{e}_{3}=-r}^{0}\sum_{{e}_{1}=-1}^{v-r}
\sum_{{e}_{2}=v}^{1}(-1)^{v+\sum_{k=1}^{3}{e}_{k}}~
2^{(v-r)(1-|{e}_{1}|)}|{e}_{1}+\bar\zeta|~
|{e}_{2}+\bar\eta|^{1-v}~|{e}_{3}+\bar\zeta|^{r}~.
\label{MUL.01a}
\ee
The integral representation (\ref{LA.03}) is the main result of the paper.

Building on the latter representation, one concludes that the pattern of the
$\theta\neq{0}$ expansion (\ref{CO.01})
shows, especially in the limit (\ref{LI.01}), a number of features which are
in sharp contrast with the $1/N$ expansion of the average in the $\theta=0$
case. Indeed, in the latter
case, the Nambu-Goto pattern (\ref{1.41b}) provides the
exact result $<W(\Box)>_{U(N)}=<W(\Box)>^{(0)}_{U(N)}$ for an arbitrary
non-self-intersecting loop $C$, and the
corresponding subleading terms are vanishing: $<W(C)>_{U(N)}^{(G)}=0$ for
$G\geq{1}$. Furthermore, for
self-intersecting contours $C$, nonvanishing subleading $G\geq{1}$ terms
$<W(C)>_{U(N)}^{(G)}$ all possess \cite{Kaz&Kost} the area-law asymptote
like in Eq. (\ref{1.41b}) in the limit $\bar{A}\rightarrow{\infty}$.

When $\theta\neq{0}$, even for a rectangular loop $C$, the 
pattern of $<W(C)>_{U_{\theta}(N)}$ is characterized by an infinite
$1/N-$series, each $G\geq{1}$ term of which nontrivially depends both on
$\bar\theta$ and on $\bar{A}(C)$. In addition, we present simple arguments
that, in contradistinction with Eq. (\ref{1.41b}), the asymptote
(\ref{CO.03f}) of the next-to-leading term  exhibits a power-like (rather
than exponential) decay for areas $\sigma^{-1}<<A(\Box)<<{\theta}$ much larger
than the string tension $\sigma$.  This
asymptote is evaluated in \cite{ADM05b} with the result
\be
\frac{1}{N^{2}}<W(\Box)>_{U_{\theta}(1)}^{(1)}~\longrightarrow~
\frac{4}{\pi^2\left(\sigma\theta N\right)^{2}}~
\frac{ln(\sigma{A})}{\sigma{A}}~~~~~,~~~~~
\sigma\theta,~\sigma{A}~\longrightarrow{~\infty}~,
\label{FA.13b}
\ee
that can be traced back to the (infinite, in the
limit $\theta\rightarrow{\infty}$) {\it nonlocality}\/ of the star-product
(\ref{1.8}) emphasized in the discussion \cite{Minw} of the $UV/IR$ mixing.
Due to the generality of the reasoning, all the subleading $G\geq{1}$
coefficients $<W(C)>_{U_{\theta}(1)}^{(G)}$ are as well expected to show,
irrespectively of the form of $C$, a power-like decay for
$\sigma^{-1}<<A(C)<<{\theta}$. In particular, it precludes a straightforward
stringy interpretation of the subleading terms of the expansion (\ref{CO.01})
in the spirit of Gross-Taylor proposal \cite{Gr&Tayl} for the $\theta=0$
commutative $D=2$ gauge theory.

In Section \ref{gen}, we put forward a concise form (\ref{1.31b})
of the perturbative $2n-$point functions, the loop-average
$<W(C)>_{U_{\theta}(1)}$ is composed of in the $D=2$ $U(1)$ theory
(\ref{1.7}). In Section \ref{deform1}, it is sketched how these functions are
modified under the two auxiliary (genus-preserving) deformations of a given
diagram to be used for the derivation of the decomposition (\ref{SU.01z}).
To put the deformations into action, in Section \ref{parameter}, we introduce
a finite number of the judiciously selected elementary genus-one graphs and
propose their $\gamma jrv-$parameterization.

Then, any remaining nonelementary $G=1$ perturbative diagram can be obtained
through the appropriate multiple application of the latter deformations to one of
thus selected elementary graphs. When a
particular elementary diagram with a given
$\gamma jrv-$assignment is dressed by all its admissible deformations, the
corresponding perturbative $2n-$point function is replaced by the effective
one, as it is shown in Section \ref{repres1}. The replacement is implemented
in such a way that certain $n-v$ propagators of the are superseded by their
effective counterparts (\ref{KEY.01}). The integral representation of the
effective $2n-$point functions, is completed in Section \ref{effective}.

In Section \ref{effective1}, we express the $G=1$ term
$<W(\Box)>_{U_{\theta}(1)}^{(1)}$ of the expansion (\ref{CO.01})
as a superposition of the effective amplitudes (\ref{FR.01}) that are obtained
when the arguments of the above $2n-$point functions are integrated over the
rectangle $C=\Box$. The effective amplitudes can be
collected into the three $rv-$superpositions
${\cal Z}_{rv}(\bar{A},\bar{\theta}^{-1})$ associated
to the corresponding protographs parameterizing the decomposition
(\ref{SU.01z}). The explicit expression (\ref{MUL.04}) for 
${\cal Z}_{rv}(\cdot)$ is then derived. It is observed
that, for a fixed $rv-$specification, this expression
can be deduced directly through the appropriate dressing of the
$rv-$protograph. The derivation of the large $\theta$ representation
(\ref{LA.03}) is sketched in Section \ref{prescr}.
Conclusions, a brief discussion
of the perspectives, and implications for $D=3,4$ gauge theory (\ref{1.7})
are sketched in Section \ref{conclus}. Finally, the Appendices contain
technical details used in the main text.

\section{Generalities of the perturbative expansion}
\label{gen}

Building on the integral representation
of the $U_{\theta}(1)$ average, we begin with a sketch of the derivation of
the relevant perturbative $2n-$point functions.

\subsection{Average of the noncommutative Wilson loop}

To this aim, consider the perturbative expansion of the average of the
noncommutative Wilson loop~\cite{Ish}
\be
W(C)={\mathcal{P}}e_{\star}^{
i\oint_{C} dx_{\mu}(s) {\mathcal{A}}_{\mu}({\bf x}(s))}~.
\label{1.3a}
\ee
in the $U_{\theta}(N)$ noncommutative gauge theory on the 2D plane
${\Bbb R^{2}}$. For this purpose, it is sufficient to use the path-integral
representation~\cite{ADM04} of the $U_{\theta}(1)$ average
\be                                                          
<W(C)>_{U_{\theta}(1)}=\Bigg<exp\left(-\frac{1}{2}
\oint\limits_{C} dx_{\mu}(s)\oint\limits_{C} dx_{\nu}(s')
D_{\mu\nu}({\bf x}(s)-{\bf x}(s')+{\bf \xi}(s)-
{\bf \xi}(s'))\right) \Bigg>_{\xi(\tilde{s})}~,
\label{1.1}
\ee
as it follows from the $N-$independence of
the quantities $<W(C)>^{(G)}_{U_{\theta}(N)}$ which are, therefore, replaced
by $<W(C)>^{(G)}_{U_{\theta}(1)}$ in Eq. (\ref{CO.01}). In Eq. (\ref{1.1}),
$D_{\mu\nu}({\bf z})$ is the standard $D=2$ photon's propagator
in the axial gauge ${\mathcal{A}}_{1}=0$,
\be
D_{\mu\nu}({\bf z})=<{\mathcal{A}}_{\mu}({\bf z})
{\mathcal{A}}_{\nu}({\bf 0})>_{U(1)}=
-\frac{g^{2}}{2}~\delta_{\mu 2}\delta_{\nu 2}~|z_{1}|~\delta(z_{2})~,
\label{1.2}
\ee
and the functional averaging over the auxiliary $\xi_{\mu}(s)$ field 
(parameterized by
the proper time $s\in[0,1]$ chosen to run clockwise starting
with the left lower corner of $C=\Box$) is to be performed according to the
prescription
\be
\Bigg<{\mathcal{B}}[{\bf \xi}(s)]\Bigg>_{\xi(\tilde{s})}=
\int {\mathcal{D}}\xi_{\mu}(s)~e^{\frac{i}{2}(\theta^{-1})_{\mu\nu}
\int ds ds' \xi^{\mu}(s)G^{-1}(s,s')
\xi^{\nu}(s')}~{\mathcal{B}}[{\bf \xi}(s)]~.
\label{1.4}
\ee
Here, ${\mathcal{D}}\xi_{\mu}(s)$ denotes the standard flat measure so that
$<\xi^\mu (s) \xi^\nu (s')>= i\theta^{\mu\nu} {\rm sign}\,(s-s')/2$,
where, prior to the regularization, we
are to identify $G^{-1}(s,s')=\dot \delta(s-s')$.

Let us also note that Eq. (\ref{1.4}) is based on the integral
representation   
\be
exp\left(-\frac{i}{2}~{\theta}_{\mu\nu}
{\partial^{{\bf x}}_{\mu}}
{\partial^{{\bf y}}_{\nu}}\right)f_{1}({\bf x})~f_{2}({\bf y})=
\int e^{2i ({\theta}^{-1})_{\mu\nu} \xi_1^\mu\xi_2 ^\nu }
f _1 ({\bf x}+{\bf \xi}_1)  f _2 ({\bf y}+{\bf \xi}_2)~
\prod_{j=1}^{2}\frac{d^{2}\xi^{\mu}_{j}}{w(\theta)}
\label{IR.01}
\ee
of the star-product (\ref{1.8}), where $w(\theta)=
(\pi^2|\det {\theta}|)^{1/2}$. In consequence, the noncommutative Wilson
loop~\rf{1.3a} itself can be represented as~\cite{Oku99},
\be
W(C)=\Bigg<exp\left(i\oint\limits_{C} dx_{\mu}(s)
{\mathcal{A}}_{\mu}({\bf x}(s)+{\bf \xi}(s))\right) \Bigg>
_{\xi(\tilde{s})}~.
\label{1.3}
\ee

Finally, the coupling $g^{2}$ of the $U_{\theta}(N)$
noncommutative gauge theory is related with the string tension $\sigma$, 
entering \eq{1.41b},  by the formula
\be
\sigma=g_{U_{\theta}(N)}^{2}N/2~.
\label{1.41d}
\ee

\subsection{Perturbative $\theta$-dependent $2n-$point functions}

Take any given $n$th order diagram of the weak-coupling expansion of
the average (\ref{1.1}) that, being applied to the $1/N$ series
(\ref{CO.01}) can be rewritten in the form
\be
<W(C)>_{U_{\theta}(N)}^{(G)}=\sum_{n=0}^{\infty}\lambda^{2n}~
<W(C)>_{U_{\theta}(N)}^{(G,n)}
\label{NE.02}
\ee
with $\lambda=g^{2}N$. For a particular $n\geq 2G$,
$<W(C)>_{U_{\theta}(N)}^{(G,n)}$ is given by the multiple contour integral
of the $\xi$-average applied to the
corresponding product of $n$ $\xi$-dependent propagators
$D_{\mu\nu}({\bf y}_{l}+\xi(s_{l})-\xi(s'_{l}))$, where
\be
{\bf y}_{l}={\bf x}(s_{l})-{\bf x}(s'_{l})~,
\label{1.38aa}
\ee
with $l=1,2,...,n$. Then, any diagram can be topologically visualized as the
collection of the {\it oriented}\/ (according to the proper-time
parameterization) lines so that the $q$th propagator-line starts at a given
point ${\bf x}(s'_{q})\in C$ and terminates at the corresponding 
${\bf x}(s_{q})\in C$.
When the $\xi$-averaging of the product is performed, the perturbative
$2n-$point function can be rewritten~\cite{ADM04} in the form
\be
V^{(n)}_{U_{\theta}(1)}({\bf y}_{1},...,{\bf y}_{n})=o_{n}~
\prod_{1\leq l<j}^{n}exp\left(\frac{i}{2}~{\cal C}_{lj}\breve{ \theta}_{\mu\nu}
{\partial^{{\bf z}_{l}}_{\mu}}
{\partial^{{\bf z}_{j}}_{\nu}}\right){D}_{22}({\bf z}_{1})~
{D}_{22}({\bf z}_{2})~...~{D}_{22}({\bf z}_{n})
\Big|_{\{{\bf z}_{k}={\bf y}_{k}\}}~,
\label{1.31b}
\ee
where\footnote{In the computation of any $2n$th order perturbative diagram,
the factor $o_{n}$ disappears. The subfactor $2^{-n}$ is exactly cancelled by
the symmetry factor responsible for the interchange of two different
end-points of each of the $n$ lines. By the same token, the subfactor factor
$1/n!$ is precisely cancelled by the symmetry factor corresponding to all
possible permutations of the $n$ different (non-oriented) lines. Finally,
$(-1)^{-n}$ is to be combined with the implicit factor $(-1)^{-n}$ that
arises when one  pulls  the minus sign out of each propagator (\ref{1.2})
entering $V^{(n)}_{U_{\theta}(1)}(\cdot)$.}
$o_{n}=(-1/2)^{n}/n!$, and the {\it intersection}\/ matrix
${\mathcal{C}}_{lj}=-{\mathcal{C}}_{ji}$, being defined algebraically as
\be
{\mathcal{C}}_{lj}=\frac{1}{2}\left(sign(s_{l}-s_{j})+
sign(s'_{l}-s'_{j})-sign(s_{l}-s'_{j})-
sign(s'_{l}-s_{j})\right)~,
\label{1.34}
\ee
counts the number of times the $l$th oriented line
crosses over the $j$th oriented line (and, without loss of generality,
we presume that $s_{l}\geq s'_{l}$ for $\forall{l}$).
As for the relevant noncommutative parameter $\breve{ \theta}_{\mu\nu}$, it is
{\it twice}\/ larger
\be
\breve{ \theta}_{\mu\nu}=2{\theta}_{\mu\nu}~,
\label{1.26}
\ee
compared to the parameter ${\theta}_{\mu\nu}$ defining the original
star-product (\ref{1.8}). Being rewritten in the momentum space,
Eq. (\ref{1.31b}) implies that, compared to the commutative case, a given
$\theta\neq 0$ perturbative $2n-$point function is assigned with the extra
$\theta-$dependent factor
\be
\Bigg<\prod_{k=1}^{n}
e^{i{\bf p}_{k}\cdot({\bf \xi}(s_{k})-{\bf \xi}(s'_{k}))}
\Bigg>_{\xi(\tilde{s})}=
exp\left(-i\sum_{l<j}{\mathcal{C}}_{lj}~\theta_{\mu\nu}~
p_{l}^{\mu}~p_{j}^{\nu}\right)~.
\label{1.33}
\ee
where the momentum ${\bf p}_{l}$ is canonically conjugated to the $l$th
coordinate (\ref{1.38aa}). In turn, the r.h. side of Eq. (\ref{1.33})
reproduces the existing formula \cite{Filk,Minw} obtained
in the analysis of the partition function in the noncommutative
field-theories.

Finally, the pattern of Eqs. (\ref{1.33}) and (\ref{1.34}) suggests the
natural definition of (dis)connected diagrams. Algebraically, a particular
$n$th order graph is to be viewed as disconnected in the case when the
associated $n\times n$ matrix ${\mathcal{C}}_{lj}$ assumes a
{\it block-diagonal}\/ form ${\mathcal{C}}_{lj}=
\otimes_{k}{\mathcal{C}}^{(k)}_{l_{k}j_{k}}$, with $\sum_{k}n_{k}=n$,
so that the nonvanishing entries of ${\mathcal{C}}_{lj}$ are reproduced
exclusively by smaller $n_{k}\times n_{k}$ matrices
${\mathcal{C}}^{(k)}_{l_{k}j_{k}}$, where $n_{k}<n$ for $\forall{k}$.
Conversely, when a nontrivial implementation of this decomposition
of a particular ${\mathcal{C}}_{lj}$ is impossible, the corresponding diagram
is called connected. As the rank $r[{\cal C}]=2G(\{{\bf y}_{l}\})$ 
of the matrix
${\cal C}_{ij}$ is known to be equal to the doubled genus
$G(\{{\bf y}_{l}\})$ of the diagram, one expects that the order $n$ of a
connected genus $G$ graph complies with the inequality
$n(\{{\bf y}_{l}\})\geq{2G(\{{\bf y}_{l}\})}$.

\section{The two deformations and the irreducible diagrams}
\label{deform1}

The aim of this Section is to present the central elements of
the exact resummation%
\footnote{The details of this resummation procedure 
will be published elsewhere.}  
of the weak-coupling series applied to
the noncommutative Wilson loop average that, in turn, leads to the
decomposition (\ref{SU.01z}) introducing the parameters $r$ and $v$. For this
purpose, observe first that the complexity of the perturbative expansion of
the considered average roots in the complexity of the perturbative
$2n-$point functions (\ref{1.31b}) associated to the connected graphs (of
an arbitrary large order) discussed in the end of the previous Section.
In consequence, for a connected graph of order $m\geq 2G$, the $2m-$point
function (\ref{1.31b}) can be expressed in the simplest cases
as a multiple irreducible star-product
$f_{1}\star{f_{2}}\star ...\star{f_{m}}$, where the
quantities $f_{k}(\cdot)$ are composed of the propagators (\ref{1.2}). (In
general, the pattern of the connected $2n-$point function can be deduced
according to the prescription discussed in the beginning of subsection
\ref{deform1c}.) In particular, it can be shown that
$2\leq m\leq{3}$ for
$G=1$, while Eq. (\ref{1.31b}) for a generic $G=1$ diagram
can be represented in the form of an ordinary product of a single $m$th order
star-product and a number of the propagators (\ref{1.2}).

To put this representation into use, we introduce the two
{\it genus-preserving}\/ deformations to be called ${\cal R}_{a}^{-1}-$ and
$\bar{\cal R}_{b}^{-1}-$deformations. Increasing the order $n$ of a
$G\geq{1}$ graph by one, they relate the corresponding pairs of functions
(\ref{1.31b}) in a way that does {\it not}\/ change the multiplicities of all
the irreducible star-products involved. Correspondingly, with respect to the
inverse ${\cal R}_{a}-$ and $\bar{\cal R}_{b}-$deformations, one introduces
${\cal R}_{a}\otimes\bar{\cal R}_{b}-$irreducible Feynman diagrams.

The advantage of the construction is that
nonvanishing amplitudes (\ref{1.31b}) are associated only to the {\it finite}\/
number of the irreducible diagrams depicted in figs. 1, 2, 7a, and 7e
(which are postulated to fix the {\it topology}\/ of the attachment of the
lines' end-points to the upper and lower horizontal sides
of $C=\Box$). Then, the complete set of the 
{\it connected}\/ genus-one diagrams can be generated applying all possible
$\bar{\cal R}_{b}^{-1}-$deformations to all $m$ lines of these irreducible
diagrams. In the end of the Section, we discuss a reason for a further
refinement of the resummation algorithm. Being implemented via certain
dressing of the lines of the so-called elementary (rather than irreducible)
diagrams, the algorithm prescribes that, for any of the dressed lines of
the latter diagram, (in the relevant amplitude) one replaces the perturbative
propagator by a concise effective propagator (\ref{KEY.01}).


\subsection{The ${\cal R}_{a}^{-1}-$deformations}
\label{deform1a}

The first one is what we call
${\cal R}_{a}^{-1}-$deformation when a given elementary graph is modified by
the addition of an extra $i$th propagator-line that does
not intersect\footnote{The condition (\ref{1.50a}) refers to such
implementation of the decomposition ${\mathcal{C}}_{lj}=
\otimes_{k}{\mathcal{C}}^{(k)}_{l_{k}j_{k}}$ when, except for a single factor
${\mathcal{C}}^{(q)}_{l_{q}j_{q}}$, all the remaining factors are 
one-dimensional.} any line in the original $\{k\}$-set of the
elementary diagram:
\be
{\cal C}_{ik}=0~~~~~~~~,~~~~~~~~\forall{k\neq{i}}~.
\label{1.50a}
\ee
Starting with a given $2n-$point function $V^{(n)}_{U_{\theta}(1)}(\cdot)$ and
identifying $i=n+1$, one readily obtains that, modulo the numerical constant,
the considered deformation of
$V^{(n)}_{U_{\theta}(1)}(\cdot)$ merely multiplies it by the extra
propagator,
\be
V^{(n+1)}_{U_{\theta}(1)}({\bf y}_{1},...,{\bf y}_{n},{\bf y}_{n+1})=
-\frac{1}{2(n+1)}~V^{(n)}_{U_{\theta}(1)}({\bf y}_{1},...,{\bf y}_{n})~
{D}_{22}({\bf y}_{n+1})~.
\label{DEF.01}
\ee
Correspondingly, one defines the inverse of the
${\cal R}_{a}^{-1}-$deformation as the ${\cal R}_{a}-$deformation which
eliminates such an $i$th line of a (${\cal R}_{a}-$reducible)
diagram that complies with Eq. (\ref{1.50a}) for some $k$. In the absence of
such a line (for any $k$), the graph is called
${\cal R}_{a}-$irreducible\footnote{Note
that any connected diagram is necessarily ${\cal R}_{a}-$irreducible.}.

Concerning an $m-$fold application of the ${\cal R}_{a}^{-1}-$deformation,
the corresponding generalization of Eq. (\ref{DEF.01}) is routine: the single
factor $-{D}_{22}({\bf y}_{n+1})/2(n+1)$
is replaced by the product
$\prod_{k=1}^{m}(-1){D}_{22}({\bf y}_{n+k})/2(n+k)$.
I.e. all of thus generated extra lines are assigned
(as well as in the $\theta=0$ commutative gauge theory) with the ordinary
perturbative propagator (\ref{1.2}). For example, a generic
non-elementary  ${\cal R}_{a}^{-1}-$deformation
of the graph in fig. 1a is described by the diagram in fig. 3a.
In the latter figure the additional lines
are depicted by dotted lines which are vertical
(i.e. characterized by vanishing relative time) owing to
the pattern of the latter propagator. More generally, the
vertical lines in figs. 5, 6, 8, and 9 are also generated by the admissible
multiple ${\cal R}_{a}^{-1}-$deformations of the corresponding elementary
graphs.

Finally, by the same token as in the $\theta=0$
case, it is straightforward to obtain that the ${\cal R}_{a}^{-1}-$dressing
of a given connected graph results in the multiplication of the amplitude,
associated to this graph, by a factor to be fixed by Eq. (\ref{EX.01h})
below.

\subsection{The $\bar{\cal R}_{b}^{-1}-$deformations}
\label{deform1b}

The second one is what we denote as the $\bar{\cal R}_{b}^{-1}-$deformation of a
given $k$th line of the elementary graph (when the remaining lines of this
graph are defined as the $\{q\}_{k}-$set) that introduces an extra line,
labeled by $i$, so that the following twofold condition is fulfilled. To
begin with, one requires that the $k$th and the $i$th
lines, being mutually non-intersecting, intersect the
$\{q\}_{k}-$set in the topologically equivalent way (modulo possible
reversion of the orientation). E.g. the $\bar{\cal R}_{b}^{-1}-$copies of the
right and left solid horizontal lines in fig. 1a are
depicted by parallel (owing to the condition (\ref{FA.09}) below)
dotted lines in figs. 4b and 4c respectively. In general, it can be formalized
by the condition
\be
{\cal C}_{ik}=0~~~~~,~~~~~
{\cal C}_{iq}=\alpha_{ik}~{\cal C}_{kq}~~~~~~~~,~~~~~~~~\forall{q}\neq{k,i}~,
\label{1.50}
\ee
where, depending on the choice of the relative orientation of the $i$th line,
the $q$-independent constant $\alpha_{ik}\equiv{\alpha_{i,k}}$ is equal to $1$
or $-1$ (with $\alpha^{(r)}_{11}=1$). Additionally, it is convenient
to impose that thus introduced extra line should {\it not}\/ be
horizontal, i.e., both its end-points are not attached to the same horizontal
side (along the second axis) of the rectangle $C$. 
As for the inverse transformation, the
$\bar{\cal R}_{b}-$deformation deletes such an $i$th line of a
diagram that Eq. (\ref{1.50a}) holds true. Correspondingly, any line of a
$\bar{\cal R}_{b}-$irreducible graph has {\it no}\/
$\bar{\cal R}_{b}^{-1}-$copies in the sense of the above twofold condition.

Next, identifying $\alpha_{n,n+1}=\alpha^{(n)}$, one obtains that the
$\bar{\cal R}_{b}^{-1}-$deformation of (\ref{1.31b}) results in the
$2(n+1)-$point function
\be
V^{(n+1)}_{U_{\theta}(1)}({\bf y}_{1},...,{\bf y}_{n},{\bf y}_{n+1})=
\label{DEF.02}
\ee
$$
=o_{n+1}~
\prod_{1\leq l<j}^{n}e^{\frac{i}{2}~{\cal C}_{lj}\breve{ \theta}_{\mu\nu}
{\partial^{{\bf z}_{l}}_{\mu}}
{\partial^{{\bf z}_{j}}_{\nu}}}{D}_{22}({\bf z}_{1})~
...{D}_{22}({\bf z}_{n-1})\left[{D}_{22}({\bf z}_{n})~
{D}_{22}(({\bf z}_{n}-{\bf y}_{n})\alpha^{(n)}+{\bf y}_{n+1})\right]
\Big|_{\{{\bf z}_{k}={\bf y}_{k}\}}~.
$$
which is expressed through the original $n\times n$ intersection matrix
${\cal C}_{lj}$. In view of the pattern (\ref{1.2}) of the propagator, Eq.
(\ref{DEF.02}) implies that
\be
\alpha^{(n)}y^{2}_{n}=y^{2}_{n+1}
\label{DEF.03}
\ee
In turn, it entails that the $\bar{\cal R}_{b}^{-1}-$copy of a given line
spans the same time-interval (fixed by the second component $y^{2}_{l}$ of
the relative distance (\ref{1.38aa})) as the latter
line does. In full generality, this property is expressed by 
Eq.~(\ref{GP.01}) that crucially simplifies the computations.

Next, the multiple\footnote{In what follows, a composition of multiple
${\cal R}_{a}^{-1}-$ and $\bar{\cal R}_{b}^{-1}-$deformations (associated to
some lines of an elementary graph) is, in short,
denoted as ${\cal R}_{a}^{-1}\otimes\bar{\cal R}_{b}^{-1}-$deformation of a
graph.} application of the $\bar{\cal R}_{b}^{-1}-$deformations (\ref{1.50}),
introduces an extra $\{i_{a}\}-$set of the lines which,
intersecting nether each other nor the $k$th line, fulfill the
$i\rightarrow i_{a}$ option of Eq. (\ref{1.50}).
Then, to reproduce the replacement (\ref{KEY.01}), one should take advantage
of the following reduction.
When applied to an elementary graph, the (multiple)
$\bar{\cal R}_{b}^{-1}-$deformations result in diagrams described by
vanishing amplitudes unless they are constrained by a particular
$\{\alpha^{(r)}\}-$assignment. The amplitude (\ref{1.31b})
may be nonvanishing only when, for any given $r$th line of the
elementary graph, all its $\bar{\cal R}_{b}^{-1}-$copies (if any) are
assigned with {\it one and the same}\/ value of the parameter
\be
\alpha^{(r)}_{l1}=\alpha^{(r)}~~~~~,~~~~~\forall{l}\geq{2}~,~\forall{r}~,
\label{GP.01}
\ee
where $\alpha^{(r)}_{lk}=\pm 1$ enters the implementation of Eq. (\ref{1.50})
corresponding to the $r$th line.

Let us also note that another useful property of the
$\bar{\cal R}_{b}^{-1}-$deformations which generalizes the relation
(\ref{DEF.03}). The corresponding implementations of the $2n-$point function
(\ref{1.31b}) enforce that
\be
y^{2}_{k}=\alpha^{(k)}_{l1}y^{2}_{k,l}~~~~~,~~~~~
\forall{l=2,...,n_{k}}~~~,~~~\forall{k}~,
\label{FA.09}
\ee
where ${\bf y}_{k,l}$ denotes the relative distance (\ref{1.38aa})
corresponding to the $l$th $\bar{\cal R}_{b}^{-1}-$copy of the $k$th line
(described by ${\bf y}_{k}\equiv{\bf y}_{k,1}$).
That is why, for $C=\Box$, the latter copies are depicted by such straight
dotted lines which are mutually {\it parallel}\/ like dotted lines in figs. 4b
and 4c.

\subsubsection{The necessity for a further refinement}
\label{deform1c}

According to above, given a generic connected graph, the pattern of the
corresponding  $2n-$point function can be deduced from Eq. (\ref{1.31b}) via
the replacement ${D}_{22}({\bf z}_{k})\rightarrow{f_{k}({\bf z}_{k},...)}$.
Here, $f_{k}({\bf z}_{k},...)$ takes into account possible
$\bar{\cal R}_{b}^{-1}-$dressing of the $k$th line of the irreducible diagram
(described by the matrix ${\cal C}_{lj}$ of rank
$n$) which is associated to the connected graph in question via a sequence
of $\bar{\cal R}_{b}-$deformations. Conversely, once a line of an irreducible
diagram may be dressed by conglomerates of
$\bar{\cal R}_{b}^{-1}-$copies characterized only by an unambiguous value of
the corresponding $\alpha^{(k)}$, (in the computation of the amplitude) the
overall $\bar{\cal R}_{b}^{-1}-$dressing
of this line results in the replacement of the associated perturbative
propagator by its effective counterpart to be fixed by the $f_{k}=1$ option of
Eq. (\ref{KEY.01}).

Still, the shortage of the resummation algorithm, built on the irreducible
diagrams, is that some of the time-ordered components of the latter diagrams
possess a single line which may be assigned with {\it different}\/ signs of
$\alpha^{(r)}$. In consequence, the concise prescription of the modification
(\ref{KEY.01}) of the propagator can not be directly applied to such a line.
To circumvent this problem, we use an alternative
prescription to reproduce the complete set of the
connected $G=1$ diagrams. The idea is to introduce the larger set of the
elementary {\it time-ordered}\/ graphs (belonging to the three $rv-$varieties
in accordance with the decomposition (\ref{SU.01z})) and properly change the
algorithm of their $\bar{\cal R}_{b}^{-1}-$dressing so that a single line of
some elementary graphs is not dressed at all. As for the overall dressing of
each of the remaining lines, being characterized by an {\it unambiguous}\/ sign
of the corresponding $\alpha^{(k)}$, it is as previously fixed by the
$f_{k}=1$ option of the replacement (\ref{KEY.01}).

In this way, the set of the genus-one diagrams
(generated by the perturbative expansion of the average (\ref{1.1}))
can be unambiguously decomposed into a finite number of subsets parameterized
by the elementary graphs. Then, each subset is described by the associated
effective $2n-$point function that, therefore, accumulates the overall
${\cal R}_{a}^{-1}\otimes \bar{\cal R}_{b}^{-1}-$dressing of the corresponding
{\it elementary}\/ connected $G=1$ graph.

\section{The parameterization of the elementary graphs}
\label{parameter}


Let us realize the program formulated in subsection \ref{deform1c} and
introduce such parameterization of the elementary graphs that is
as well applicable after the overall
${\cal R}_{a}^{-1}\otimes \bar{\cal R}_{b}^{-1}-$dressing of these graphs.
To this aim, the set of the elementary time-ordered graphs is postulated to
include not only all time-ordered components of the 
${\cal R}_{a}\otimes\bar{\cal R}_{b}-$irreducible Feynman
diagrams in figs. 1, 2, 7a, and 7e, but also a variety of a few connected
$\bar{\cal R}_{b}-${\it reducible}\/ graphs associated to certain components of
the diagrams in figs. 1c and 2e. The additional graphs are obtained from
the (time-ordered components of the) diagrams in figs. 7a and 7e via the
vertical reattachments applied to the leftmost or/and rightmost end-points of
each of the latter diagrams. 
{\it Preserving}\/ both the time-coordinates of the
latter end-points
and the intersection-matrix (modulo possible change of the sign of its
entries), the reattachments replace the single end-point of one or both
their horizontal lines from one horizontal side of $C$ to another.
Modulo the reflection interchanging the horizontal 
sides of $C=\Box$, the additional diagrams
are depicted in the remaining figs. 7. Note that all these
extra diagrams\footnote{Observe also that, in view of the constraint
(\ref{DEF.03}), the geometry of these diagrams implies the additional
constraint on the relative time-ordering of their end-points. E.g., in figs.
7c and 7g the lower leftmost point must be to the right with respect to the
upper leftmost point.} possess exactly one pair of the lines which, being
labeled by $i$ and $k$, comply with the condition (\ref{1.50}).
Also, the discussion below implicitly takes into account that both the
elementary graphs in the figs. 2a and 2b and all their deformations, being
assigned with {\it vanishing}\/ amplitudes (\ref{1.31b}), can be
therefore excluded from the analysis.

\subsection{$S(4)-$symmetry and reflection-invariance}
\label{S(4)}

To properly enumerate the elementary graphs and introduce their
$\gamma jrv-$parameterization, we should first discuss
two types of the transformations which relate the elementary graphs in such a
way that the structure of the overall
${\cal R}_{a}^{-1}\otimes\bar{\cal R}_{b}^{-1}-$dressing is kept intact.
In turn, to facilitate the application of these transformations, we are to
postulate the following convention. When the elementary graphs
(or protographs, see subsection \ref{assignment1}) are associated
to one and the same time-ordered component of a given Feynman diagram, they
are nevertheless considered to be {\it different}, provided the topology of
the attachment of their lines' end-points (to the upper and lower horizontal
sides of $C=\Box$) is different. For example, the pairs of distinct graphs
are depicted in figs. 1a, 1b and figs. 1c, 1d respectively.

Turning to the transformations of the elementary graphs, the first type is
implemented through the vertical reattachments which can be
combined to generate $S(4)-${\it multiplets}\/ of the latter graphs.
Consisting of four graphs, each such multiplet implements the discrete space
of the $S(4)-$group\footnote{When applied simultaneously to this $4-$set of
the graphs, the reattachments can be used to generate the $4!$ elements of
the group itself.} of permutations. Note that not only the
elementary graphs but also their deformations, included into the subsets
described by the corresponding effective amplitudes, are
unambiguously splitted into a finite number of distinct (non-overlapping)
$S(4)-$multiplets. An example is given by the diagrams in figs. 5c and 6a--6c,
where the bold lines depict the associated elementary graph while
the nonvertical dotted lines represent the $\bar{\cal R}_{b}^{-1}-$copies
introduced by the $\bar{\cal R}_{b}^{-1}-$deformations.
As it is illustrated by the latter four figures, the required symmetry of the
dressing is maintained by the condition that the positions of the end-points
of all the $\bar{\cal R}_{b}^{-1}-$copies are left intact.

As for the second type of the transformations, the $S(4)-$multiplets
of the elementary graphs may be related
via the $S(2)-$reflection that (mapping
the contour $C=\Box$ onto itself) mutually interchanges the two
horizontal (or, what is equivalent in the $v=1$ case, vertical)
sides of $C$.

Finally, one can implement the $S(4)\otimes S(2)-$transformations to combine
the {\it dressed}\/ (by all admissible
${\cal R}_{a}^{-1}\otimes\bar{\cal R}_{b}^{-1}-$deformations) elementary
graphs into the $S(4)\otimes S(2)-$multiplets.
The prescription reads that both
the vertical reattachments and the reflections are as previously applied only
to the lines associated to the elementary graph, leaving intact the
positions of the end-points of all ${\cal R}_{a}^{-1}-$ and
$\bar{\cal R}_{b}^{-1}-$copies of these
lines. (E.g., figs. 5c, 6a--6c represent the four members of the 
$S(4)-$multiplet of the dressed graphs to be assigned with $r=v=0$, see
below.)

\subsection{The $\gamma jrv-$parameterization of the multiplets}
\label{assignment}


Both prior and after their dressing by all admissible
${\cal R}_{a}^{-1}\otimes\bar{\cal R}_{b}^{-1}-$deformations, the elementary
graphs are convenient to collect into the $\gamma jrv-$varieties of the
$S(4)-$multiplets which consist of $h_{rv}=2+r-v$ $S(4)-$multiplets
related
via one of the two types of $S(2)-$reflections discussed above. Then, the four
integer numbers $\gamma,~j,~r$, and $v$ parameterize
$S(4)\otimes S(2)-$multiplets of the diagrams so that
the relevant geometry of the multiple deformations of the graphs in each such
multiplet is specified in the reflection- and
reattachment-{\it invariant}\/ way.

As a result, the algorithm of the resummation can be decomposed into the two
steps. At the first step, for given values of $\gamma$, $j$, $r$, and $v$, one
constructs the $h_{rv}=2+r-v$ effective amplitudes parameterized
by certain special elementary graphs related (when $h_{rv}=2$) via the
$S(2)-$reflections. In turn, possessing the {\it maximal}\/ number
(equal to $h_{rv}$) of the horizontal lines attached to $1+r$ different
horizontal sides of $C=\Box$, each of the latter
$h_{rv}$ graphs enters the corresponding $S(4)-$multiplet in the
$\gamma jrv-$variety. In turn, these are precisely the
lines that constitute the associated protograph
(relevant for the decomposition (\ref{SU.01z})) which, in the $rv-$variety of
the $S(4)-$multiplets (with different $\gamma$ and $j$), has the maximal
amount of the horizontal lines. This construction
is sketched below (see also Appendix \ref{enumer1}).

At the second step, the remaining three elementary graphs of each
$S(4)-$multiplet (as well as the rest of the protographs) can
be then reproduced are obtained via the vertical reattachments of the
leftmost or/and rightmost end-points of the above $2+r-v$ horizontal lines.
Modulo possible change of the sign of its entries, the intersection-matrix is
invariant under these $S(4)-$transformations since, by construction of the
elementary graphs, the leftmost and rightmost points of the entire graph
necessarily belong to the $h_{rv}$ lines defining the corresponding
protograph.

\subsubsection{The topological $jrv-$parameterization and the protographs}
\label{assignment1}

Consider first the integers $j,~r,$ and $v$ which can be interpreted
directly in terms of the relevant topological properties {\it common}\/ for all
the graphs in a given $S(4)-$multiplet. To begin with, $r=0,1$ is equal to
the number of the ${\cal R}_{b}^{-1}-$copies\footnote{The definition of this
type of the deformation can be obtained from the
one of the $\bar{\cal R}_{b}^{-1}-$deformation omitting the requirement that
the extra line, introduced according to (\ref{1.50}), is necessarily
nonhorizontal.} of the graph with the maximal number of a single line (to be
identified below) available in a given elementary graph.
Correspondingly, all the graphs associated to figs. 7 are assigned with
$r=1$ (while $v=1$ since $1\geq v\geq{r}$), while the remaining 
diagrams in figs. 1 and 2, are parameterized by $r=0$.

Next, the number of the lines is equal to $n=1+j+r$. In turn,
for a given $r$, $j+1=n-r=2,3$
yields the multiplicity of the irreducible star-product the form of which
assumes both the corresponding perturbative $2n-$point function (\ref{1.31b}),
and, owing to the replacement (\ref{KEY.01}), its effective counterpart
considered in
subsection \ref{bdeform}. Therefore, the graphs in figs. 1 and 7a--7d are
assigned with $j=1$, while the remaining elementary diagrams are assigned
with $j=2$.

As for $v=0,1$ (with $0\leq r\leq v\leq 1$ in compliance with 
Eq.~(\ref{SU.01z})), the $n$th order elementary graph has $h_{rv}=2+r-v$
lines which may be involved into the vertical $S(4)-$reattachments
{\it without}\/ changing (the module of) 
the entries of the intersection-matrix.
Furthermore, only $2+r-2v$ of these reattached lines
are to be dressed, together with the remaining $n-h_{rv}$ lines, by the
$\bar{\cal R}_{b}-$copies in compliance with
the $f_{k}=1$ prescription (\ref{KEY.01}). 
In particular, for $v=r=1$ graphs of figs. 7, it
is the single line, devoid of the latter type of the dressing, that is
considered to possess one ${\cal R}_{b}-$copy. (Alternatively, one may state
that both the latter line and its copy share the same
common $\bar{\cal R}_{b}-$dressing.)

For each of the reattached lines, $2-(v-r)$ of its end-points
are $S(4)-$transformed. Therefore, for cases other than $r=1-v=0$, the
reattachments can be faithfully represented by the parameters%
\footnote{One is to identify
$R={x}^{1}(s_{l})-{x}^{1}(s'_{l})$,
$s_{l}>s'_{l}$, with the spatial component of the relative distance
(\ref{1.38aa}) such that ${\bf x}(s_{l})$ and ${\bf x}(s'_{l})$ belong
respectively to the lower and upper horizontal sides of the contour $C=\Box$.
In turn, in view of the proper time parameterization fixed prior to Eq.
(\ref{1.4}), it implies that the vertical $1-$axis is to be directed from the
upper to the lower horizontal side of the rectangle $\Box$.}
$a_{k}=0,1$ so that $y^{1}_{k}=a_{k}R$, with $k$ assuming $h_{rv}=2$
different values. In the $r=1-v=0$ case (when $h_{rv}=1$), in addition to
$a_{1}$ we have to introduce the extra parameter $\tilde{a}_{1}$,
\be
x^{1}(s'_{1})=(1-a_{1})\tilde{a}_{1}R~~~~~,~~~~~
x^{1}(s_{1})=\tilde{a}_{1}R+(1-\tilde{a}_{1})a_{1}R~,
\label{ST.01}
\ee
which is equal to $1$ and $0$ depending on whether or
not the reattachment involves the left(most) end-point of the horizontal
line of fig. 2e (while $x^{1}(s_{1})-x^{1}(s'_{1})=a_{1}R$). To simplify the
notations, the pair of the parameters, used to represent
the reattachments, is denoted as $\{a_{k}\}\equiv \{a_{k}\}_{rv}$ for all
$0\leq r\leq v\leq 1$.


Next, the parameters $r$ and $v$ can be used to enumerate the protographs
which are time-ordered as well. A particular protograph, can be
reconstructed eliminating all the $n-h_{rv}$ lines of the corresponding
elementary graph except for the $2+r-v$ lines affected by the
$S(4)-$reattachments. Modulo the $S(4)-$reattachments,
thus separated protographs are depicted by bold lines in
figs. 3a--3e for those protographs which, for a given $rv-$assignment possess
the maximal number $2+r-v$. The figs. 3a, 3b and 3d,3e are in one-to-one 
correspondence
with the pairs of the $S(4)-$multiplets which, being related via the
reflection (interchanging the horizontal sides of the contour $C$), are
characterized by $r=v=0$ and $r=v=1$ respectively. It should be stressed that,
to avoid double-counting, one is to consider only such reflections of the
protographs which can not be alternatively reproduced
by the vertical $S(4)-$reattachments. Correspondingly,
fig. 3c refers to the single $r=v-1=0$
multiplet\footnote{The $v=1$ protograph in fig. 3c
should {\it not}\/ be accompanied by the reflection-partner 
which, being defined 
by the requirement that both end-points of the single line are
attached to the lower side of C, can be alternatively obtained via the
composition of the two vertical reattachments. Note also is that the genus of
the $v=1$ protographs is zero rather than one which explains
why we have to start from the elementary graphs rather than directly from the
protographs.}.

In sum, there are
precisely $h_{rv}=2+r-v$ $S(4)-$multiplets of the protographs which, being
parameterized by a particular $rv-$assignment, are
related via the $S(4)-$reflections.
In compliance with Appendix \ref{enumer1}, in each such multiplet the labels
of the $\bar{\cal R}_{b}-$dressed lines assume $n-v=r+j-v+1$ different values
in the set $\Omega_{jrv}$ obtained from the sequence
$1+v,~2,~1+j,~2+2r$ via the identification of the $v+(2-j)+(1-r)=4+v-n$
coinciding entities (all being equal to 2) so that
$n-v=\sum_{k\in{\Omega_{jrv}}}1$. Correspondingly, to parameterize
the entire set of $n$ lines, we reliable these lines introducing the
the set $\tilde\Omega_{jrv}$ obtained from the sequence $1,~2,~1+j,~2+2r$
via the identification of the $(2-j)+(1-r)=4-n$
coinciding entities so that $n=\sum_{k\in \tilde{\Omega}_{jrv}}1$.
In turn, the labels of the $2+r-v$ lines, involved into the
$S(4)-$reattachments, assume values in the set ${\cal S}_{rv}$
obtained from the sequence $1,~c_{rv}=2+3r-v$ (with $c_{rv}=1,2,4$)
via the identification of the $v-r$ coinciding entities.





\subsubsection{The residual $\gamma-$parameterization of the elementary
graphs}
\label{gamspecif}

The necessity to complete the $jrv-$parameterization and introduce one more
parameter $\gamma$, additionally specifying the
$S(4)\otimes S(2)-$multiplets, is motivated by the
geometry of the pairs of the elementary graphs depicted by bold lines in
figs. 8a, 8b (both characterized by $j=r+1=v=1$) and 9a, 9b 
(both characterized by
$j=r=v=1$).
In general, with the help of this parameter $\gamma=1,...,f_{jrv}$, 
one is to enumerate distinct $S(4)\otimes S(2)-$multiplets of the graphs
which are separated when one fixes the parameters $j$, $r$, and $v$.

The previous discussion suggests that, to find the number $f_{jrv}$ of such
multiplets, it is sufficient to consider only those
elementary graphs which, representing the corresponding
multiplet, possess the {\it maximal}\/ number $h_{rv}$ of the horizontal lines
for a particular $jrv-$specification. Then, $f_{jrv}$ is equal to the number
of such graphs which, being different time-ordered components of the same
Feynman diagram, are {\it not}\/ related via the $S(4)\otimes 
S(2)-$transformations.
In turn, the latter number can be found in the following way.

To begin with, for a particular $jrv-$assignment and the matrix
${\cal C}_{ik}$, one fixes generic positions both of the $2+r-v$ horizontal
lines (involved into the $S(4)-$reattachments) and of the upper end-points of
the remaining $n-(2+r-v)=j-1+v$ nonhorizontal lines. Actually, it is
straightforward to infer from Eqs.~(\ref{DEF.03}) and (\ref{1.50}) that
$f_{jrv}$ is $r-$independent, $f_{jrv}\equiv f_{jv}$, which allows to deduce
$f_{jv}$ restricting our analysis the $r=0$ cases. 
Next, we should take into account that
both by the perturbative and the associated effective $2n-$point functions
impose $n-2$ specific constraints (see Eq. (\ref{1.40kk}) below) on
admissible combinations of
the temporal components $y^{2}_{l}$ of the relative distances (\ref{1.38aa}).
In consequence, among the $j-1+v$ lower end-points of the nonhorizontal lines,
only $v$ points remain to be independent degrees of freedom (in addition to
the $2h_{rv}|_{r=0}+(j-1+v)$ ones already fixed above).

When $v=0$, obviously
$f_{j0}=1$ for $j=1,2$. As for $v=1$, when we vary the position of the lower
end-point of $k$th nonhorizontal line representing the residual $v=1$ degree
of freedom, the resulting time-ordered components of the transformed diagram
are distinguished by the $\left[\otimes_{i=1}^{j}
Sign(y^{2}_{i})\right]/S(j)-$assignment\footnote{$Sign(y^{2}_{i})$ denotes
the sign-function depending on the relative time $y^{2}_{i}$ (associated to
the $i$th nonhorizontal line) which may be changed through the variation of
the lower end-point in question.} defined {\it modulo}\/ possible
$S(j-1+v)-$permutations of the labels $i$ of the $(j-1+v)|_{v=1}$
nonhorizontal lines. Therefore, $f_{j1}=(j-1)+2(2-j)$, where it is formalized
that for $f_{11}=(v+1)|_{v=1}$, while $f_{21}=f_{11}-1=1$ (as it is
clear from fig. 2e). Summarizing, one arrives at the formula
\be
f_{jv}=(1-v)+(3-j)v~~~~~;~~~~~f_{j0}=f_{2v}=1~~,~~\forall{j,v}~,
\label{SU.01l}
\ee
so that $1\leq f_{jv}\leq 2$, and $f_{jv}=2$ only when $j=v=1$.

Note that, although the construction
of the parameter $f_{jv}$ is obviously reflection-invariant, the action of
the reflections on the $S(4)-$multiplets of the elementary graphs
is still nontrivial in the cases when $f_{jv}=2$. The reflections, introduced
in subsection \ref{assignment1}, in these cases relate those of the latter
multiplets which, being endowed with the same $jrv-$assignment, are
described by the two different values of $\gamma$.
Summarizing, there are precisely $h_{rv}=2+r-v$
$S(4)-$multiplets of the elementary graphs which, being
parameterized by a particular $\gamma jrv-$assignment,
are related via the $S(4)-$reflections in the $j-${\it independent}\/ way.

\section{Dressing of the elementary graphs and protographs}
\label{repres1}

To derive the representation (\ref{SU.01z}), the first step is
to express $<W(\Box)>_{U_{\theta}(1)}^{(1)}$ in terms of the $S(4)-$multiplets
of the effective $2n-$point functions. Each of these functions describes
the corresponding elementary graph together with all its admissible
${\cal R}_{a}^{-1}-$ and $\bar{\cal R}_{b}^{-1}-$deformations
according to the algorithm sketched in the previous Section.

In view of the factorization (\ref{DEF.01}), for $C=\Box$ it is convenient to
represent the effective functions as the product
${\cal I}^{(n)}(\{{\bf y}_{k}\})
\tilde{V}_{U_{\theta}(1)}^{(n)}(\{{\bf y}_{k}\})$, where $\{{\bf y}_{k}\}$
denotes the set of the relative coordinates (\ref{1.38aa}) characterizing the
corresponding time-ordered elementary graph of a given order $2n$. In
particular, the factor ${\cal I}^{(n)}(\cdot)$ (to be defined in
Eq. (\ref{EX.01h})) accumulates the overall ${\cal R}_{a}^{-1}-$dressing of
the latter graph. As for
$\tilde{V}_{U_{\theta}(1)}^{(n)}(\cdot)$, it
describes a given elementary graph together with the entire its
$\bar{\cal R}_{b}^{-1}-$dressing in the way consistent with the
$S(4)\otimes S(2)-$symmetry. In turn, the quantity
$\tilde{V}_{U_{\theta}(1)}^{(n)}(\cdot)$ can be introduced as the 
concise modification of the corresponding elementary $2n-$point function
(\ref{1.31b}). For this purpose, the perturbative propagators of certain
$n-v$ lines should be replaced by the effective ones defined by the $f_{k}=1$
option of Eq. (\ref{KEY.01}).

Parameterizing the effective functions, the elementary graphs can be viewed as
the {\it intermediate}\/ collective coordinates which are useful in the
computation of the corresponding individual effective amplitudes
\be
\frac{1}{(2\pi\bar{\theta})^{2}}~
{\cal Z}^{(\gamma)}_{jrv}(\{a_{k}\},\bar{A},\bar{\theta}^{-1})=
\oint\limits_{C}\prod_{l=1}^{n}dx^{2}(s_{l})dx^{2}(s'_{l})~
{\cal I}^{(n)}(\{{\bf y}_{k}\})
\tilde{V}_{U_{\theta}(1)}^{(n)}(\{{\bf y}_{k}\})\Big|^{\gamma}_{jrv}
\label{FR.01w}
\ee
where the vertical $S(4)-$reattachments of the $h_{rv}=2+r-v$ lines are
described by the set of the parameters $\{a_{k}\}\equiv\{a_{k}\}_{rv}$
introduced in subsection \ref{assignment1}. By
construction, the $G=1$ term $<W(\Box)>_{U_{\theta}(1)}^{(1)}$
of the expansion (\ref{CO.01}) can be represented as the superposition of
the amplitudes (\ref{FR.01w})
which, as we will see, should be combined into the $rv-$superposition
${\cal Z}_{rv}(\{a_{k}\},\cdot)$. These superpositions are obtained
summing up the amplitudes (\ref{FR.01w}) corresponding to all $\sum_{j=1}^{2}
f_{jv}$ elementary graphs which are associated to a given
$rv-$protograph according to the prescription
discussed in subsection \ref{assignment1}. Then, Eq.~(\ref{SU.01z}) is
reproduced provided
\be
{\cal Z}_{rv}(\bar{A},\bar{\theta}^{-1})=\sum_{\{a_{l}\}_{rv}}
{\cal Z}_{rv}(\{a_{k}\},\bar{A},\bar{\theta}^{-1})~~~~~,~~~~~
{\cal Z}_{rv}(\{a_{k}\},\bar{A},\bar{\theta}^{-1})=
\sum_{j=1}^{2}\sum_{\gamma=1}^{f_{jv}}
{\cal Z}^{(\gamma)}_{jrv}(\{a_{k}\},\bar{A},\bar{\theta}^{-1})~,
\label{SU.01}
\ee
where $f_{jv}$ is defined in Eq. (\ref{SU.01l}), and the sum over
$\{a_{l}\}_{rv}$ includes
the contribution of the four
terms related through the $S(4)-$reattachments applied to the 
end-points of those lines which, being associated to the corresponding
protograph, are parameterized by the label $l\in {\cal S}_{rv}$ (where the set
${\cal S}_{rv}$ is introduced in the end of subsection \ref{assignment1}). 
In this
way, ${\cal Z}_{rv}(\bar{A},\bar{\theta}^{-1})$ yields the contribution of
the $S(4)-$multiplet of the protographs endowed with the $rv-$assignment.
(Eq. (\ref{SU.01}) takes into account that, when $h_{rv}=2$, the dressed
elementary diagrams can be collected into the pairs related via the
reflection
which leaves the amplitudes
${\cal Z}_{rv}(\bar{A},\bar{\theta}^{-1})$ invariant, as it is verified in
Appendix \ref{symmetry3}.)

Then, building on the pattern of the perturbative functions
(\ref{1.31b}), the amplitude ${\cal Z}_{rv}(\{a_{k}\},\cdot)$ can be
reproduced in a way which reveals the important reduction resulting in the
final set of the collective coordinates that, in turn, supports
the relevance of the protographs.
We postpone the discussion of this issue till subsection \ref{completeness}.

\subsection{Introducing an explicit time-ordering}
\label{ordering}


To proceed further, it is convenient to
reformulate both ${\cal I}^{(n)}(\{{\bf y}_{i}\})$ and
$\tilde{V}_{U_{\theta}(1)}^{(n)}(\{{\bf y}_{i}\})$ in terms of a minimal
amount of independent variable arguments instead of the $n-$set
$\{{\bf y}_{i}\}$ of the relative distances (\ref{1.38aa}). Consider a
rectangular contour $C=\Box$ such that $R$ and
$T$ denote the lengths of its vertical and horizontal sides
which, in the notations of Eq. (\ref{1.2}), are parallel
respectively to the first and the second axis. Then, the subset
$\{y^{1}_{i}\}$ can be reduced to $\{a_{k}\}$, defined after 
Eq.~(\ref{SU.01}). Concerning the reduction of the remaining subset
$\{y^{2}_{i}\}$, it can be shown
that, among the $\delta-$functional constraints imposed by the $G=1$ effective
$2n-$point function $\tilde{V}_{U_{\theta}(1)}^{(n)}(\{{\bf y}_{k}\})$, there
are $n-2$ constraints which are the same as in the case of the
perturbative $2n-$point function of the associated elementary diagram,
\be
\tilde{V}_{U_{\theta}(1)}^{(n)}(\cdot)\sim
{V}_{U_{\theta}(1)}^{(n)}(\{{\bf y}_{l}\})
\sim \prod_{p=1}^{n-2G}
\delta\left(\sum_{l=1}^{n}\lambda^{(p)}_{l}~y^{2}_{l}\right)~,
\label{1.40kk}
\ee
where the $n-2G$ $n$-vectors $\lambda^{(p)}_{l}$,
depending only on the topology of the associated elementary graph, span the
subspace of those eigenvectors of the intersection matrix
${\mathcal{C}}_{kl}$ which possess vanishing eigenvalue:
$\sum_{l=1}^{n}{\mathcal{C}}_{kl}\lambda^{(p)}_{l}=0$ for $p=1,2,...,n-2G$.

In view of the latter constraints imposed on the
temporal components $y^{2}_{l}$,
there are only $m=n+2$ independent time-ordered
parameters $\tau_{k}\geq{\tau_{k-1}}$ which can be 
chosen to replace the set of the
$2n$ temporal coordinates $x^{2}(s_{l})$ and
$x^{2}(s'_{l})$ assigned to the line's end-points of a given elementary graph.

Then, as it is shown in Appendix \ref{freedom}, there is a simple
geometrical prescription to introduce $\tau_{k}$ (with $\tau_{0}=0,~
\tau_{m+1}=T$). In particular, there are $m-v$ parameters $\tau_{i}$ that
are directly identified with the properly associated coordinates
$x^{2}(\cdot)$ in the latter $2n-$set. E.g., in the simplest cases
to figs. 1a  and 2c, the $m$ parameters $\tau_{i}$ simply relabel,
according to the relative {\it time-ordering}\/ (see Eq. (\ref{FA.02y})), the
set of the $m$ end-points attached in these two figures to the upper side
of $C=\Box$. Then, combining the $S(4)-$reattachments and the reflection, this
prescription can be generalized to deal with the two reflection-pairs of the
$r=v=0$ $S(4)-$multiplets of the graphs endowed with $j=1$ and $j=2$.
Correspondingly, for $v=1$ in Appendix \ref{freedom} we propose,
besides $m-1$ such reidentifications associated to certain side of $\Box$, a
simple geometrical operation to represent the remaining quantity
$\tau_{i_{0}}$ as a superposition of three temporal coordinates
$x^{2}(\cdot)$. Furthermore,
the proposed $v=1$
prescription can be formulated in the way invariant both under the
$S(4)-$reattachments and under the reflection.

Next, on the upper (or, alternatively, lower) horizontal side
of the rectangle $C$, $\tau_{q}$ and $\tau_{q-1}$ can be viewed as the
bordering points of the $n+3$ connected nonoverlapping intervals
\be
\Delta \tau_{k-1}=\tau_{k}-\tau_{k-1}~\geq{~0}~~~~~,~~~~~
\sum_{k=0}^{m}\Delta \tau_{k}=T~~~~~,~~~~~m=n+2=j+r+3~,
\label{FA.02w}
\ee
the overall time $T$ is splitted into. Consequently, the previously introduced
relative times\footnote{As the temporal intervals $y^{2}_{l}$ are overlapping in
general, it hinders a resolution the $G=1$ constraints (\ref{1.40kk})
directly in terms of
these intervals. On the other hand, the $n>2$ splitting (\ref{FU.02}) yields
the concise form to represent, for any given time-ordered component of a
Feynman graph, the latter resolution employing the {\it nonoverlapping}\/
intervals $\Delta\tau_{k}$.}
\be
{t}^{(\gamma)}_{p}\equiv y^{2}_{p}\Big|^{\gamma}_{jrv}=
(-1)^{s_{p}(\{a_{k}\})}\sum_{l=0}^{m}d^{(\gamma)}_{jrv}(p,l)~
\Delta{\tau}_{l}~~~~~,~~~~~
d^{(\gamma)}_{jrv}(p,0)=d^{(\gamma)}_{jrv}(p,m)=0~,
\label{FU.02}
\ee
can be represented as the superpositions of $\Delta{\tau}_{k}$, and for
simplicity we omit the superscript $jrv$ in the notation ${t}^{(\gamma)}_{p}$.
In Eq. (\ref{FU.02}), $d^{(\gamma)}_{jrv}(p,l)=0,\pm 1$, while
$s(\{a_{k}\})$ denotes an integer-valued $p-$dependent function (fixed by Eq.
(\ref{LAT.01}) in Appendix \ref{symmetry3}) which additionally depends on the
set $\{a_{k}\}$ of the variables introduced after Eq. (\ref{ST.01}).

It is noteworthy that, provided $f_{jv}=2$, the
$\gamma-$dependence of ${t}^{(\gamma)}_{p}$ (together with the implicit
$\gamma-$dependence of the auxiliary parameter
$\omega^{(\gamma)}_{k}$ introduced in Eq. (\ref{KEY.01b})) is the
only source of the $\gamma-$dependence of the effective amplitude
${\cal Z}^{(\gamma)}_{jrv}(\cdot)$, see
Eqs. (\ref{SR.01g}) and (\ref{SR.01f}) below, associated to the elementary
graphs in a $S(4)-$multiplet with a given $\gamma jrv-$assignment.
(E.g., the geometry of figs. 8a and 8b implies
that $t_{2}^{(1)}<0$ and $t_{2}^{(2)}>0$ respectively.)
In turn, the $S(4)-$symmetry of the dressing of the elementary graphs
guarantees that the pattern of $t_{p}^{(\gamma)}$ (as well as that of
$\omega^{(\gamma)}_{k}$ in Eq. (\ref{KEY.01b})) is the same for
all members of each $S(4)-$multiplet.


\subsection{Accumulating the ${\cal R}_{a}^{-1}-$deformations}
\label{adeform}

Given a particular effective amplitude, the associated
${\cal R}_{a}^{-1}-$deformations a given elementary graph are generated via
{\it all}\/ possible inclusions of such extra lines that, in accordance with
Eq. (\ref{1.50a}), intersect neither each other nor the original lines of the
graph. In figs. 3--9 the latter extra lines are depicted as vertical (due to
the $\delta-$function in the perturbative propagator (\ref{1.2})) and dotted.
In view of Eq. (\ref{DEF.01}), the inclusion of the
deformations of this type merely multiplies
the amplitude, describing the original elementary graph, by a factor
${\cal I}^{(n)}(\{{\bf y}_{k}\})$.
To deduce this factor, we note that the temporal coordinates of the
upper 
end-points of the ${\cal R}_{a}^{-1}-$copies may span only the first and
the last intervals $\Delta \tau_{0}$ and $\Delta \tau_{n+2}$ respectively.
Then, akin to the commutative case, it is straightforward to obtain
that, when the superposition of all the admissible
${\cal R}_{a}^{-1}-$copies is included, it result in 
\be
{\cal I}^{(n)}(\{{\bf y}_{k}\})=
exp\left(-\sigma |R|\left[\Delta \tau_{0}+\Delta \tau_{n+2}\right]\right)~,
\label{EX.01h}
\ee
the $\tau_{j}-$dependence of which matches the pair of the conditions
(\ref{FU.02}) imposed on $d^{(\gamma)}_{jrv}(p,l)$.

\subsection{The $\bar{\cal R}_{b}^{-1}-$deformations and the effective
propagators}
\label{propag}

Next, consider the block $\tilde{V}_{U_{\theta}(1)}^{(n)}(\cdot)$ that
results after the $\bar{\cal R}_{b}^{-1}-$dressing of a given elementary
graph with $n$ lines. The short-cut way to
reconstruct this block is to specify those $n-v$ (with $v=0,1$) lines of
the latter graph where the corresponding propagator (\ref{1.2}) is replaced,
in the relevant implementation of Eq. (\ref{1.31b}), by its effective
counterpart so that the $S(4)-$symmetry
of the overall dressing is maintained. When the $k$th line
is dressed by all
admissible $\bar{\cal R}_{b}^{-1}-$deformations, the replacement is fixed by
the $f_{k}=1$ option of the substitution
\be
{D}_{22}({\bf z}_{k})~\longrightarrow~\left(-\sigma |z^{1}_{k}|\right)^
{f_{k}}\delta\left(z^{2}_{k}\right)~
exp\left(-\sigma |R+(z^{1}_{k}-y^{1}_{k})\alpha^{(k)}|
\Delta T^{b}_{k}(f_{k},\gamma) \right)~~~~,~~~~k\in{\Omega_{2rv}}~,
\label{KEY.01}
\ee
that reduces to the ordinary multiplication of the propagator
${D}_{22}(\cdot)$ by the $k-$dependent exponential factor (with the set
$\Omega_{jrv}$ of the $n-v$ different labels being defined in the end of
subsection \ref{assignment1}). In this factor,
the parameter $\alpha^{(k)}=\pm 1$ is traced back to Eq.
(\ref{DEF.02}), and each elementary graph may
be endowed only with a single $\{\alpha^{(i)}\}-$assignment (with
$i\in{\Omega_{jrv}}$) which renders the global topological characteristic
(\ref{GP.01}) of the dressing (\ref{KEY.01}) {\it unambiguous}. Also,
\be
\Delta T^{b}_{k}(f_{k},\gamma)=\Delta \tau_{q(k)}+
[1-f_{k}]\Delta \tau_{q(k)+2\omega^{(\gamma)}_{k}-1}~~~~~~~,~~~~~~~~
\Delta T^{a}_{q}=\Delta \tau_{q}~~,~~q=0,n+2~,
\label{KEY.01b}
\ee
(i.e., $f_{0}=f_{n+2}=1$), while the extra subscripts $a$ and $b$ are
introduced to indicate the type of the associated deformations. In turn,
in the $f_{k}=1$ case the interval $\Delta T^{b}_{k}(1,\gamma)=
\Delta \tau_{q(k)}$ is spanned by the end-points of the
$\bar{\cal R}_{b}^{-1}-$copies of the $k$th line (see Appendix \ref{enumer1}
for particular examples). As for the function $q(k)$ defining the label of
the corresponding interval (\ref{FA.02w}), it formally determines an
embedding
of an
element of the $S(n-v)$ group of permutations into the $S(n+1)$ group:
$0<q(k)<n+2$ for all different $n-v$ values of $k\in{\Omega_{jrv}}$. In
Appendix \ref{freedom}, we sketch
a simple rule which allows to reconstruct $q(k)$ so that
this function is common for any given $S(4)-$multiplet of
elementary graphs.

\subsubsection{The completeness of the $\bar{\cal R}_{b}^{-1}-$dressing of
the protographs}
\label{completeness}

To explain the relevance for a certain $f_{k}\neq{1}$ option of the
replacement (\ref{KEY.01}), we should take into account that, in the
evaluation of the amplitude
${\cal Z}_{rv}(\{a_{k}\},\cdot)$ defined by Eq. (\ref{SU.01}), there are
important cancellations between the contributions of
the individual effective amplitudes (\ref{FR.01w}).
To obtain ${\cal Z}_{rv}(\{a_{k}\},\cdot)$ for a particular $rv-$assignment,
it is sufficient to take the {\it single}\/ $j=2$ elementary graph
(unambiguously associated to the corresponding protograph) and apply,
according to a judicious $\{f_{k}\}-$assignment, the
replacement (\ref{KEY.01}) to the same $n-v$ lines of this graph as
previously. But, contrary to the computation of the amplitudes (\ref{FR.01w}),
the $f_{k}=1$ variant of the replacement (\ref{KEY.01}) remains to be applied
only to the $2+r-2v$ lines involved into the $S(4)-$reattachments.
The point is that $f_{k}=0$ for all the $(v+j-1)|_{j=2}$ lines which, being
not affected by the reattachments (while labeled by
$k\in\Omega_{2rv}/{\cal S}_{rv}$, i.e., $k=3$ and, when $v=1$,
$k=3-v$), therefore do {\it not}\/ belong to the
protograph. Furthermore, it is accompanied by such reduction of the measure
that, implying the a specific fine-tuning (\ref{KEY.01a}), retains relevant 
variable arguments which, at least in the simpler $r=v=0$ case, are
associated only to the corresponding protograph. It is further discussed in
subsection \ref{colcoor}, where a more subtle situation for other values of
$r$ and $v$ is also sketched.

An explicit ($rv-$dependent) construction of the auxiliary
parameters $\omega^{(1)}_{k}=0,1$ with $k=3-v,~3$ (and $\gamma=1$ owing to
the $j=2$ option of Eq. (\ref{SU.01l}))
is presented in subappendix \ref{omega} so
that $0<q(k)+2\omega^{(1)}_{k}-1<n+2$, i.e.,
$\Delta T^{b}_{k}(f_{k},1)\cap \Delta T^{a}_{0}=
\Delta T^{b}_{k}(f_{k},1)\cap \Delta T^{a}_{2+n}=0$, $\forall{k
\in{\Omega_{2rv}}}$.
Then, the emphasized above cancellations result, for any admissible
$\{f_{k}\}-$assignment, in the important completeness condition (to be
verified in subappendix \ref{omega}) fulfilled by the $n-v=
\sum_{k\in{\Omega_{2rv}}}1$ intervals $\Delta T^{b}_{k}(f_{k},1)$:
\be
\sum_{k\in{\Omega_{2rv}}} \Delta T^{b}_{k}(f_{k},1)=T-\Delta T^{a}_{0}-
\Delta T^{a}_{n+2}~~~~~~~,~~~~~~~\sum_{k\in{\Omega_{2rv}}}f_{k}=2+r-2v~,
\label{KEY.01a}
\ee
where each $\Delta T^{b}_{k}(f_{k},1)$ is spanned by the end-points of
the $\bar{\cal R}_{b}^{-1}-$copies which, in a given $rv-$protograph, are
associated to the $k$th line of the corresponding $j=2$ elementary diagram
according to Eq. (\ref{MUL.04y}). To interpret Eq. (\ref{KEY.01a}), we first
note that the residual time-interval $T-\Delta T^{a}_{0}-\Delta T^{a}_{n+2}$
results from the overall temporal domain $T$ after the exclusion of its left-
and rightmost segments $\Delta T^{a}_{0}$ and $\Delta T^{a}_{n+2}$ which
(entering the factor (\ref{EX.01h})) are spanned by the end-points of the
${\cal R}_{a}^{-1}-$copies. Also, the $n-v$ open intervals
$\Delta T^{b}_{k}(f_{k},1)$ are mutually nonoverlapping,
$\Delta T^{b}_{k}(f_{k},1)\cap \Delta T^{b}_{q}(f_{q},1)=0$ for
$\forall{k\neq{q}}$, which means that
$q(p)\neq q(k)+2\omega^{(1)}_{k}-1$ $\forall{k}\in
\Omega_{2rv}/{\cal S}_{rv},~\forall{p}\in \Omega_{2rv}$. Then, the
completeness condition (\ref{KEY.01a}) geometrically implies therefore that,
once a particular $rv-$protograph is fully dressed,
the {\it entire}\/ residual 
time-interval is covered by the $n-v=3+r-v$ mutually
nonoverlapping intervals $\Delta T^{b}_{k}(f_{k},1)$. Examples are
described by figs. 5c (with $r=v=0$), 8f (with $r=v-1=0$), and 9f (with
$r=v=1$).

On the other hand, in the evaluation of the individual effective amplitudes
(\ref{FR.01w}), the sum in the l.h. side of Eq. (\ref{KEY.01a}) is replaced by
the sum $\sum_{k\in{\Omega_{jrv}}} \Delta \tau_{q(k)}<T-\Delta T^{a}_{0}-
\Delta T^{a}_{n+2}$ which generically is less than the residual time-interval.
In turn, this inequality follows from the fact that the number
$n+1$ of the relevant elementary intervals $\Delta \tau_{k}$ (the residual
interval is decomposed into so that $\sum_{i=1}^{n+1}
\Delta \tau_{i}=T-\Delta T^{a}_{0}-\Delta T^{a}_{n+2}$) is always less than
the number $n-v=\sum_{k\in{\Omega_{2rv}}}1$ of the lines involved into the
$f_{k}=1$ dressing
(\ref{KEY.01}). Examples are presented in the $j=2$
case by figs. 5b, 8c, and 9c.

Finally, in order to introduce and properly utilize the basic formula
(\ref{SR.01f}) below, in the $v=1$ cases it is convenient to define both
$\omega^{(\gamma)}_{k}$ and $\Delta T^{b}_{k}(f_{k},\gamma)$ not only for
$j=2$ but for $j=1$ as well. As it is verified in
subappendix \ref{omega}, the extension is fixed by the
prescription\footnote{It matches Eq. (\ref{FI.02}).}:
$\omega^{(\gamma)}_{2}|_{j=1}=\omega^{(1)}_{4-\gamma}|_{j=2}$
with $\gamma=1,2$, while $\Delta T^{b}_{k}(f_{k},\gamma)$ is to be defined by
the same Eq. (\ref{KEY.01b}).

\section{The structure of the effective amplitude
$\tilde{V}_{U_{\theta}(1)}^{(n)}(\{{\bf y}_{k}\})$}
\label{effective}


The convenient representation (\ref{SR.01f}) of the
factor $\tilde{V}_{U_{\theta}(1)}^{(n)}(\cdot)$, describing a given $2n$th
order elementary graph together with the
entire its $\bar{\cal R}_{b}^{-1}-$dressing, can be
deduced from the integral representation of the elementary $2n-$point
function (\ref{1.31b}) through the simple prescription. For this purpose, the
product of the concise exponential factors (\ref{SR.01a})
is to be included under the integrand of such representation of the
function (\ref{1.31b}) that generalizes Eq.
(\ref{IR.01}). In subsection \ref{mcontact}, we present a brief verification
that this prescription matches the result of the appropriate application of
the $n-v$ replacements (\ref{KEY.01}) with $f_{k}=1$.

\subsection{The $\bar{\cal R}_{b}^{-1}-$deformations of the elementary
$2n-$point functions}
\label{bdeform}

Let us introduce the effective functions
$\tilde{V}_{U_{\theta}(1)}^{(n)}(\{{\bf y}_{k}\})$ in a
way that makes manifest the relations between those functions which are
parameterized by the elementary graphs with a given $rv-$assignment.  
For this purpose, we first get rid\footnote{It is admissible because the
effective amplitude (\ref{FR.01w}) anyway involves the contour
integrals over the $2n$ temporal coordinates $x^{2}(\cdot)$ of the lines'
end-points which define the set $\{{\bf y}_{k}\}$. Also, the lines are
labeled in Appendix \ref{enumer1}) so that the fourth line, being present only
in the $r=v=1$ cases, is the ${\cal R}_{b}^{-1}-$copy of the first line. As
for the third line, being present only in the $j=2$ cases, it is {\it not}\/
involved into the $S(4)-$reattachments as well as the second line (present in
all cases).} of the $n-2=r+j-1$ $\delta(\cdot)-$functions (defined by
the $G=1$ Eq.~(\ref{1.40kk})),
starting with the $(r+j-1)-$fold integral 
\be
\int\limits_{-T}^{T} d^{j-1}t^{(\gamma)}_{3}~d^{r}t^{(\gamma)}_{4}~
\tilde{V}_{U_{\theta}(1)}^{(n)}(\{{\bf y}_{k}\})\Big|^{\gamma}_{jrv}=
J_{jrv}\left[\frac{(-1)^{\omega^{(\gamma)}_{3}-1}\partial}
{\partial \tau_{q(3)+\omega^{(\gamma)}_{3}}}\right]^{j-1}
\left[\frac{(-1)^{\omega^{(\gamma)}_{2}-1}\partial}
{\partial \tau_{q(2)+\omega^{(\gamma)}_{2}}}\right]^{v}
\tilde{\cal V}^{(\gamma)}_{jrv}(\{a_{i}\},\{\Delta \tau_{q(k)}\}),
\label{SR.01f}
\ee
where $J_{jrv}=(-1)^{v+j-1}{\sigma^{2+r-v}}/{(2\pi\theta)^{2}}$,
$\omega^{(\gamma)}_{k}=0,1$ is introduced in subsection \ref{completeness} on
the basis of Eq. (\ref{KEY.01b}), and (for the sake of
generality) we temporarily formulate the integration
in terms of the relative times (\ref{FU.02}) (rather than $x^{2}(\cdot)$),
postulating that $\int d^{0}x \tilde{V}(y)=\tilde{V}(y)$. Then,
$$
\tilde{\cal V}^{(\gamma)}_{jrv}(\{a_{i}\},\{\Delta \tau_{q(k)}\})=
\int d\zeta d\eta~
e^{i\left(\eta t^{(\gamma)}_{1}-\zeta t^{(\gamma)}_{2}\right){\cal C}_{21}/\theta}~
{\cal K}_{rv}(\zeta,\eta,\{a_{i}\})~\times
$$
\be
\times 
\left[{\cal F}(\alpha^{(1)}\zeta,\Delta \tau_{q(1)})\right]^{1-v}
{\cal F}(\alpha^{(2)}\eta,\Delta \tau_{q(2)})~
\left[{\cal F}(\alpha^{(3)}T_{{\cal C}_{ij}}(\eta,\zeta),\Delta \tau_{q(3)})\right]^{j-1}
\left[{\cal F}(\alpha^{(1)}\zeta,\Delta \tau_{q(4)})\right]^{r}~,
\label{SR.01g}
\ee
with\footnote{Actually, the quantities $\alpha^{(k)}\equiv
\alpha^{(k)}(\{a_{k}\})$,
${\cal C}_{ij}\equiv {\cal C}_{ij}(\{a_{k}\})$, and $t^{(\gamma)}_{p}\equiv
t^{(\gamma)}_{p}(\{a_{k}\})$ implicitly depend (see Appendix \ref{symmetry3}) 
on the set
$\{a_{k}\}$ of the variables introduced after Eq. (\ref{ST.01}).}
$\alpha^{(k)}$, ${\bf y}_{k}$, ${\cal C}_{ij}$, and $t^{(\gamma)}_{p}$
being given by Eqs. (\ref{GP.01}), (\ref{1.38aa}), (\ref{1.34}), and
(\ref{FU.02}) respectively, while
\be
\tilde{\cal K}_{rv}(\zeta,\eta,\{a_{i}\})=|a_{1}R+\zeta|~
|a_{2}R+\eta|^{1-v}|a_{4}R+\alpha^{(1)}\zeta|^{r}~~~~,~~~~y^{1}_{k}=a_{k}R~,
\label{SR.01s}
\ee
\be
{\cal F}(\eta,\Delta \tau_{q(k)})=
exp\left(-\sigma|R+\eta|\Delta \tau_{q(k)}\right)~,
\label{SR.01a}
\ee
where
$\Delta \tau_{q(k)}$ is the same
$k\rightarrow q(k)$ option of the time-interval (\ref{FA.02w}) as in the
$f_{k}=1$ variant of Eq. (\ref{KEY.01}), and
\be
T_{{\cal C}_{ij}}(\eta,\zeta)=
({\cal C}_{31}/{\cal C}_{21})\eta-({\cal C}_{32}/{\cal C}_{21})\zeta~,
\label{1.24h}
\ee
with $|{\cal C}_{21}|=1$.

Finally, Eq. (\ref{SR.01f}) is to be augmented by the $r+j-1$ constraints
(imposed by thus resolved $\delta-$functions of Eq. (\ref{1.40kk}))
that results in the relations
\be
(j-1)\left(t^{(\gamma)}_{3}-
T_{{\cal C}_{ij}}(t^{(\gamma)}_{2},t^{(\gamma)}_{1})\right)=0~~~~~,~~~~~
r\left(t^{(\gamma)}_{1}-\alpha^{(1)}t^{(\gamma)}_{4}\right)=0~,
\label{SR.01y}
\ee
where, the second condition yields (when $r=1$) the implementation of the
general constraint (\ref{FA.09}), while the first one will be interpreted
geometrically in Appendix \ref{dMUL.04y}. It also noteworthy that the
r.h. side of Eq. (\ref{SR.01f}) depends on $\gamma$ 
{\it only}\/ through the $\gamma-$dependent quantities $\omega^{(\gamma)}_{p}$
together with the $\gamma-$dependent decomposition (\ref{FU.02}) of the
parameters $t^{(\gamma)}_{1}$ and $t^{(\gamma)}_{2}$ entering Eq.
(\ref{SR.01g}).

\subsection{Relation to the elementary $2n-$point functions}
\label{mcontact}

Before we discuss how to reinterpret the partial integrand of the effective
function (\ref{SR.01f}) in
compliance with the replacement (\ref{KEY.01}), the intermediate step is to
establish the relation between the latter function and its counterpart
associated to the corresponding elementary graph. Also, we point out
a preliminary indication of the relevance of the parameterization in terms
of the protographs.

For this purpose, we first take into account that, as it will be proved
in Appendix \ref{dMUL.04y}, in the r.h. side of Eq. (\ref{SR.01f}) the
partial derivative $\partial/\partial \tau_{q(p)+\omega^{(\gamma)}_{p}}$ acts
only on the $p$th factor (\ref{SR.01a}) (with $p=3-v,3$) of the expression
(\ref{SR.01g}). In consequence, these
derivatives merely insert,
under the integrand, the $r+j-1$ factors
$-\sigma|R+{\cal G}_{k}(\zeta,\eta)|$
entering the exponent of Eq. (\ref{SR.01a}):
\be
\left[\frac{(-1)^{\omega^{(\gamma)}_{3}-1}\partial}
{\partial \tau_{q(3)+\omega^{(\gamma)}_{3}}}\right]^{j-1}
\left[\frac{(-1)^{\omega^{(\gamma)}_{2}-1}\partial}
{\partial \tau_{q(2)+\omega^{(\gamma)}_{2}}}\right]^{v}
~~~
\longrightarrow~~~(-\sigma)^{j-1+v}|R+T_{{\cal C}_{ij}}(\eta,\zeta)|^{j-1}
|R+\eta|^{v}~,
\label{VAR.02}
\ee
where we have used that $\alpha^{(2)}=1$ when $v=1$, while $\alpha^{(3)}=1$
when $j=2$. Once the replacement
(\ref{VAR.02}) is performed, the general rule states
that, for any admissible $n=1+j+r$, the integral representations of a given
elementary $2n-$point function ${V}_{U_{\theta}(1)}^{(n)}(\{{\bf y}_{k}\})$
can be deduced from the corresponding effective one
$\tilde{V}_{U_{\theta}(1)}^{(n)}(\{{\bf y}_{k}\})$ through the replacement
\be
{\cal F}(\cdot,\Delta \tau_{q(k)})~\longrightarrow~1~~,~~\forall{k}
~~~~~~~\Longrightarrow~~~~~~~
\tilde{V}_{U_{\theta}(1)}^{(n)}(\{{\bf y}_{k}\})~\longrightarrow~
{V}_{U_{\theta}(1)}^{(n)}(\{{\bf y}_{k}\})~,
\label{RED.01}
\ee
with ${\cal F}(\cdot)$ being defined by Eq. (\ref{SR.01a}).
In particular, (provided ${\cal C}_{12}=-1$) the reduced option
(\ref{RED.01}) of the $r=v=j-1=0$ implementation of the function
(\ref{SR.01f}),
being associated to the diagrams of fig. 1, assumes the form of
integral representation of the $f_k ({\bf z})\rightarrow D_{22}({\bf z}),~
k=1,2$ implementation of the star-product (\ref{IR.01}),
where the propagator $D_{22}({\bf z})$ is introduced in Eq. (\ref{1.2}). It
is manifest after the identifications: $\zeta\rightarrow{\xi^{1}_{1}},~
\eta\rightarrow{\xi^{1}_{2}}$.
Correspondingly, the $k$th factor $({\cal F}(\cdot,\Delta \tau_{q(k)})-1)$
accumulates the contribution of all admissible $\bar{\cal R}_{b}^{-1}-$copies
of the $k$th line so that the interval\footnote{It is noteworthy that,
underlying the solvability of the problem, the {\it local}\/ in
$\Delta \tau_{q(k)}$ pattern of the factor (\ref{SR.01a}) is traced back to
the specific constraints (\ref{FA.09}) imposed by the perturbative amplitudes
(\ref{1.31b}).} $\Delta \tau_{q(k)}$ is spanned by the temporal coordinates
of the upper (or, equivalently, lower) end-points of the latter copies.

Altogether, in thus reduced Eq. (\ref{SR.01f}) the integral
representation ${V}_{U_{\theta}(1)}^{(n)}(\{{\bf y}_{k}\})$ of a given $2n$th
order elementary graph includes, besides the exponential factor
$e^{i\left(\eta t_{1}-\zeta t_{2}\right){\cal C}_{21}/\theta}$ (inherited from
Eq. (\ref{IR.01})) and the $G=1$ product (\ref{1.40kk}) of $n-2$ different
$\delta(\cdot)-$functions, the product
\be
\tilde{\cal K}_{n+2}(\zeta,\eta,R)=
\prod_{i=1}^{4}|a_{i}R+{\cal G}_{i}(\zeta,\eta)|^{w_{i}}~~~~~,~~~~~
{\cal G}_{k}(\zeta,\eta)=\tilde{b}_{k}\zeta+\tilde{c}_{k}\eta~,
\label{SUB.01j}
\ee
composed of $n=1+j+r$ factors $|a_{i}R+{\cal G}_{i}(\zeta,\eta)|$ each of
which represents (when $w_{i}\neq{0}$, i.e., for
$i\in \tilde{\Omega}_{jrv}$) the $i$th propagator of the latter
graph so that $w_{1}=w_{2}=1,~w_{3}=j-1=0,1,~w_{4}=r=0,1$ with
$\sum_{i=1}^{4}w_{i}=n$. Here, $a_{k}$ is defined in Eq. (\ref{SR.01s}),
$\tilde{b}_{q},\tilde{c}_{q}=0,\pm 1$, while $R$ is
defined in the footnote prior to Eq. (\ref{ST.01}), and $2\leq n\leq 4$. In
turn, it is the
$\tilde{\cal K}_{rv}(\cdot)-$part (\ref{SR.01s}) 
of $\tilde{\cal K}_{n+2}$ which,
being associated to the $2+r-v$ lines involved into the $S(4)-$reattachments
(when $a_{k}$ assumes both of the admissible values), refers to the
protographs. The remaining $v+j-1$ lines, corresponding to
the $f_{k}=0$ option of the replacement (\ref{KEY.01}),
are not affected by the reattachments
so that the corresponding $a_{k}$ are equal to unity which matches the
$a_{k}=1$ pattern of the exponents (\ref{SR.01a}) {\it necessarily}\/
associated to all these lines via the replacement (\ref{VAR.02}).

In conclusion, it is routine to convert the inverse of
the prescription (\ref{RED.01}) into the composition of the replacements
(\ref{KEY.01}). In view of Eq. (\ref{VAR.02}), the general pattern
(\ref{SR.01f}) is such that any particular effective $2n-$point function
$\tilde{V}_{U_{\theta}(1)}^{(n)}(\{{\bf y}_{k}\})$ can be deduced from the
associated elementary one through the corresponding option of the replacement
\be
\prod_{k=1}^{4}|a_{k}R+{\cal G}_{k}(\zeta,\eta)|^{w_{k}}
~\longrightarrow~\prod_{k=1}^{4}
|a_{k}R+{\cal G}_{k}(\zeta,\eta)|^{w_{k}}~
e^{-\sigma v_{k}|R+{\cal G}_{k}(\zeta,\eta)\alpha^{(k)}|
\Delta \tau_{q(k)}}~,
\label{SUB.01a}
\ee
where $v_{1}=1-v=0,1$, $v_{k}=w_{k}$, for $k=2,3,4$ (with $v_{k}\neq 0$ only
when $k\in \Omega_{jrv}$, $\sum_{k=1}^{4}v_{i}=n-v$)
that matches the pattern of Eq.~(\ref{SR.01g}). Then, it takes a
straightforward argument to verify that the substitution
(\ref{SUB.01a}) is indeed equivalent to the $f_{k}=1$ prescription
(\ref{KEY.01}) applied, with the identification $\alpha^{(4)}=\alpha^{(1)}$,
to the corresponding $\Omega_{jrv}-$subset of the $n-v$ perturbative
propagators.

\section{Integral representation of the effective amplitudes}
\label{effective1}

At this step, we are ready to obtain
the explicit form of the amplitude ${\cal Z}_{rv}(\bar{A},\bar{\theta}^{-1})$
which, being introduced in Eq. (\ref{SU.01}), defines the decomposition
(\ref{SU.01z}) of  the
$G=1$ term $<W(\Box)>_{U_{\theta}(1)}^{(1)}$ of the $1/N$ expansion
(\ref{CO.01}). For this purpose, we first put forward the general
representation (\ref{FR.01}) of the individual effective amplitudes
(\ref{FR.01w}) which, being
parameterized by the corresponding elementary graphs, are
evaluated {\it non-perturbatively}\/ both in $g^{2}$ and in
$\theta$. The latter amplitudes arise when the
$2n-$point function $\tilde{V}_{U_{\theta}(1)}^{(n)}(\{{\bf y}_{k}\})$,
being multiplied by the factor (\ref{EX.01h}), is integrated over the
$n$ pairs of the relative coordinates (\ref{1.38aa}) (defining the set
$\{{\bf y}_{k}\}$), all restricted to the contour $C=\Box$.


Then, building on this representation,
the superposition (\ref{SU.01}) of ${\cal Z}^{(\gamma)}_{jrv}(\cdot)$ is
evaluated collecting
together the contributions associated to all the elementary graphs with the
same $rv-$assignment. In turn, the specific cancellations, taking place
between the different terms of the superposition, support
the pattern of the relevant collective coordinates.
In particular, it verifies the 
representation of $\tilde{V}_{U_{\theta}(1)}^{(n)}(\cdot)$ (discussed in
subsection \ref{completeness}) formulated in terms of the properly dressed
protographs. We also clarify the relation between the latter dressing and the
structure of the collective coordinates.

\subsection{General pattern of the individual effective amplitudes}
\label{key}


Synthesizing the factors (\ref{EX.01h}) and (\ref{SR.01f}), one concludes
that the individual effective amplitudes (\ref{FR.01w}) assume the form
$$
(-1)^{\sum_{l\in {\cal S}_{rv}}a_{l}}~
{\cal Z}^{(\gamma)}_{jrv}(\{a_{k}\},\bar{A},\bar{\theta}^{-1})=
$$
\be
=\bar{A}^{2+h_{rv}}
\int d^{n+2}\bar{\tau}
e^{-\bar{A}\left(\Delta \bar\tau_{0}+\Delta \bar\tau_{n+2}\right)}
\left[\frac{(-1)^{\omega^{(\gamma)}_{3}-1}\partial}
{\partial \tau_{q(3)+\omega^{(\gamma)}_{3}}}\right]^{j-1}
\left[\frac{(-1)^{\omega^{(\gamma)}_{2}-1}\partial}
{\partial \tau_{q(2)+\omega^{(\gamma)}_{2}}}\right]^{v}
{\cal V}^{(\gamma)}_{jrv}(\{a_{i}\},\{\Delta \bar\tau_{q(k)}\}),
\label{FR.01}
\ee
where the sum in the l.h. side runs over the labels $l$ of the $2+r-v$ lines
involved into the $S(4)-$reattachments in the first relation of
Eq. (\ref{SU.01}) (with the set ${\cal S}_{rv}$ being
specified in the end of subsection \ref{assignment1}), and
we introduce the compact notation
\be
\int\limits_{0\leq \bar{\tau}_{k}\leq
\bar{\tau}_{k+1}}^{\bar{\tau}_{m}\leq 1} \prod_{l=1}^{m} d\bar{\tau}_{l}
...\equiv
\int d^{m}\bar{\tau}...~,
\label{MEA.01}
\ee
for the integrations over the $m=n+2$ ordered
times $\bar\tau_{j}$, and, for a rectangular contour $C=\Box$ of the size
$R\times T$, it is convenient to utilize the change of the variables
\be
\tau_{k}=T\bar{\tau}_{k}~~~~~,~~~~~t_{k}=T\bar{t}_{k}~~~~~,~~~~~
\zeta=R\bar{\zeta}~~,~~\eta=R\bar{\eta}~,
\label{BR.02a}
\ee
which introduces the dimensionless quantities $\bar{\tau}_{k}$ (with
$\Delta \bar\tau_{k-1}=\bar\tau_{k}-\bar\tau_{k-1}\geq{0}$),
$\bar{\eta}$, $\bar{\zeta}$, and $\bar{t}_{k}$ so that
$R^{4+r-v}{\cal V}_{jrv}(\{a_{i}\},\{\Delta \bar\tau_{q(k)}\})=
\tilde{\cal V}_{jrv}(\{a_{i}\},\{T\Delta \bar\tau_{q(k)}\})$. In
particular, this change makes it manifest that, besides the dependence on
$\{a_{i}\}$ and $\bar{\theta}=\sigma\theta$, the considered effective
amplitudes are certain functions
of the dimensionless area of the rectangle $C$
\be
\bar{A}=\sigma A(C)\Big|_{C=\Box}=\sigma RT
\label{BR.02}
\ee
rather than of $R$ and $T$ separately.

Note also that in the r.h. side of Eq. (\ref{FR.01}) the $m=n+2$ species of
the $d\bar\tau_{j}-$integrations reproduce\footnote{To make use of Eq.
(\ref{SR.01f}), we utilize the fact that, owing to the $G=1$ pattern
(\ref{1.40kk}), the integrations $\int d^{j-1}t_{3}~d^{r}t_{4}$ can be
reformulated as $r+j-1$ integrations with respect to the temporal
coordinates $x^{2}(\cdot)$ of those end-points which are {\it not}\/ involved
into the definition of $\tau_{k}$.}, according to the discussion of subsection
\ref{ordering}, the $2n-$fold contour-integral (\ref{FR.01w}) which
runs over the time-coordinates $dx^{2}(s_{l})$ and $dx^{2}(s'_{l})$
constrained by the $G=1$ product (\ref{1.40kk}) 
of the $\delta(\cdot)-$functions.
Correspondingly, this transformation of the measure has the jacobian which is
equal to $(-1)^{n_{jrv}^{-}}$, where $n_{jrv}^{-}$ and $n_{jrv}^{+}$ denote
the numbers of the line's end-points attached, for a given elementary graph,
respectively to the lower and upper horizontal side of the rectangle 
$C=\Box$ so that
\be
(-1)^{n_{jrv}^{-}}=\prod_{i\in \tilde{\Omega}_{jrv}}(-1)^{a_{i}}~~~~~~~,~~~~~~~
\frac{1}{2}(n_{jrv}^{+}+n_{jrv}^{-})=n_{jrv}\equiv n=
\sum_{k\in \tilde{\Omega}_{jrv}}1~,
\label{SIG.01}
\ee
where $a_{k}$ is defined in Eq. (\ref{SR.01s}), while the set
$\tilde{\Omega}_{jrv}$ is specified in the end of subsection \ref{assignment1}.
In turn, to justify the $\{a_{k}\}-$dependent sign-factor in the l.h. side of
Eq. (\ref{FR.01}), it remains to notice that
$\sum_{k\in \tilde{\Omega}_{jrv}}a_{k}=v+j-1+\sum_{l\in {\cal S}_{rv}}a_{l}$
since $a_{k}=1$ for $\forall{k}\in \tilde{\Omega}_{jrv}/{\cal S}_{rv}$.

\subsection{Derivation of the combinations
${\cal Z}_{rv}(\bar{A},\bar{\theta}^{-1})$}
\label{tuning}

The effective amplitudes (\ref{FR.01}), parameterized by the individual
elementary graphs,
are still intermediate quantities.
To say the least, for generic $\bar{A}$,
they are {\it singular}\/ for $\bar{\theta}^{-1}\rightarrow{0}$.
To arrive at amplitudes which are already continuous in $\bar{\theta}^{-1}$
in a vicinity of $\bar{\theta}^{-1}=0$, our aim is to evaluate
the combinations (\ref{SU.01})
of the latter amplitudes entering the decomposition (\ref{SU.01z}).


Then, to reveal the cancellations between different terms of the sum
(\ref{SU.01}), in the r.h. side of Eq. (\ref{FR.01}) one is to perform
$v+j-1$ integrations to get rid of the corresponding number of the partial
derivatives (employing that $0<q(k)<n+2$ for $\forall{k}=1,..,n$).
In Appendix \ref{dMUL.04y}, it is shown that, matching the prescription
(\ref{KEY.01}) formulated in subsection \ref{completeness},
a straightforward computation yields
\be
{\cal Z}_{rv}(\bar{A},\bar{\theta}^{-1})=\bar{A}^{2+h_{rv}}
\sum_{\{a_{l}\}_{rv}}(-1)^{\sum_{l\in {\cal S}_{rv}}a_{l}}
\int d^{2+h_{rv}}\bar\tau~
e^{-\bar{A}\left(\Delta \bar\tau_{0}+\Delta \bar\tau_{2+h_{rv}}\right)}~
{\cal V}_{2rv}(\{a_{k}\},\{\Delta \bar{T}^{b}_{k}\})~,
\label{MUL.04y}
\ee
where the sum over $a_{l}$ is the same as in Eq. (\ref{SU.01}),
$\tilde{\cal V}_{2rv}(\cdot)\equiv{\tilde{\cal V}^{(1)}_{2rv}(\cdot)}$
is defined in Eq. (\ref{SR.01g}), and we have omitted the subscript $\gamma=1$
since, in view of Eq. (\ref{SU.01l}), $\gamma$ assumes the single value for
$j=2$ irrespectively of the values of $r$ and $v$.
Note that, in the exponent, $\Delta \bar\tau_{n+2}$ is replaced by
$\Delta \bar\tau_{2+h_{rv}}\equiv T^{-1}\Delta T^{a}_{2+h_{rv}}$ while,
in the quantity ${\cal V}_{2rv}(\cdot,\cdot)$, the
set $\{\Delta \bar\tau_{q(k)}\}$ is superseded by
$\{\Delta \bar{T}^{b}_{k}\}\equiv\{\Delta \bar{T}^{b}_{k}(f_{k},1)\}$,
where the intervals $\Delta \bar{T}^{b}_{k}(f_{k},1)=
T^{-1}\Delta T^{b}_{k}(f_{k},1)$, being
constrained by the condition (\ref{KEY.01a}), are introduced in Eq.
(\ref{KEY.01b}). Altogether, omitting the subscripts $a$ and $b$, the
relevant $3+h_{rv}$ intervals
$\Delta \bar{T}_{i}=\bar\tau_{i+1}-\bar\tau_{i}\geq 0$ are expressed through
$2+h_{rv}$ ordered quantities $\bar\tau_{i}$ characterized by the
$m=2+h_{rv}$ option of the measure (\ref{MEA.01}).

Finally, according to Appendices \ref{dMUL.04} and \ref{symmetry3},
the r.h. side of Eq. (\ref{MUL.04y}) can be rewritten in the form
\be
{\cal Z}_{rv}(\bar{A},\bar{\theta}^{-1})=
\bar{A}^{2+h_{rv}}
\int d^{2+h_{rv}}\bar\tau \int\limits_{-\infty}^{+\infty}d\bar\zeta d\bar\eta~
e^{i\left(\bar\eta \bar{t}_{1}-
\bar\zeta \bar{t}_{2}\right)\bar{A}/\bar{\theta}}~
{\cal K}_{rv}(\bar\zeta,\bar\eta)~
{\cal Y}_{rv}(\bar\zeta,\bar\eta,\{\Delta \bar\tau_{k}\}),
\label{MUL.04}
\ee
where $\bar{t}_{p}\equiv\bar{t}^{(1)}_{p}$, $p=1,2$,
\be
{\cal Y}_{rv}(\cdot)=
e^{-\bar{A}\left(\Delta \bar\tau_{0}+\Delta \bar\tau_{2+h_{rv}}\right)}~
exp\left(-\bar{A}\left[(1+r-v)|1-\bar\zeta|\Delta \bar\tau_{3}+
|1+\bar\eta|\Delta \bar\tau_{1+v}
+|1-\bar\zeta+\bar\eta|\Delta \bar\tau_{2-v}\right]\right)~,
\label{MUL.02a}
\ee
${\cal K}_{rv}(\cdot)$ is given by Eq. (\ref{MUL.01a}), and the sum over
$\{e_{k}\}$ supersedes the one over $\{a_{l}\}_{rv}$ (combining four
different implementations ${\cal Z}_{rv}(\{a_{k}\},\cdot)$) so that the
dressing-weight ${\cal Y}_{rv}(\cdot)$ is manifestly $S(4)-$invariant, i.e.,
$\{e_{k}\}-$independent. For the particular $\{a_{k}\}-$assignments,
${\cal Z}_{rv}(\{a_{k}\},\cdot)$ is
diagrammatically depicted in figs. 5c ($a_{1}=a_{2}=0$), 8f ($a_{1}=0$), and
9f ($a_{1}=a_{4}=0$) which are associated to
$r=v=0$, $r=v-1=0$, and $r=v=1$ respectively. 

Let us also note that the representation (\ref{MUL.04}) readily allows to
demonstrate that, for $\forall{\bar{A}}>0$,
${\cal Z}_{rv}(\bar{A},\bar{\theta}^{-1})$ is indeed continuous
in $\bar{\theta}^{-1}$ in a vicinity of $\bar{\theta}^{-1}=0$.
This property, implied in the transformation of Eq. (\ref{SU.01z})
into Eq. (\ref{CO.03f}), will be explicitly derived in \cite{ADM05b}.

\subsection{A closer look at the pattern of the collective coordinates}
\label{colcoor}

In conclusion, let us clarify the following subtlety concerning
the pattern of the collective coordinates relevant for the dressing of
the $rv-$protograph. The point is that, in the $v=1$ Eq. (\ref{MUL.04}),
both the measure $d^{2+h_{rv}}\bar\tau$ and the relative time $\bar{t}_{2}
\equiv \bar{t}^{(\gamma)}_{2}$
can not be fully determined only on the basis of the configuration of the
$rv-$protograph itself (postulated to be constrained, in the $r=1$ case, by
the second of the conditions (\ref{SR.01y})). The general reason is traced
back to the fact that
the $v=1$ protographs are not of genus-one and, therefore, their dressing
necessarily encodes certain structure inherited from the associted $j=2$
elementary diagrams.

In consequence, the above measure includes integration over one more
parameter\footnote{This parameter supersedes, after the two integrations (over
$\tau_{q(p)+\omega^{(1)}_{p}}$ with $p=1,2$), the parameter $\bar\tau_{3+r}$
defined by Eq. (\ref{REP.04}) in the case of the $j=2$ amplitude
(\ref{FR.01}).} $\bar\tau_{2+r}$ in addition to the $2h_{rv}-r=2-v+h_{rv}$
parameters which are directly identified
(see Appendices \ref{dMUL.04y} and \ref{freedom} for the details), with the
independent temporal coordinates of the end-points of the protographs' lines:
\be
\int d^{2+h_{rv}}\bar\tau~...=
\int d^{2-v+h_{rv}}\bar\tau \int d^{v}\bar{t}_{2}...~,
\label{CLO.01}
\ee
where we have used that $\bar{t}_{2}=\bar\tau_{2+r}-\bar\tau_{1}$
(with $\tau_{1}=x^{2}(s'_{1})$ as it is depicted in figs. 8f and 9f).
Then, as it is discussed in Appendix \ref{freedom}, the 
presence of $\tau_{2+r}$ is tightly related to the first of the constraints
(\ref{SR.01y}) fulfilled by the three parameters $t_{p},~p=1,2,3$.
In turn, as it is sketched in Appendix \ref{dMUL.04y}, the
latter constraints underlie the completeness condition (\ref{KEY.01a}) for
$\forall{r,v}$.

Note also the reduction $\int d^{2+n}\bar\tau~..
\rightarrow \int d^{4+r-v}\bar\tau~..$ of the relevant measure, formalized by
the transition from the combination of the individual amplitudes
(\ref{FR.01}) to Eq. (\ref{MUL.04y}), entails the relevant 
$f_{k}=0$ replacements (\ref{KEY.01}) applied to the $j=2$ Eq. (\ref{FR.01}).
Indeed, the latter replacements result after such integration over
$n-(2+r-v)|_{j=2}=v+1$ parameters $\tau_{q(p)+\omega^{(\gamma)}_{p}}$ (with
$p=3-v,3$), in the process of which the corresponding intervals
$\Delta \tau_{q(k)}$ vary in the domains $[0,
\Delta T^{b}_{k}(0,\gamma)]$ (see Appendix \ref{dMUL.04y} for more details).

\section{The large $\theta$ limit}
\label{prescr}

At this step we are ready to put forward the prescription
(\ref{LA.02c}) to implement the large $\theta$ limit in Eq. (\ref{MUL.04}).
By virtue of the $1/\theta^{2}$ factor in front of the sum in the r.h. side of
eq. (\ref{CO.03f}), the asserted large $\theta$ scaling
$<W(\Box)>_{U_{\theta}(1)}^{(1)}\sim \theta^{-2}$
is a consequence of the important property of the combinations
${\cal Z}_{rv}(\bar{A},\bar{\theta}^{-1})$. For any finite $\bar{A}\neq 0$,
the relevant large $\theta$ limit (\ref{LI.01})
can implemented directly through the substitution
\be
e^{i\left(\bar\eta \bar{t}_{1}-\bar\zeta \bar{t}_{2}\right)\bar{A}
/\bar{\theta}}~\longrightarrow~1~~~\Longrightarrow~~~
{\cal Z}_{rv}(\bar{A},\bar{\theta}^{-1})~\longrightarrow~
{\cal Z}_{rv}(\bar{A},0)~,
\label{LA.02c}
\ee
to be made in the integrand of the representation (\ref{MUL.04}) of the
quantity ${\cal Z}_{rv}(\bar{A},\bar{\theta}^{-1})$ that
replaces the latter quantity by its reduction ${\cal Z}_{rv}(\bar{A},0)$.
In turn, provided Eq. (\ref{CO.03}) is valid, the prescription (\ref{LA.02c})
yields the integral represetation (\ref{CO.03f})
for the next-to-leading term of the $1/\theta$ expansion (\ref{CO.02})
(with $<{\cal W}(\Box)>_{N}^{(1)}=0$).

The self-consistency of the deformation (\ref{LA.02c}) is maintained
provided ${\cal Z}_{rv}(\bar{A},\bar{\theta}^{-1})$ is continuous in
$\bar{\theta}^{-1}$ in a vicinity of $\bar{\theta}^{-1}=0$. In turn, 
it can be shown that, for $\bar{A}\neq 0$, the latter
property is valid provided this
deformation does not violate the {\it convergence}\/ of the $(m+2)-$dimensional
integral over $\bar{\tau}_{k}$, $\bar\zeta$, and $\bar\eta$ defining the
representation
(\ref{MUL.04}) of ${\cal Z}_{rv}(\bar{A},\bar{\theta}^{-1})$, where
$m=2+h_{rv}$. To demonstrate
the convergence, it is convenient first to get rid of the explicit
$m-$dimentional ordered integration over $\bar{\tau}_{j}$. For this purpose,
it is useful to perform the Laplace transformation of
${\cal Z}_{rv}(\bar{A},\bar{\theta}^{-1})$ with respect to the
dimensionless area (\ref{BR.02}) that results in
\be
\tilde{\cal Z}_{rv}(\beta,\bar{\theta}^{-1})=\int_{0}^{+\infty}
d\bar{A}~{\cal Z}_{rv}(\bar{A},\bar{\theta}^{-1})~e^{-\beta \bar{A}}~.
\label{LA.02}
\ee
The advantage of this trick is that, in the integral representation of the
image $\tilde{\cal Z}_{rv}(\beta,\bar{\theta}^{-1})$,
the $\bar{\tau}_{j}-$integrations can be easily performed
using the general relation
\be
\prod_{j=0}^{m}\frac{1}{\beta+B_{j}}=
\int\limits_{0}^{+\infty} d\bar{A}~
e^{-\beta \bar{A}}\int\limits_{0\leq \breve{\tau}_{k}\leq
\breve{\tau}_{k+1}}^{\breve{\tau}_{m}\leq \bar{A}} \prod_{k=1}^{m}
d\breve{\tau}_{k}\prod_{j=0}^{m}
exp\left(-B_{j}\Delta\breve{\tau}_{j}\right)~,
\label{LA.01}
\ee
where $\breve{\tau}_{j}$ is to be identified with $\bar{A}\bar{\tau}_{j}$,
while $\Delta\breve{\tau}_{j-1}=\breve{\tau}_{j}-\breve{\tau}_{j-1}$ with
$\breve{\tau}_{0}\equiv{0}$ and $\breve{\tau}_{m+1}\equiv{\bar{A}}$.
In particular, in this way one proves that the Laplace image
$\tilde{\cal Z}_{rv}(\beta,0)$ of the large $\theta$ asymptote
${\cal Z}_{rv}(\bar{A},0)$ of the amplitude (\ref{MUL.04}) assumes
the form (\ref{LA.03}).

As for the self-consistency
of the prescription (\ref{LA.02c}), it can be
verified provided the double integral (\ref{LA.03}) is convergent for
$\forall{\beta}>0$ so that
$\tilde{\cal Z}_{rv}(\beta,\bar{\theta}^{-1})$ is continuous in
$\bar{\theta}^{-1}$ in a vicinity of $\bar{\theta}^{-1}=0$.
A direct inspection verfies that the convergence indeed takes place.

Also, it should be stressed that, due to the infrared
singularities of the propagators,
the prescription (\ref{LA.02c}) is not applicable directly to each
individual perturbative diagram. This property may be inferred from the
integral representations of the elementary amplitudes given by the reduction
(\ref{RED.01}) of the effective amplitudes considered in subsection
\ref{bdeform}. Actually, even the individual effective amplitudes
(\ref{FR.01}) still are not suitable for this purpose either that can be
traced back to the violation
of the completeness condition (\ref{KEY.01a}).
It takes certain specific cancellations between the latter amplitudes that,
resulting in the latter condition, makes the substitution (\ref{LA.02c})
applicable to the combinations (\ref{MUL.04}).
We shall continue the discussion of this issue in \cite{ADM05b}.

\section{Conclusions}
\label{conclus}

In the present paper
we obtain the exact integral representation (\ref{SU.01z}) of
the next-to-leading term $<W(\Box)>_{U_{\theta}(1)}^{(1)}$ of the $1/N$
expansion (\ref{CO.01}) of the average in the $D=2$ gauge theory
(\ref{1.7}).
It provides the
rigorous non-perturbative\footnote{It is specifically important in the large
$\theta$ limit (\ref{LI.01}), where the truncated perturbative series of
$<W(C)>_{U_{\theta}(1)}^{(G)}$ is shown \cite{ADM04} to result in the false
asymptotical $\theta-$scaling that is supposed to take place not only for
$D=2$ but for $D=3,4$ as well.} computation made, from the first principles,
in the noncommutative gauge theory.

The Laplace image (\ref{LA.02}) of
the large $\theta$ asymptote of $<W(\Box)>_{U_{\theta}(1)}^{(1)}$ assumes the
particularly concise form (\ref{CO.03f}). In turn, the latter asymptote
is argued to be directly related (\ref{CO.03}) to the next-to-leading
term of the $1/\theta$ expansion (\ref{CO.02}) of $<W(C)>_{U_{\theta}(1)}$.   
It is noteworthy that the considered asymptote reveals
the power-like decay which is in sharp contrast with the exponential area-law
asymptote (\ref{1.41b}) valid in the leading order
of the $1/N-$ (or, equivalently, $1/\theta-$) expansion. Furthermore, as the
origin of the power-like decay can be traced back to the
(infinite, in the limit $\theta\rightarrow{\infty}$) nonlocality of
the star-product, similar decay is supposed to persist for all $G\geq{1}$
subleading\footnote{Contrary to the $G\geq{1}$ terms, the leading $G=0$ term
is insensitive to the star-product structure that matches its
$\theta-$independence (\ref{1.41b}).} terms $<W(C)>_{U_{\theta}(1)}^{(G)}$ 
of the large $\theta$ $1/N$ expansion.

In consequence, it precludes an apparent extension of the
stringy representation of the latter expansion in the spirit of the
Gross-Taylor proposal \cite{Gr&Tayl} formulated for the
commutative $D=2$ gauge theories. Another subtlety, concerning possible
stringy reformulation of the noncommutative observables, is that the
noncommutative gauge invariance is also maintained \cite{Ish} for certain
combinations of the Wilson lines associated to the {\it open}\/ contours
$C=C_{xy}$ with ${\bf x}\neq{\bf y}$. Nevertheless, the optimistic point of
view could be that all these subtleties may suggest a hint for
a considerable extension of the stringy paradigm conventionally utilized in
the context of two-dimensional gauge (or, more generally, matrix) systems.

As the developed here methods are general enough, we hope that our analysis
makes a step towards a derivation of an arbitrary two-dimensional
average $<W(C)>_{U_{\theta}(1)}$. Most straightforwardly, they can be applied
to consider the $G=1$ term of the average (\ref{1.1}) for a generic
rectangular contour $C=\Box$ with a nontrivial number $n\geq{2}$ of windings. 
E.g.,
it would be interesting to adapt the pattern (\ref{FR.01}) to the case when
$n>>1$ and estimate its asymptotical dependence on $n$. Also, the $G\geq{2}$
terms $<W(\Box)>_{U_{\theta}(1)}^{(G)}$ could be
in principle evaluated akin to the $G=1$ case that is expected to lead to
a generalization of Eq. (\ref{FR.01}). In particular, we expect that
there should be $2G$ parameters $\zeta_{q}$, $\eta_{q}$ with $q=1,...,G$,
while the factor in front of the integral becomes $\bar{A}^{m}/
(\sigma\theta)^{2G}$.

More subtle open question is to generalize our approach to a
(non-self-intersecting) contour of a generic geometry. In the commutative
$\theta=0$ case, the crucial simplification takes place by virtue of the
invariance of the partition function under the group of (simplectic)
area-preserving diffeomorphisms which guarantees
that $<W(C)>_{U(1)}$ depends only on the area $A(C)$ irrespectively of the
form of $C$. On the other hand, the representation (\ref{1.1}) does
not make manifest if there is a symmetry that relates the averages
$<W(C)>_{U_{\theta}(1)}$ with different geometries of the contour $C$.
Furthermore, the lowest order perturbative computation \cite{ADM04}
indicates that the simplectic invariance may be lost in the
non-commutative case. Nevertheless, the explicit $A(\Box)-$ (rather than
twofold $R-$ and $T-$) dependence of the derived $G=1$
term $<W(C)>_{U_{\theta}(1)}^{(1)}$ 
looks like a promising sign. 
Also, it would be interesting to make contact with
the noncommutative Loop equations \cite{ANMS99,LE/NCYM} which might be
an alternative approach to the above problems.

Finally, among other new questions raised by the present analysis,
we would like to mention the following one important in the context of the
$D=4,3$ noncommutative Yang-Mills theory (\ref{1.7}). We conjecture that
in this case the minimal area-law asymptote, presumably valid for a generic
closed fundamental Wilson loop in the $N\rightarrow\infty$ limit, fades away
at the level of the subleading $G\geq{1}$
terms similarly to what happens in the $D=2$ case.

\subsection*{Acknowledgments}

This work was supported in part by the grant INTAS--00--390.
The work of J.A.\ and Y.M.\ was supported in part by
by the Danish National Research Foundation.
A.D.\ and Y.M.\ are partially supported by 
the Federal Program of the Russian Ministry of Industry,
Science and Technology No 40.052.1.1.1112 and by the Federal Agency for
Atomic Energy of Russia.

\vspace{24pt}

\section*{Appendices}

\app{Elementary graphs and their deformations}
\label{enumer1}


To complete the discussion of subsections \ref{assignment1} and
\ref{gamspecif}, let us first explicitly separate, for any given
$jrv-$assignment, the elementary graphs with the maximal amount of the
horizontal lines and sketch the pattern of their
$\bar{\cal R}_{b}^{-1}-$deformations. Also, we note that the
$S(4)\otimes S(2)-$transformations can be consistently applied to the latter
graphs both prior
and after the ${\cal R}_{a}^{-1}\otimes\bar{\cal R}_{b}^{-1}-$dressing.

In the $r=v=0$ case when $f_{jv}|_{v=0}=1$, the two
$j=1$ and $j=2$ $S(4)-$multiplets can be generated from the graphs in figs.
1a, 1b and 2c, 2d respectively so that, for each $j$, the two corresponding
figures may be related via the reflection mutually interchanging the
horizontal sides of $C=\Box$. (When $j=2$, we take into account that 
both the elementary graphs in the figs. 2a, 2b and
all their ${\cal R}_{a}^{-1}\otimes\bar{\cal R}_{b}^{-1}-$deformations
are assigned with {\it vanishing} amplitudes (\ref{1.31b}).)
As for the
parameterization of the lines, the left and the right horizontal lines in
figs. 1a and 2c are assigned with
labels $1$ and $2$ respectively so that ${\cal C}_{21}=1$. The remaining
nonhorizontal line in fig. 2c attains the label $3$. 
A for the $\bar{\cal R}_{b}^{-1}-$dressing, it applies to
all $n=j+1$ lines of the considered $r=v=0$ elementary graphs.
Being depicted by the corresponding bunch of (nonvertical) parallel dotted
lines, these $\bar{\cal R}_{b}^{-1}-$dressed graphs are described by figs.
5a (with $j=1$) and 5b (with $j=2$).

In the $r=v-1=0$ case, the graphs with $2+r-v=1$
horizontal line are depicted by bold lines in figs. 8a--8c, where the 
horizontal line is assigned with labels $1$, with the nonhorizontal line(s)
being parameterized by the label(s) $2,1+j$ so that ${\cal C}_{23}=1$ when
$j=2$.
The constraint, separating these $v=1$ components, is
that the $j+1$ end-point at the lower side do belong to the time-interval
bounded by the end-points of the remaining horizontal line attached to the
upper side.
In turn, to make the $\bar{\cal R}_{b}^{-1}-$dressing of the latter
graphs unambiguous, in the $v=1$ case we should specify those of their lines
which are accompanied by their $\bar{\cal R}_{b}^{-1}-$deformations.
For $r=v-1=0$, all of the line possess their individual dressings except for
the single line (assigned with the label 1), involved into the vertical
reattachments, which is {\it not} dressed: see figs. 8a, 8b (with $j=1$)
and 8c (with $j=2$). As previously, the $\bar{\cal R}_{b}^{-1}-$dressing of a
given nonhorizontal (bold) line is depicted by the bunch of (nonvertical)
dotted lines which, in the $1+r=v=1$ case at hand, are all parallel to the
latter line.

In the remaining $r=v=1$ case, the graphs with $2+r-v=2$
horizontal lines are given by the entire decomposition of the
Feynman diagrams in figs. 7a (with $j=1$) and 7e (with $j=2$) into the
time-ordered components parameterized by $j=1,2$ and $\gamma=1,f_{jv}$.
In turn, for a given $j$ and $\gamma$, these components can be
collected into the pairs which are comprised of the two graphs related via
the reflection mutually interchanging the horizontal (or, equivalently,
vertical) sides of $C=\Box$. Correspondingly, the labels 1 and 4 are assigned
to the left and right horizontal lines, while (in the case when the first line
is attached to the upper side of $C$) the remaining two nonhorizontal
lines are parameterized similar to the corresponding figs. 8a--8c.
are assigned with {\it vanishing} amplitudes (\ref{1.31b}).)
Concerning the pattern of the $\bar{\cal R}_{b}^{-1}-$dressing,
all of the line possess their individual dressings except for
the two horizontal lines (assigned with the labels 1 and 4 respectively).
As it is clear from figs. 9a, 9b (with $j=1$) and 9c (with $j=2$),
the latter two lines share the same $\bar{\cal R}_{b}^{-1}-$dressing
which, in Eq. (\ref{SR.01g}), is formally associated to the fourth
line.

Note also that, given these rules, a direct inspection
demonstrates that each graph (with $2+r-v$ horizontal lines) is
{\it unambiguously} endowed with the unique $\{\alpha^{(k)}\}-$assignment
which matches the aim formulated subsection \ref{deform1c}.

Finally, it is straightforward to reproduce the remaining three members
of each $S(4)-$multiplet of the elementary graphs, employing the
$S(4)-$reattachments defined in subsection \ref{S(4)}.
Then, for $h_{rv}=2$ one readily combines the latter $jrv-$multiplets into 
the pairs related via the $S(2)-$reflections interchanging the horizontal
(or, what is equivalent in the $v=1$ case, vertical) sides of the contour
$C$.


\app{The $\{\Delta \tau_{q(k)}\}-$assignment}
\label{freedom}

By virtue of the $S(4)\otimes S(2)-$symmetry
implemented in Section \ref{parameter} and Appendix \ref{enumer1},
there is the following
short-cut way to introduce the prescription that fixes the
$\{\Delta \tau_{q(k)}\}-$assignment (entering Eq. (\ref{KEY.01}))
unambiguously for all the effective
amplitudes collected into the $S(4)\otimes S(2)-$multiplets. For all
inequivalent values of $j$, $r$, and $v$, we first fix the
prescription\footnote{In certain cases, this assignment may be imposed in a
few alternative ways without changing the corresponding effective amplitude.
The prescription fixes this freedom in the $S(4)-$invariant way.} for a
single graph in a particular $S(4)\otimes S(2)-$multiplet with given
$\gamma jrv-$assignment. Then, it is verified that
the pattern of the prescription is not changed when adapted
to the remaining graphs obtained employing the
$S(4)-$reattachments
combined with the $S(2)-$reflections.

In turn, given an elementary graph representing such a multiplet, there are
two steps to implement the $\{\Delta \tau_{q(k)}\}-$assignment.
The first step, discussed in the present Appendix, is to perform such a
change of the
variables that replaces $2n$ temporal coordinates\footnote{Recall that $l$
labels the $l$th line of a given graph, $s'_{l}<s_{l}$ for $\forall{l}$, and
the proper-time parameterization goes clockwise starting with the left lower
coner of $C=\Box$.} $x^{2}(s_{l})$ and $x^{2}(s'_{l})$,
constrained by $G=1$ Eq. (\ref{1.40kk}), by $n+2$ independent
parameters $\tau_{i}$. At the second step, one is to determine the function
$q(k):~k\rightarrow{q}$. The latter step is established in Appendix
\ref{dMUL.04}.

\sapp{The $r=v=0$ case}

Both of the steps are most straightforward in the case of the $r=v=0$
multiplets when the realization of the two
relevant symmetries of the assignment in question is routine as well.
Presuming that $s_{k}\geq s'_{k}$ for $\forall{k}$,
the first step can be formalized by the prescription
\be
x^{2}(s'_{k})=\tau_{k}~~~~~,~~~~~x^{2}(s_{k})=\tau_{k+j+1}~~~,~~~k=1,2
~~~~~,~~~~~(j-1)\left(x^{2}(s'_{3})-{\tau}_{3}\right)=0~,
\label{FA.02y}
\ee
where $x^{\mu}(s'_{k}),~x^{\mu}(s_{k})$ are the end-points of the left
($k=1$) and right ($k=2$) lines in fig. 1a ($j=1$) and 2c ($j=2$).

\sapp{The $v=1$ cases}

Concerning the $v=1$ cases\footnote{Recall that, one is to restrict the
admissible positions of the lower end-points of those $j$ nonhorizontal lines
which are not involved into the $S(4)-$reattachments. In the $r=v-1=0$
and $r=v=1$ cases, it is fixed by Eqs. (\ref{RE.02a}) and (\ref{INN.04})
correspondingly.}, consider first the
$j=1$ graphs which, being depicted by bold lines in
figs. 8a, 8b and 9a, 9b, are associated to $r=0$ and $r=1$ respectively, where
$\gamma=1$ and $\gamma=2$ are assigned to figs. 8a,9a and
8b, 9b correspondingly. In all figures, $\tau_{1}$ and $\tau_{4+r}$ should be
identified respectively with the temporal coordinates of the leftmost and
rightmost end-points of the elementary graph, belonging to $1+r$ bold lines
(defining the associated protograph). Next, the remaining $1+r$ end-points
of the latter lines
can be as well directly identified with the
corresponding parameters $\tau_{i}$ so that it can be
summarized by equations
\be
x^{2}(s'_{1})=\tau_{1}~~,~~
x^{2}(s_{1})\delta_{1\gamma}+x^{2}(s'_{4})\delta_{2\gamma}=\tau_{4+r}~~,~~
x^{2}(s_{2})=\tau_{1+r+\gamma}~~~,~~~
\delta_{2r}\cdot\left(x^{2}(s_{4})-\tau_{2}\right)=0~,
\label{REP.01}
\ee
where $\delta_{nm}$ denotes the standard Kronecker delta-function with
$\delta_{nn}=1$ and $\delta_{nm}$ for $\forall n\neq m$.

For a given $n+2=3+j+r$, the direct reidentification (\ref{REP.01}) allows
to define only $n+1$ parameters $\tau_{i}$. The remaining $(n+2)$th parameter
$\tau_{4+r-\gamma}$ has to be introduced via the following procedure
which is also used to determined the corresponding interval\footnote{In turn,
 $\Delta \tau_{q(2)}$ is to be identified with the interval spanned
by the lower end-point of the second bold line
in the process of this parallel transport: $q(2)=2+r$.}
$\Delta \tau_{q(2)}$. The proposal is to identify $\tau_{4+r-\gamma}$ with the
new position
\be
\left(x^{2}(s_{1})\delta_{1\gamma}+x^{2}(s'_{1})\delta_{2\gamma}\right)+
(-1)^{\gamma}t_{2}
=\tau_{4+r-\gamma}
\label{REP.02}
\ee
of the lower end-point of the second bold line resulting after
the judicious {\it parallel transport} of this line. Namely, the line is
transported, until its upper end-point hits the corresponding end-point of
the first bold line, to the right in the $\gamma=1$ case of figs. 8a, 9a and
to the left in the $\gamma=2$ case of figs. 8b, 9b.
Note also that
$\bar\tau_{4+r-\gamma}$ describes the collective coordinates defining the
measure (\ref{CLO.01}).

Turning to the $j=2$ case of figs. 8c and 9c (both assigned with $\gamma=1$),
we first note that the addition
of the extra bold line (compared to figs. 8a, 8b and 9a, 9b) 
results in the one more
delta-function in the $G=1$ factor (\ref{1.40kk}). 
In consequence, compared to the
associated $j=1$ cases, only a single additional parameter $\tau_{i}$ is 
introduced which can be directly identified with the
the temporal coordinate $x^{2}(s_{3})$ of the lower end-point of this extra
line (which, being nonhorizontal, is not involved into the reattachments).
As for the remaining $n+1=4+r$ parameters $\tau_{k}$, they are defined
in the way similar to the previous $j=1$ discussion.

Actually, it can be reformulated in the more geometrically clear way. For this
purpose, in all figures, $\tau_{1}$ and $\tau_{5+r}$ should be
identified correspondingly with the temporal coordinates of the leftmost and
rightmost end-points of the elementary graph depicted by the bold
lines. Additionally, the $2+r$ end-points (of the latter lines)
can be as well directly identified with the
corresponding parameters $\tau_{i}$ that can be summarized in the form
\be
x^{2}(s'_{1})=\tau_{1}~,~
x^{2}(s_{1})\delta_{1\gamma}+x^{2}(s'_{4})\delta_{2\gamma}=\tau_{5+r}~,~
x^{2}(s_{2})=\tau_{4+r}~,~
x^{2}(s_{3})=\tau_{2+r}~,~
\delta_{1r}\left(x^{2}(s_{4})-\tau_{2}\right)=0.
\label{REP.03}
\ee
In this way, we define the $n+1$ parameters while
the so far missing $(n+2)$th parameter
$\tau_{3+r}$ can be determined through the following procedure
utilizing the double parallel transport which, geometrically, can be
visualized the triangle-rule (most transparent in figs. 8f and 9f).
The proposal is to identify $\tau_{3+r}$ with the
position
\be
\tau_{1}+t_{2}=\tau_{3+r}
\label{REP.04}
\ee
where the two lower end-points of the second and the third bold lines
coalesce when these two lines are transported until their upper end-points
{\it simultaneously} hit the corresponding end-points of the first
(horizontal)
bold line. In turn, it implies the algebraical fine-tuning maintained
by the first of the conditions (\ref{SR.01y}) which, geometrically, means
that (when properly transported and reoriented) the three vector
${\bf y}_{k},~k=1,2,3,$ can be combined into a triangle\footnote{A direct
inspection of fig. 5b and figs. 6 reveals that, with a minor modification,
a similar triangle-rule can be formulated in the $v=r=0$ case
as well.} in the $a_{1}=0$ case of figs. 8c and 9c.

\sapp{The $S(4)-$ and reflection-invariance}
\label{assisym}

Evidently, the proposed algorithm to introduce the
$\{\Delta \tau_{q(k)}\}-$assignment is not changed after a generic
combination of the $S(4)-$reattachments. Indeed, it readily follows from the
fact that, keeping the temporal coordinates of the
end-points intact, they are applied only to the right- or/and leftmost
end-points of elementary graphs. 

Concerning the reflection-invariance, consider first the $r=v=0$ case. Then
the reflection (interchanging the horizontal sides of the rectangle $C$)
is applied to the two $S(4)-$multiplets corresponding to the figs. 1a (with
$j=1$) and 2c (with $j=2$). In the reflection-partners represented by figs. 1b
and 2d respectively, the time-intervals $\Delta\bar{\tau}_{k}$ be associated
to the lower horizontal side of $C$. In the latter two figures, we
parameterize the left and the right horizontal lines by label 1 and 2
correspondingly (so that, in fig. 2d, the remaining nonhorizontal line is
assigned with the label 3). Introducing the parameters $\tau_{i}$ by the same
token as previously, it guarantees that the function
$q(k)$ is reflection-invariant. Also, compared to the case of fig. 1a and 2c,
the figs. 1b and 2d can be characterized via the replacements
$t_{p}\rightarrow{-t_{p}}$, ${\cal C}_{il}\rightarrow
-{\cal C}_{il}$ with $p=1,2$ and $i,l=1,2,3$. In turn, the latter replacements
follow from the definitions (\ref{1.38aa}) and (\ref{1.34}) which are
augmented by the convention
to implement the proper-time parameterization (implying, in particular, that
$s_{l}\geq s'_{l}$ for $\forall{l}$).

Finally, consider the remaining case of the three pairs of the $r=v=1$
$S(4)-$multiplets (assigned with $\gamma=1,2$ for $j=1$ and $\gamma=1$ for
$j=2$) which, within a particular pair, are related through the
reflection interchanging the vertical sides of $C=\Box$.

In each of the latter multiplets it is sufficient to consider the single
elementary graph with the two horizontal lines. E.g., see figs. 9g and 9h
which are the reflection-partners of figs. 9b and 9c respectively.
For concreteness, we restrict\footnote{The remaining two pairs, associated to
figs. 9a and 9b, are handled in a similar way.}
the discussion to figs. 9c and 9h, associating
the time-intervals $\Delta\bar{\tau}_{k}$ to
the lower horizontal side of $C$. In the latter two figures, we
parameterize the left and the right horizontal lines by label 1 and 4
correspondingly. Then, to maintain the reflection-covariance of the algorithm
(introduced in the previous subappendix), in the case of fig. 9h one is to
perform the additional change of the variables
$\bar{\tau}_{k}\rightarrow \bar{\tau}_{n+3-k}$ with $k=1,...,n+2$
(possessing the jacobian equal to unity) that results in the 
reidentification
$\Delta\bar{\tau}_{k}\rightarrow \Delta\bar{\tau}_{n+2-k}$ applied to
$k=0,...,n+2$. (As previously, 
we require that $s_{l}\geq s'_{l}$ for $\forall{l}$.)
This reidentification evidently implies the transformation
$q(k)\rightarrow q(n+2-k)$, provided the labels 2,3 are assigned to the
remaining nonhorizontal lines
so that ${\cal C}_{32}\rightarrow -{\cal C}_{32}$ (while
${\cal C}_{1p}\rightarrow -{\cal C}_{1p}$ for $p=2,3$). In turn, a direct
inspection demonstrates that, after this transformation, the function $q(k)$
assumes the same form as in the case of fig. 9c which verifies its
reflection-covariance. As for the splitting
(\ref{FU.02}), the reflection-partners 
can be characterized through the replacements 
$t_{i}\rightarrow{-t_{i}}$ for $i=1,2,3,4$.

\app{Justifying Eq. (\ref{MUL.04y})}
\label{dMUL.04y}

To transform the superposition (\ref{SU.01}) into the form of Eq.
(\ref{MUL.04y}), in the integral representation (\ref{FR.01})
of ${\cal Z}^{(\gamma)}_{jrv}(\cdot)$ one is to first perform
(for each $v+j-1>0$ term) the change of the variables
\be
\int d^{2+n}\bar\tau~...~~~\longrightarrow~~~\int d^{2+h_{rv}}\bar\tau~
\int_{\bar\tau_{q_{\gamma}(3)-1}}^
{\bar\tau_{q_{\gamma}(3)+1}}
d^{j-1}\bar\tau_{q_{\gamma}(3)}
\int_{\bar\tau_{q_{\gamma}(2)-1}}^
{\bar\tau_{q_{\gamma}(2)+1}}
d^{v}\bar\tau_{q_{\gamma}(2)}~...
\label{FI.03}
\ee
that manifestly separates the $2+h_{rv}$ collective coordinates combined
into the measure (\ref{CLO.01}), provided $q_{\gamma}(p)=q(p)+
\omega^{(\gamma)}_{p}$, where $\omega^{(\gamma)}_{p}=0,1$ is explicitly
constructed in subappendix \ref{omega} so that the prescription,
formulated in the end of subsection \ref{completeness}, is valid.
In turn, due to the presence
of the corresponding number of the derivatives in the r.h.side of Eq.
(\ref{FR.01}), the remaining $v+j-1$ integrations\footnote{Recall
that these integrations are associated to those lines (of a given elementary
graph) which, being non-horizontal, are {\it not} involved into the
$S(4)-$reattachments.} (with respect to $\bar\tau_{q_{\gamma}(p)}\in
[\bar\tau_{q_{\gamma}(p)-1},\bar\tau_{q_{\gamma}(p)+1}]$)
are readily performed. The computation is simplified by the
fact\footnote{It is this fact that verifies the prescription (\ref{VAR.02}).}
that, by construction of $\omega^{(\gamma)}_{p}$, both
$\bar{t}^{(\gamma)}_{1},~\bar{t}^{(\gamma)}_{2}$ and
$\Delta\bar\tau_{q(i)}$ are {\it independent} of
$\bar\tau_{q_{\gamma}(p)}$ for $\forall{i\neq{p}}$, $\forall{p}=3-v,3$, and
$\forall{\gamma}=1,f_{jv}$, while $(-1)^{\omega^{(\gamma)}_{3}-1}
\partial/\partial \bar\tau_{q_{\gamma}(p)}$ can be replaced by
$\partial/\partial \Delta\bar\tau_{q(p)}$ when it acts on the
$\Delta\bar\tau_{q(p)}-$dependent factor (\ref{SR.01a}).
In consequence, in the expression (\ref{SR.01g}) for
$\tilde{\cal V}^{(\gamma)}_{jrv}(\cdot)$, the dependence on
$\tau_{q_{\gamma}(p)}$ is localized in the corresponding $k=p$ implementation
of the factor (\ref{SR.01a}). Furthermore, the interval
$\Delta \bar\tau_{q(p)}$ varies in the domain
$[0,\Delta \bar{T}^{b}_{p}(0,\gamma)]$ (where
${T}^{b}_{p}(0,\gamma)=\tau_{q_{\gamma}(p)+1}-\tau_{q_{\gamma}(p)-1}$
is defined in Eq. (\ref{KEY.01b})) when
$\bar\tau_{q_{\gamma}(p)}$ spans the domain 
$[\bar\tau_{q_{\gamma}(p)-1},\bar\tau_{q_{\gamma}(p)+1}]$.

Altogether, the amplitude (\ref{FR.01}) can be rewritten in the form which
can be obtained from Eq. (\ref{MUL.04y}) through the replacement
\be
{\cal V}_{2rv}(\{a_{k}\},\{\Delta \bar{T}^{b}_{k}\})
~~~\longrightarrow~~~\sum_{j=1}^{2}\sum_{\gamma=1}^{f_{jv}}
\left[\sum_{\breve\tau_{3}=0}^{1}
(-1)^{\breve\tau_{3}}\right]^{j-1}
\left[\sum_{\breve\tau_{2}=0}^{1}
(-1)^{\breve\tau_{2}}\right]^{v}
{\cal V}^{(\gamma)}_{jrv}(\{a_{i}\},\{\Delta \bar\tau_{q(k)}\})
\Big|_{\{\breve\tau_{p}\}}~,
\label{FI.04}
\ee
where sum over $\breve\tau_{p}=
(\Delta \bar{T}^{b}_{p}(0,\gamma)-\Delta \bar\tau_{q(p)})/
\Delta \bar{T}^{b}_{p}(0,\gamma)$
reproduces the sum over the boundary values of the relevant intervals
$\Delta \bar\tau_{q(p)}$, while ${\cal V}_{2rv}(\cdot,\cdot)\equiv
{\cal V}^{(1)}_{2rv}(\cdot,\cdot)$ as well as in Eq. (\ref{MUL.04y}).

Next, in the r.h. side of Eq. (\ref{FI.04}), there are mutual
cancellations (see Eqs. (\ref{FI.01}) and (\ref{FI.02}) below) which,
due to the $S(4)-$invariance of the $\bar{\cal R}^{-1}_{b}-$dressing,
are maintained between the $j=1$ and $j=2$ terms
considered {\it separately} for any admissible $\{a_{l}\}-$assignment.
As a result, survives only the single $j=2$ term
\be
{\cal V}_{2rv}(\{a_{i}\},\{\Delta \bar\tau_{q(k)}\})\Big|_
{\{\Delta\bar\tau_{q(p)}=\Delta \bar{T}^{b}_{p}(0)\}}
={\cal V}_{2rv}(\{a_{k}\},\{\Delta \bar{T}^{b}_{k}\})
\label{FI.05}
\ee
(with $\bar{T}^{b}_{p}(0)\equiv \bar{T}^{b}_{p}(0,\gamma)|_{\gamma=1}$)
characterized by the condition
\be
\Delta\bar\tau_{q_{\gamma}(p)-1}=0~~~~\Longrightarrow~~~~
\Delta\bar\tau_{q(p)}=\Delta \bar{T}^{b}_{p}(0,1),~~~~~~,~~~~~
\forall{p}=3-v,3~,
\label{REI.01}
\ee
that reduces the number $2+n$ of the original variables $\bar\tau_{i}$,
entering Eq. (\ref{FR.01}), to the smaller amount $2+h_{rv}$ associated to
Eq. (\ref{MUL.04}). In consequence, for fixed values of those
$\bar\tau_{k}$ which define the collective coordinates entering the measure
(\ref{CLO.01}), it maintains the {\it maximal} value of
$\sum_{p=3-v}^{3} \Delta \tau_{q(p)}$,
where $p$
labels those $v+j-1$ lines of a given elementary graph which, being associated
to $f_{k}=0$ replacement (\ref{KEY.01}), are not
involved into the $S(4)-$reattachments. In turn, by virtue of the $j=2$
constraints (\ref{SR.01y}), it supports the completeness
condition (\ref{KEY.01a}).
Altogether, it verifies Eq. (\ref{MUL.04y}).

As for the asserted mutual cancellations, the simplest situation takes place
in the $r=v=0$ case when the parameter $\gamma$, assuming the
singe value (since $f_{j0}=1$ according to Eq. (\ref{SU.01l})), can be safely
omitted. Therefore, for each admissible values of $a_{1}$ and $a_{2}$
(involved in the $v=0$ summation in Eq. (\ref{MUL.04y})), the fine-tuning
takes place between the pairs of effective amplitudes
${\cal Z}_{j00}(\{a_{k}\},\bar{A},\bar{\theta}^{-1})$ with $j=1,2$.
In this case, due to
the identity ${\cal F}(z,0)=1$ valid for $\forall{z}$ (as it is clear from
the definition (\ref{SR.01a})), the very
pattern (\ref{SR.01g}) of ${\cal V}_{jrv}(\cdot)$ ensures the relation
\be
{\cal V}_{200}(\{a_{k}\},\{\Delta \bar\tau_{q(i)}\})
\Big|_{\Delta \bar\tau_{q(3)}=0}=
{\cal V}_{100}(\{a_{k}\},\{\Delta \bar\tau_{q(i)}\}),
\label{FI.01}
\ee
so that the reduction $\Delta \bar\tau_{q(k)}=0$ converts the
$6-$set $\{\Delta \bar\tau_{k}\}$ (associated to the $j=2$ l.h. side of the
identity) into its counterpart (in the $j=1$ l.h. side) consisting of the
$5$ intervals $\Delta \bar\tau_{k}$. In turn, it proves the inverse of the
$v=0$ replacement (\ref{FI.04}) and, in consequence, the $v=0$ option of the
prescription (\ref{KEY.01}) endowed with the $\{f_{k}\}-$specification in
compliance with subsection \ref{completeness}. For the particular case of
$a_{1}=a_{2}=0$, ${\cal Z}_{00}(\{a_{k}\},\cdot)$ (resulting from the
cancellation between ${\cal Z}_{200}(\{a_{k}\},\cdot)$, fig. 5b, and
${\cal Z}_{100}(\{a_{k}\},\cdot)$, fig. 5a) is diagrammatically depicted by
fig. 5c. The remaining options of ${\cal Z}_{00}(\{a_{k}\},\cdot)$ are
represented by figs. 6a--6c.


Concerning the $v=1$ cases, the inverse of the $v=1$ replacements
(\ref{FI.04}) follow from the pair of the relations\footnote{In the derivation
of Eq. (\ref{FI.02}), we utilize that
$\Delta T^{b}_{2}(0,\gamma)|_{j=1}=\Delta T^{b}_{4-\gamma}(0,1)|_{j=2}$
provided $t^{(\gamma)}_{2}|_{j=1}=t^{(1)}_{4-\gamma}|_{j=2}$.}
\be
{\cal V}_{2r1}(\cdot,\{\Delta \bar\tau_{q(i)}\})
\Big|_{\Delta \bar\tau_{q(p)}=0}=
{\cal V}^{(p-1)}_{1r1}(\cdot,\{\Delta \bar\tau_{q(i)}\})
\Big|_{\Delta \bar\tau_{q(2)}=\bar{T}^{b}_{2}(0,p-1)}~~~~,~~~~p=2,3~,
\label{FI.02}
\ee
\be
{\cal V}_{2r1}(\cdot,\{\Delta \bar\tau_{q(i)}\})
\Big|_{\Delta \bar\tau_{q(2)}=0}^{\Delta \bar\tau_{q(3)}=0}=
{\cal V}^{(\gamma)}_{1r1}(\cdot,\{\Delta \bar\tau_{q(i)}\})
\Big|_{\Delta \bar\tau_{q(2)}=0}=0~,
\label{FI.02a}
\ee
where Eq. (\ref{FI.02}) can be deduced essentially by
the same token as Eq. (\ref{FI.01}) (while Eq. (\ref{FI.02a}) is
proved
in \cite{ADM05b}). The only new element is to take into
account that, contrary to the $r=v=0$ case (\ref{FI.01}), there are two
$p=2,3$  options to implement the $j=2\rightarrow j=1$ reduction (of the
$(6+r)-$set $\{\Delta \bar\tau_{k}\}$ into the corresponding $(5+r)-$set)
so that the $p$th option is
associated to the $\gamma=p-1$ implementation of
${\cal V}^{(\gamma)}_{1r1}(\cdot)$.
Geometrically, for the particular $\{a_{k}\}-$assignments, the latter
identification is clear from the comparison of the $j=2$ figs. 8c
and 9c with the $j=1$ pairs of the figs. 8a, 8b and 9a, 9b respectively. 
(In the derivation of this representation of
${\cal Z}^{(1)}_{1r1}(\cdot)$, we also utilize the
change of the variables $\bar\eta\rightarrow{\bar\eta+\bar\zeta}$
$\bar\zeta\rightarrow{\bar\zeta}$ that, in the combination
$\bar\eta \bar{t}^{(1)}_{1}-\bar\zeta \bar{t}^{(1)}_{2}$ entering the
relevant option of Eq. (\ref{SR.01g}), replaces $\bar{t}^{(1)}_{2}$
by $\bar{t}^{(2)}_{2}$.)

Finally, it is possible to diagrammatically visualize the $v=1$
replacement (\ref{FI.04}), in the
form similar to the $r=v=0$ one.
For simplicity, we as previously restrict the discussion
to the case of the $\{a_{k}\}-$assignments with $a_{1}=0$ and, when $r=1$,
$a_{4}=0$. Then, observe
first that (in the $\gamma=1$ case) the relation
(\ref{FI.02a}) implies the equivalence of the effective amplitudes
associated to figs. 8a, 9a and 8d, 9d correspondingly. Next, the $p=3$
variant of the relation (\ref{FI.02}) guarantees that
the superposition ${\cal Z}^{(2)}_{1r1}(\{a_{k}\},\cdot)+
{\cal Z}^{(1)}_{2r1}(\{a_{k}\},\cdot)$ is diagrammatically represented
by figs. 8e and 9e when $r=0$ and $r=1$ respectively.
As for ${\cal Z}_{r1}(\{a_{k}\},\cdot)$, being depicted in figs.
8f  and 9f when $r=0$ and $r=1$ correspondingly, it results after the
residual cancellation which takes place, by the same token as in the $r=v=0$
case, between effective amplitudes of figs. 8e (9e) and 8d (9d).

\sapp{The choice of the $\{\omega^{(\gamma)}_{k}\}-$assignment}
\label{omega}

It remains to introduce the appropriate set of the parameters
$\omega^{(\gamma)}_{k}$, where $k=3-v,3$ labels
those $n-h_{rv}=v+j-1$ lines of the elementary graph which are {\it not}
associated to the corresponding protograph, i.e.,
$k\in {\cal X}_{jrv}\equiv\tilde\Omega_{jrv}/{\cal S}_{rv}$ (where the sets
$\tilde\Omega_{jrv}$ and ${\cal S}_{rv}$ are introduced in the end of
subsection \ref{assignment1}).
For this purpose, we propose the following algorithm. First, we observe
that the parameters
$\bar\tau_{q_{\gamma}(k)}$ ($k=3-v,3,~q_{\gamma}(p)=q(p)+
\omega^{(\gamma)}_{p}$) represent the temporal coordinates
which remain dynamical when one fixes both the positions of the end-points
of the corresponding protograph's line and, in the $v=1$ case, an admissible
value of $t^{(\gamma)}_{2}$. In compliance with Appendix
\ref{freedom}, for $v+j-1>0$ it leaves variable exactly $v+j-1$ independent
temporal coordinates of either upper (when $v=0$) or lower
end-points of the $v+j-1$ lines labeled by $k\in {\cal X}_{jrv}$.
Correspondingly, each of thus
introduced parameters $\bar\tau_{q_{\gamma}(k)}$ is associated to the two
adjacent intervals $\Delta\bar\tau_{q_{\gamma}(k)-i}$ with $i=0,1$ so that
$\sum_{0}^{1}\Delta\bar\tau_{q_{\gamma}(k)-i}=\Delta{T}^{b}_{p}(0,\gamma)$.
Then, it is a matter of convention to choose one of the
two possible values of $i=i_{\gamma}(k)$ in order to identify
$q(k)=q_{\gamma}(k)-i_{\gamma}(k)$ for a given $\gamma$. Having fixed this
freedom\footnote{A direct inspection of the elementary graphs verifies that
this freedom is absent for the remaining $h_{rv}$ lines which, being endowed with the
$\bar{\cal R}^{-1}_{b}-$dressing, are assigned with $k\in {\cal S}_{rv}=
\Omega_{jrv}/{\cal X}_{jrv}$, while $f_{k}=1$ in the sence of Eq.
(\ref{KEY.01b}), i.e., $\Delta{T}^{b}_{k}(1,\gamma)=
\Delta\bar\tau_{q(k)}$ for $k\in {\cal S}_{rv}$.} according to the
prescription of Appendix \ref{dMUL.04}, one is led to the
identification $\omega^{(\gamma)}_{k}=i_{\gamma}(k)$.
Given this prescription, one obtains
(in the $S(4)-$invariant and reflection-covariant way
in the sense of subappendix \ref{assisym})
that $\omega^{(1)}_{3}=0$ for $r=v=j-2=0$, $\omega^{(\gamma)}_{2}=\gamma-1$
for $v=j=1$ (and $\forall{r}=0,1$), while
$\omega^{(1)}_{p}=3-p$ for $v=j-1=1$ (and $\forall{r}=0,1$).
Also, it supports the prescription,
formulated in the end of subsection \ref{completeness}.

Next, by construction, thus introduced $n-v$ intervals
$\Delta{T}^{b}_{k}(f_{k},\gamma)$ meet the
important constraint (justified by a direct inspection of the relevant
elementary graphs): these intervals are mutually {\it nonoverlapping}.
Furthermore, in the $j=2$ case, the very number $n-v$ of the intervals ensures
that they comply with the completeness condition (\ref{KEY.01a}). (Among the
$n+1$ intervals $\Delta\bar\tau_{i}$, comprising the residual temporal
interval in the r.h. side of this condition, there are exactly $v+1$ pairs
combined into the corresponding intervals $\Delta{T}^{b}_{p}(0,1),~p=3-v,3$.)
Also, the latter constraint
guarantees that $\Delta\bar\tau_{q(i)}$ is independent of
$\bar\tau_{q_{\gamma}(p)}$ for $\forall{i\neq{p}}$, $\forall{p}=3-v,3$.
Finally, it is straightforward to argue that the same independence of
$\bar\tau_{q_{\gamma}(p)}$ holds true
for $\bar{t}^{(\gamma)}_{1},~\bar{t}^{(\gamma)}_{2}$ as well.
It is most transparent in the $r=v=0$ case where these relative times are
fully determined by the positions of the end-points of the $2+r-v$ lines
involved into the $S(4)-$reattachments. In the $v=1$ case, this argument still
applies to $\bar{t}^{(\gamma)}_{1}$, while the independence of
$\bar{t}^{(\gamma)}_{2}$ is verified by the relation (\ref{REP.02}).

\app{Explicit implementation of
$\tilde{\cal V}_{2rv}(\{a_{i}\},\{\Delta \tau_{q(k)}\})$}
\label{dMUL.04}

The aim of this Appendix to explicitly determine the
$\{a_{l}\}-${\it dependent}
parameters which define the relevant implementation of the pattern
(\ref{SR.01g}) of the quantity $\tilde{\cal V}_{2rv}(\cdot)$ entering Eq.
(\ref{MUL.04y}). In compliance with the discussion of subsection
\ref{enumer1}, our strategy is to introduce the required parameters for
generic $\{a_{l}\}-$assignment as the $\{a_{l}\}-$dependent deformation of the
parameters associated to a particular elementary graph in a given
$rv-$variety of the elementary diagrams. 
In the next Appendix, we will verify that, after an
appropriate change of the variables $\bar\zeta$ and $\bar\eta$, one can
rewrite this quantity in the form matching Eq. (\ref{MUL.04}).

\sapp{The $r=v=0$ case}

Consider first the $a_{1}=a_{2}=0$ contribution to the $r=v=0$ superposition
(\ref{MUL.04y}) which is determined by such implementation of
$\tilde{\cal V}_{200}(\{a_{i}\},\{\Delta \tau_{q(k)}\})$ that is
parameterized by the $j=2$ graph\footnote{Recall that
both the elementary and the effective amplitudes, associated to figs. 2a
and 2b, are vanishing due to the
specific implementation of the ${V}_{U_{\theta}(1)}^{(n)}(\cdot)\rightarrow
\tilde{V}_{U_{\theta}(1)}^{(n)}(\cdot)$ option of
the $G=1$ constraints (\ref{1.40kk}).} 2c
(when ${\cal C}_{12}={\cal C}_{13}={\cal C}_{23}=-1$),
the deformations of which are described in fig. 5b. Defining 
$\bar{\tau}_{k}$ and $\Delta\bar{\tau}_{k}$ according
to the $j=2$ Eq. (\ref{FA.02y}), the $\{a_{l}\}-$independent parameters
$\bar{t}_{p}$ are determined by the $z=0$ variant relations
\be
(-1)^{a_{1}+z}\bar{t}_{1}=\bar{\tau}_{4}-\bar{\tau}_{1}=
\Delta \bar{\tau}_{1}+\Delta \bar{\tau}_{2}+\Delta \bar{\tau}_{3}~~~~~,~~~~~
(-1)^{z}\bar{t}_{2}=\bar{\tau}_{5}-\bar{\tau}_{2}=
\Delta \bar{\tau}_{2}+\Delta \bar{\tau}_{3}+\Delta \bar{\tau}_{4}~.
\label{AM.05}
\ee
with $\bar{t}_{1}-\bar{t}_{2}+\bar{t}_{3}=0$, while the convention to fix the
labels $k=1,2,3$ is fixed in Appendix \ref{enumer1}. Correspondingly, it
leads to the $a_{1}=z=0$ option of the identification
\be
q(2)=1~,~q(1)=4~,~q(3)=3~~,~~\alpha^{(p)}=
(-1)^{z}{\cal C}_{32}=1~~,~~-\alpha^{(1)}=
(-1)^{z}{\cal C}_{p1}=(-1)^{a_{1}}~,~p=2,3~,
\label{AMB.01}
\ee
where, as it should, the function $q(k)$ is $\{a_{l}\}-$independent.
Also, the $\omega^{(1)}_{3}=0$ option of the $v=0$ Eq. (\ref{REI.01}) implies
that the measure of the $r=v=0$ representation (\ref{MUL.04}) is obtained through the
reidentification: $\bar\tau_{i}\rightarrow\bar\tau_{i}$ for $i=1,2$, while
$\bar\tau_{i}\rightarrow\bar\tau_{i-1}$ for $i=4,5$ so that $\bar\tau_{3}$
disappears.

As for the remaining three contributions to the $r=v=0$ superposition
(\ref{MUL.04y}), they are associated to such implementation of the
quantity $\tilde{\cal V}_{200}(\{a_{i}\},\{\Delta \tau_{q(k)}\})$ that are
parameterized by the $v=0$ components of the diagrams 2e and 2g obtained
through the vertical reattachments applied to fig. 2c. For this purpose, the
leftmost or/and rightmost end-point
of the pair of the lines in fig. 2c is/are replaced, keeping their
time-coordinates $x^{2}(s'_{1})$ and $x^{2}(s_{2})$ intact, from the upper
to the lower horizontal side of the rectangle $C$. Taking into account
that the vertical $1-$axis is directed from the upper to the lower horizontal
side of the rectangle $\Box$, it is formalized
by the relations
\be
x^{1}(s'_{1})=a_{1}R~~~~~,~~~~~x^{1}(s_{2})=a_{2}R~~~~~,~~~~~
x^{1}(s'_{2})=x^{1}(s_{1})=0~.
\label{AM.07d}
\ee
When $a_{1}+a_{2}\geq 1$, it parameterizes the three different implementations
of $\tilde{\cal V}_{200}(\cdot)$ which, being associated to figs. 6a--6c, are
described by the corresponding $\{a_{k}\}-$implementations of Eqs.
(\ref{AMB.01}) and (\ref{AM.05}), where one is to put $z=0$.

Finally, to compute the entire $r=v=0$ contribution to the decomposition
(\ref{SU.01z}), it remains to include the contribution of such $r=v=0$
superposition (\ref{MUL.04y}) that is associated to the $S(4)-$multiplet
of the $j=2$ elementary graphs specified by the graph in fig. 2d.
Alternatively, these graphs can be obtained from the previously constructed
$j=2$ $S(4)-$multiplet (specified by the graph in fig. 2c) via the reflection
interchanging the horizontal sides of the rectangle $C$.
Then, introducing the $\{\Delta\bar{\tau}_{k}\}-$assignment according to the
convention of subappendix \ref{assisym}, one
arrives at the
$r=v=0$ implementation of Eq. (\ref{MUL.04y}) fixed by the $z=1$
option of Eqs. (\ref{AM.05}) and (\ref{AMB.01}).

\sapp{The $r=v-1=0$ case}

Next, consider the $a_{1}=0$ contribution to the $r=v-1=0$ superposition
(\ref{MUL.04y}) which is determined by such implementation of
$\tilde{\cal V}_{201}(\{a_{i}\},\{\Delta \tau_{q(k)}\})$ that is parameterized
by the $v=1$ component of the $j=2$ diagram 2e
(when ${\cal C}_{12}={\cal C}_{13}={\cal C}_{23}=-1$),
the deformations of which are described in fig. 8c. 
It is geometrically evident that there is the single $v=1$
component (assigned with $\gamma=1$) of the latter diagram which is
constrained by the $p=2,3$ options of the condition
\be
x^{2}(s_{p})\in[x^{2}(s'_{1}),x^{2}(s_{1})]
\label{RE.02a}
\ee
applied to both of the nonhorizontal lines.

In this case, introducing $\bar{\tau}_{k}$ and $\Delta\bar{\tau}_{k}$
according to the $r=0$ prescription of Eqs. (\ref{REP.03}) and (\ref{REP.04}),
the decomposition of the parameters $\bar{t}_{p}$ is determined by the
$a_{1}=\tilde{a}_{1}=0$ variant of the relations
\be
(-1)^{a_{1}+\tilde{a}_{1}}\bar{t}_{1}=\bar{\tau}_{5}-\bar{\tau}_{1}=
\Delta \bar{\tau}_{1}+\Delta \bar{\tau}_{2}+
\Delta \bar{\tau}_{3}+\Delta \bar{\tau}_{4}~~~~~,~~~~~
\bar{t}_{2}=\bar{\tau}_{3}-\bar{\tau}_{1}=\Delta \bar{\tau}_{1}+
\Delta \bar{\tau}_{2}~,
\label{INN.03}
\ee
with $\bar{t}_{1}-\bar{t}_{2}+\bar{t}_{3}=0$, and
the convention\footnote{Akin to fig. 2c, the convention
${\cal C}_{23}=-1$ implies that $x^{2}(s_{3})<x^{2}(s_{2})$.}
to fix the labels $k=1,2,3$ is sketched in subsection
\ref{enumer1}. Correspondingly, it yields the $a_{1}=\tilde{a}_{1}=0$
option of the identification
\be
q(2)=3~,~q(3)=2~~~,~~~\alpha^{(3)}=\alpha^{(2)}=
{\cal C}_{32}=1~~~,~~~{\cal C}_{p1}=(-1)^{a_{1}+\tilde{a}_{1}}~,~p=2,3~.
\label{INN.02a}
\ee
Also, the $\omega^{(1)}_{p}=3-p$ option of the $v=0$ Eq. (\ref{REI.01})
implies that the measure of the $r=v-1=0$ representation (\ref{MUL.04}) is
obtained through the reidentification: $\bar\tau_{1}\rightarrow\bar\tau_{1}$,
$\bar\tau_{3}\rightarrow\bar\tau_{2}$, while $\bar\tau_{5}\rightarrow
\bar\tau_{3}$ so that $\bar\tau_{2}$ and $\bar\tau_{4}$ disappear.

Then, the remaining three contributions to the $r=v-1=0$ superposition
(\ref{MUL.04y}) are associated to such implementation of the
quantity $\tilde{\cal V}_{201}(\{a_{i}\},\{\Delta \tau_{q(k)}\})$ that are
parameterized by the $v=1$ components of the diagrams 2f and 2g. In
turn, the latter elementary graphs can be obtained from the $v=1$
component of the diagram 2e through the vertical reattachments
of the left or/and right end-point of its single horizontal line that is
formalized by Eq. (\ref{ST.01}). Together with the
already considered $w=a_{1}=0$ case, it generates the four different
implementations of $\tilde{\cal V}_{201}(\cdot)$ which are described by the
corresponding $\{a_{i}\}-$dependent implementations of
Eqs. (\ref{INN.03}) and (\ref{INN.02a}).

\sapp{The $r=v=1$ case}

Consider the $a_{1}=a_{4}=0$ contribution to the $r=v=1$
superposition (\ref{MUL.04y}) which is determined by such implementation of
$\tilde{\cal V}_{211}(\{a_{i}\},\{\Delta \tau_{q(k)}\})$ that is parameterized
by that component of the Feynman diagram 7e
(when ${\cal C}_{12}={\cal C}_{13}={\cal C}_{23}=-1$),
where the upper horizontal line is on the
left compared to the lower one.
This component, the deformations of which are described in fig. 9c, is
constrained by the $p=2,3$ options of the condition
\be
x^{2}(s_{p})\in[x^{2}(s'_{4}),x^{2}(s_{4})]~~~~~,~~~~~
x^{2}(s_{1})\leq x^{2}(s'_{4,j})\leq T~,
\label{INN.04}
\ee
where $x^{2}(s'_{4,j})$ denotes the temporal coordinate of the end-point of
the $j$th $\bar{\cal R}_{b}^{-1}-$copy that is common for both bold
horizontal lines in fig. 9c. 

Defining $\bar{\tau}_{k}$ and $\Delta\bar{\tau}_{k}$
according to the $r=0$ prescription of Eqs. (\ref{REP.03}) and (\ref{REP.04}),
the decomposition of the parameters $\bar{t}_{p}$ is determined
by the $a_{1}=z=0$ relations
\be
(-1)^{a_{1}+z}\bar{t}_{1}=\bar{\tau}_{6}-\bar{\tau}_{2}=
\Delta \bar{\tau}_{2}+\Delta \bar{\tau}_{3}+
\Delta \bar{\tau}_{4}+\Delta \bar{\tau}_{5}~~~~~,~~~~~
(-1)^{z}\bar{t}_{2}=\bar{\tau}_{4}-\bar{\tau}_{1}=\Delta \bar{\tau}_{1}+
\Delta \bar{\tau}_{2}+\Delta \bar{\tau}_{3}~,
\label{INN.09}
\ee
with $\bar{t}_{1}-\bar{t}_{2}+\bar{t}_{3}=0$.
In turn, the deformations of fig. 7e, depicted in
fig. 9c, are described by the $a_{1}=z=0$
option of the identification
\be
q(2)=4~,~q(3)=3~,~q(4)=1~~,~~\alpha^{(p)}=
(-1)^{z}{\cal C}_{32}=1~~,~~-\alpha^{(1)}=
(-1)^{z}{\cal C}_{p1}=(-1)^{a_{1}}~,~p=2,3.
\label{INN.10}
\ee
Also, the $\omega^{(1)}_{p}=3-p$ option of the $v=1$ Eq. (\ref{REI.01})
implies that the measure of the $r=v=1$ representation (\ref{MUL.04}) is
obtained through the reidentification: $\bar\tau_{i}\rightarrow\bar\tau_{i}$
for $i=1,2$, $\bar\tau_{4}\rightarrow\bar\tau_{3}$, while
$\bar\tau_{6}\rightarrow \bar\tau_{4}$ so that $\bar\tau_{3}$ and
$\bar\tau_{5}$ disappear.

Concerning the remaining contributions to the $r=v=1$ superposition
(\ref{MUL.04y}), they are associated to the implementation of the
quantity $\tilde{\cal V}_{211}(\{a_{i}\},\{\Delta \tau_{q(k)}\})$
parameterized by the three elementary graphs. Being generated through the
vertical reattachments applied to the leftmost or/and rightmost end-point of
fig. 7e (constrained by the $p=2,3$ conditions (\ref{INN.04})).
These graphs are depicted in figs.
7g, 7h and the one obtained from fig. 7g via the reflection
interchanging the horizontal sides of $C=\Box$.
It is formalized by the relations
\be
x^{1}(s'_{1})=a_{1}R~~~~~,~~~~~x^{1}(s'_{4})=a_{4}R~~~~~,~~~~~
x^{1}(s_{1})=x^{1}(s_{4})-R=0~.
\label{ST.02}
\ee
that, together with the above $a_{1}=a_{4}=0$ option, yields the four
different implementations of $\tilde{\cal V}_{211}(\cdot)$ described by the
corresponding $\{a_{k}\}-$implementations of Eqs. (\ref{INN.09}) and
(\ref{INN.10}), where one is to put $z=0$.

Finally, to compute the entire $r=v=1$ contribution to the decomposition
(\ref{SU.01z}), it remains to include the contribution of such $r=v=1$
superposition (\ref{MUL.04y}) that is associated to the $S(4)-$multiplet
of the $j=2$ elementary graphs specified by such component of fig. 7e when
the upper horizontal line is on the right compared to the lower one.
Alternatively, it can be reproduced from the $S(4)-$multiplet, specified by
the so far considered component of fig. 7e, via the reflection
interchanging the two vertical sides of $C=\Box$.
Then, introducing the $\{\Delta\bar{\tau}_{q(k)}\}-$assignment according to
the convention of subappendix \ref{assisym} and performing the auxiliary
change of the variables $\bar{\tau}_{k}\rightarrow \bar{\tau}_{n+3-k}$
(with $k=1,...,n+2$), by the same token as
previously we arrive at the $r=v=1$ implementation of Eq. (\ref{MUL.04y})
fixed by the $z=1$ option of Eqs. (\ref{INN.09}) and (\ref{INN.10}).

\app{Eq. (\ref{MUL.04}): $S(4)-$ and reflection-symmetry}
\label{symmetry3}

As for the ${\cal R}_{a}^{-1}-$deformations, the vertical nature of the
reattachments evidently implies both $S(4)-$ and reflection-symmetries of the
parameters which determine the factor (\ref{EX.01h}) (representing the latter
deformations in Eq. (\ref{MUL.04})). More generally, provided the
prescription of subappendix \ref{assisym}, these two symmetries
hold true for the algorithm (presented in Appendix \ref{freedom}) to introduce
the entire set $\{\Delta \tau_{q(k)}\}$.

Concerning the $\bar{\cal R}_{b}^{-1}-$dressing,
the situation is a little bit more tricky as it is clear from
the results discussed in the latter Appendix. The relevant parameters,
defining\footnote{In particular, it applies to the parameters
$\alpha^{(i)}$ and $y^{1}_{i}$ which determine the implementations
of the replacements (\ref{KEY.01}).} the implementation (\ref{SR.01g}) of
$\tilde{\cal V}_{2rv}(\{a_{i}\},\{\Delta \tau_{q(k)}\})$,
may be changed by a particular reattachment or reflection. In consequence,
the symmetries of the $\bar{\cal R}_{b}^{-1}-$dressing become manifest only
{\it after} the appropriate change of the variables.

To explain this point, we first accept the convention that, for a
given $rv-$specification, the $z-$ and $\tilde{a}_{1}-$dependent equations
below are
implemented according to the assignment fixed
in the previous Appendix. Then, a direct inspection (presented below)
verifies that the quantity $\tilde{\cal V}_{2rv}(\cdot)$ is
$z-${\it independent}. Furthermore, after the
corresponding implementation of the $z-$independent
change of the variables,
\be
\bar{\zeta}~\longrightarrow~(-1)^{a_{1}+m_{rv}(\tilde{a}_{1})}\bar{\zeta}~~~,~~~
\bar{\eta}~\longrightarrow~\bar{\eta}~~~~~~~,~~~~~~~
m_{rv}(\tilde{a}_{1})=v(1-r)\tilde{a}_{1}~,
\label{CHA.01}
\ee
the residual $\{a_{i}\}-$dependence (in the $r=v-1=0$ case including, by
definition given after Eq. (\ref{ST.01}), the $\tilde{a}_{1}-$dependence) of
$\tilde{\cal V}_{2rv}(\cdot)$
arises only due to the corresponding dependence of the
parameters\footnote{Eq. (\ref{LAT.02}) unifies the $r=v=0$ and $r=v=1$ cases
(characterized by $w=0$) together with the $r=v-1=0$ case. In particular,
$m_{rv}(\tilde{a}_{1})=0$ for all $0\leq r\leq v\leq 1$, except for $v-1=r=0$ when
$m_{rv}(\tilde{a}_{1})=w$.}
\be
e_{1}=(-1)^{a_{1}+m_{rv}(\tilde{a}_{1})}a_{1}~~~~~,~~~~~
e_{2}=a_{2}~~~~~,~~~~~e_{3}=-a_{4}~,
\label{LAT.02}
\ee
in terms of which one formulates the quantity
\be
{\cal K}_{rv}((-1)^{a_{1}+m_{rv}(\tilde{a}_{1})}\bar\zeta,\bar\eta,
(-1)^{a_{1}+m_{rv}(\tilde{a}_{1})}\alpha^{(1)},\{a_{l}\})=
{\cal K}_{rv}(\bar\zeta,\bar\eta,\alpha^{(1)},\{e_{l}\})
\label{LAT.02r}
\ee
where ${\cal K}_{rv}(\bar\zeta,\bar\eta,\alpha^{(1)},\{a_{l}\})=R^{v-2-r}
\tilde{\cal K}_{rv}(R\bar\zeta,R\bar\eta,\{a_{l}\})$ is obtained,
from the factor (\ref{SR.01s}) (implicitly depending on $\alpha^{(1)}$ when
$r=1$) via the change of the variables (\ref{BR.02a}), and
we take into account the transformation law
\be
\alpha^{(1)}~\longrightarrow~(-1)^{a_{1}+m_{rv}(\tilde{a}_{1})}\alpha^{(1)}~~~~~,~~~~~
\alpha^{(k)}~\longrightarrow~\alpha^{(k)}~,~\forall{k}\neq{1}~,
\label{LAT.04}
\ee
that unifies the particular implementations of this transformation which,
being given in the previous Appendix,
directly follows from definition of $\alpha^{(k)}$ defined by Eqs.
(\ref{1.50}) and (\ref{GP.01}).

Justifying the representation (\ref{MUL.04}) of ${\cal Z}_{rv}(\cdot)$, we
obtain that both the building block (\ref{MUL.02a}) and the exponential
$e^{i\left(\bar\eta \bar{t}_{1}-\bar\zeta \bar{t}_{2}\right)
\bar{A}{\cal C}_{21}/\bar{\theta}}$ are manifestly $\{e_{i}\}-$independent
(with $\bar{t}^{(1)}_{p}\equiv \bar{t}_{p}$ in the $j=2$ case at hand).
In turn, the relation (\ref{LAT.02r}) implies Eq. (\ref{MUL.01a}).
In particular, in the $v-1=r=0$ case, $e_{1}\equiv
e_{1}(a_{1},\tilde{a}_{1})$ depends on
the two independent parameters $\tilde{a}_{1}$ and $a_{1}$ which implies that
the four members of the $j-2=v-1=r=0$ $S(4)-$multiplet are specified by the
three values of $e_{1}$ so that $e_{1}=0$ appears twice (for $a_{1}=0$,
$\tilde{a}_{1}=0,1$).
In turn, it explains the origin of the factor $2^{(v-r)(1-|{e}_{1}|)}$ in Eq.
(\ref{MUL.01a}) which, being equal to
unity unless $v-1=r=0$, assumes the value 2 only when $e_{1}=0$.

To prove the asserted properties of $\tilde{\cal V}_{2rv}(\cdot)$, let us
first verify the independence of the latter
exponential. For this purpose, one is to utilize that, unifying Eqs.
(\ref{AM.05}), (\ref{INN.03}), and (\ref{INN.09}), the $\{a_{i}\}-$dependence
of the splitting (\ref{FU.02}) is defined by the replacement
\be
t_{1}~\longrightarrow~(-1)^{a_{1}+k_{rv}(z)+m_{rv}(\tilde{a}_{1})}t_{1}~~~~~,~~~~~
t_{j}~\longrightarrow~(-1)^{k_{rv}(z)}t_{j}~~~,~~~j=2,3,4~,
\label{LAT.01}
\ee
where\footnote{In order to unify the three different
$rv-$assignments, the function $k_{rv}(z)$ is chosen so that $k_{rv}(z)=z$ when
$v=r=0$ and $v=r=1$, while $k_{rv}(z)=0$ when $1-v=r=0$.}
$k_{rv}(z)=((1-v)+vr)z$.
Therefore, modulo the sign factors, the splitting is $S(4)-$ and
reflection-invariant. (In particular, the factor $(-1)^{a_{1}+m_{rv}(\tilde{a}_{1})}$
arises due to the prescription formulated in the footnote after Eq.
(\ref{SR.01a}).)
Also, the previous Appendix establishes
variables, provided the transformation properties
${\mathcal{C}}_{kl}\rightarrow~(-1)^{H_{kl}}{\mathcal{C}}_{kl}$ of the
entries of the intersection-matrix,         
\be
{\mathcal{C}}_{p1}~\longrightarrow~
(-1)^{a_{1}+k_{rv}(z)+m_{rv}(\tilde{a}_{1})}{\mathcal{C}}_{p1}~,
~\forall{p}\neq{1}~~~~~,~~~~~
{\mathcal{C}}_{il}~\longrightarrow~(-1)^{k_{rv}(z)}{\mathcal{C}}_{il}~,~
\forall{i,l}\neq{1}~,
\label{LAT.03}
\ee
where we take into account the definition (\ref{1.34}) of
${\mathcal{C}}_{il}$ combined with
the pattern of the reattachments (formalized by Eqs.
(\ref{AM.07d}), (\ref{ST.01}), and (\ref{ST.02})).
Altogether, one concludes that (in the quantity
(\ref{SR.01g})) the $\{a_{i}\}-$dependence of the factor
$e^{i\left(\bar\eta \bar{t}_{1}-\bar\zeta \bar{t}_{2}\right)
\bar{A}{\cal C}_{21}/\bar{\theta}}$ indeed disappears when Eq.
(\ref{LAT.01}) is combined with the change (\ref{CHA.01}) of the
variables, provided the transformation law (\ref{LAT.03}).

Next, let us turn to the $\{a_{i}\}-$dependence of the dressing weight
(considered prior to the change of the variables) composed of
the $n-v$ factors (\ref{SR.01a}) entering the definition
(\ref{SR.01g}) of $\tilde{\cal V}_{2rv}(\cdot)$. In view of
Eq. (\ref{LAT.03}), this dependence is determined by the
transformation law (\ref{LAT.04}) together with the 
replacement
\be
T_{{\cal C}_{ij}}(\bar\eta,\bar\zeta)~\longrightarrow~
T_{(-1)^{H_{ij}}{\cal C}_{ij}}
(\bar\eta,(-1)^{a_{1}+m_{rv}(\tilde{a}_{1})}\bar\zeta)
=\bar\eta-\bar\zeta
\label{LAT.05}
\ee
of the arguments of the combination (\ref{1.24h}).
As a result, after
the change (\ref{CHA.01}) of the variables, the considered weight
assumes the $\{a_{i}\}-$independent implementation
(\ref{MUL.02a}).

Finally, to deduce the relation (\ref{LAT.02r}), all what one needs is to
apply the replacements (\ref{LAT.03}) and
(\ref{LAT.04}) together with the change (\ref{CHA.01}) of the variables.
In particular, in the $r=v=1$ case (characterized by $m_{11}(\cdot)=0$), by
virtue of Eq. (\ref{LAT.04}), the transformation yields
$e_{4}=(-1)^{-2a_{1}}a_{4}/\alpha^{(1)}=-
a_{4}$, where $\alpha^{(1)}=-1$ is associated to figs. 7a and 7e.
Summarizing, it verifies that the decomposition (\ref{SU.01z}), 
indeed assumes the form fixed by Eq.~(\ref{MUL.04}).

\eop

\newpage
\begin{figure}
\vspace*{3mm}
\centering{
\epsfig{file=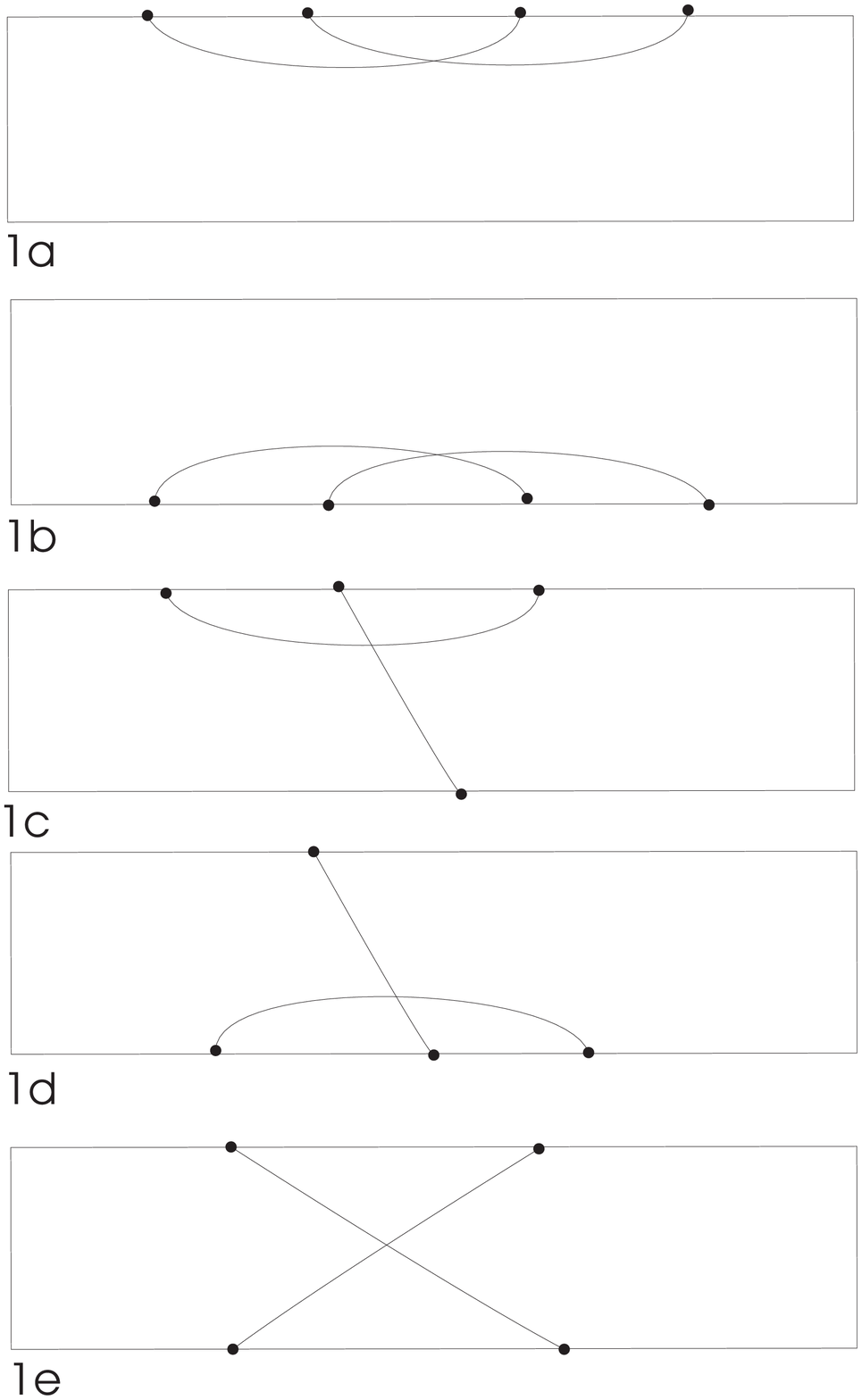,width=9.5cm}
}
\caption[]   

\end{figure}

\begin{figure}
\vspace*{3mm}
\centering{
\epsfig{file=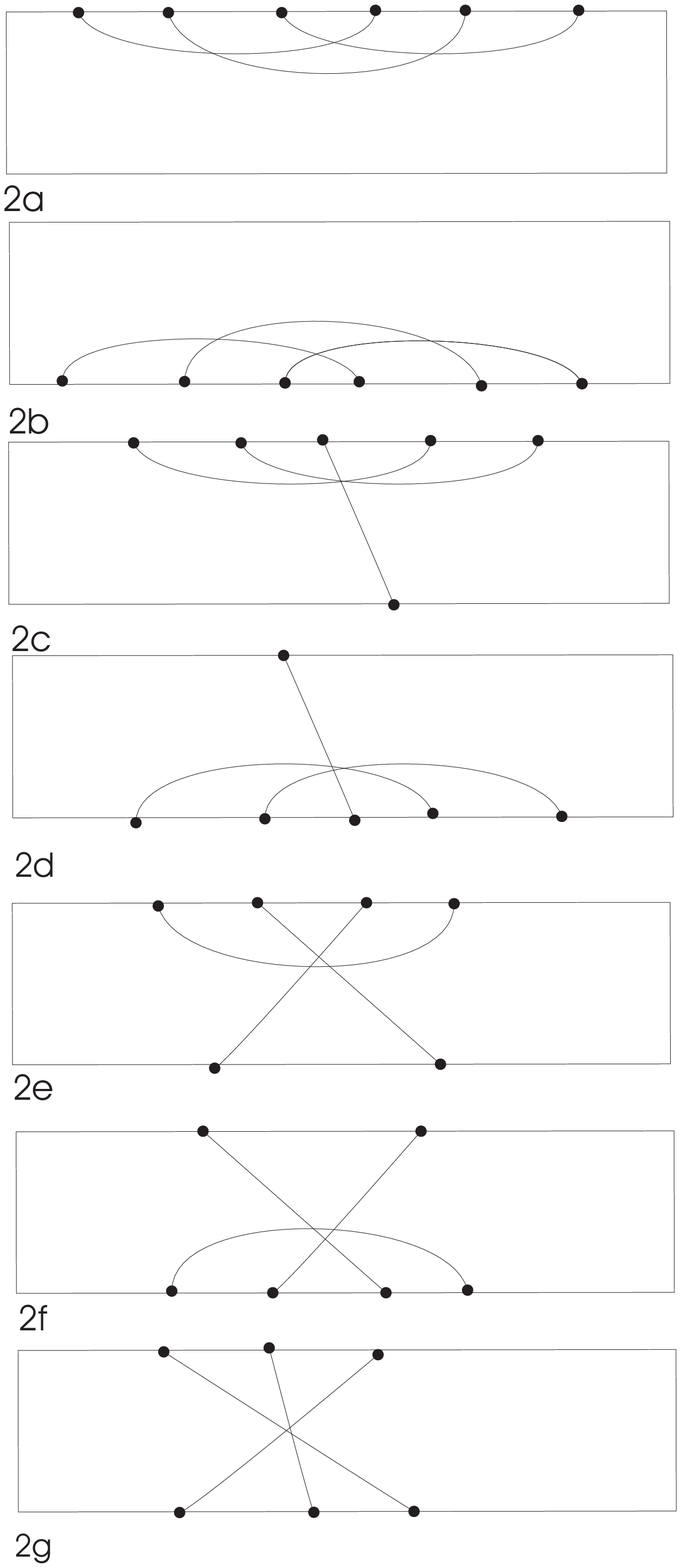,width=9.5cm}
}
\caption[]   

\end{figure}



\begin{figure}
\vspace*{3mm}
\centering{
\epsfig{file=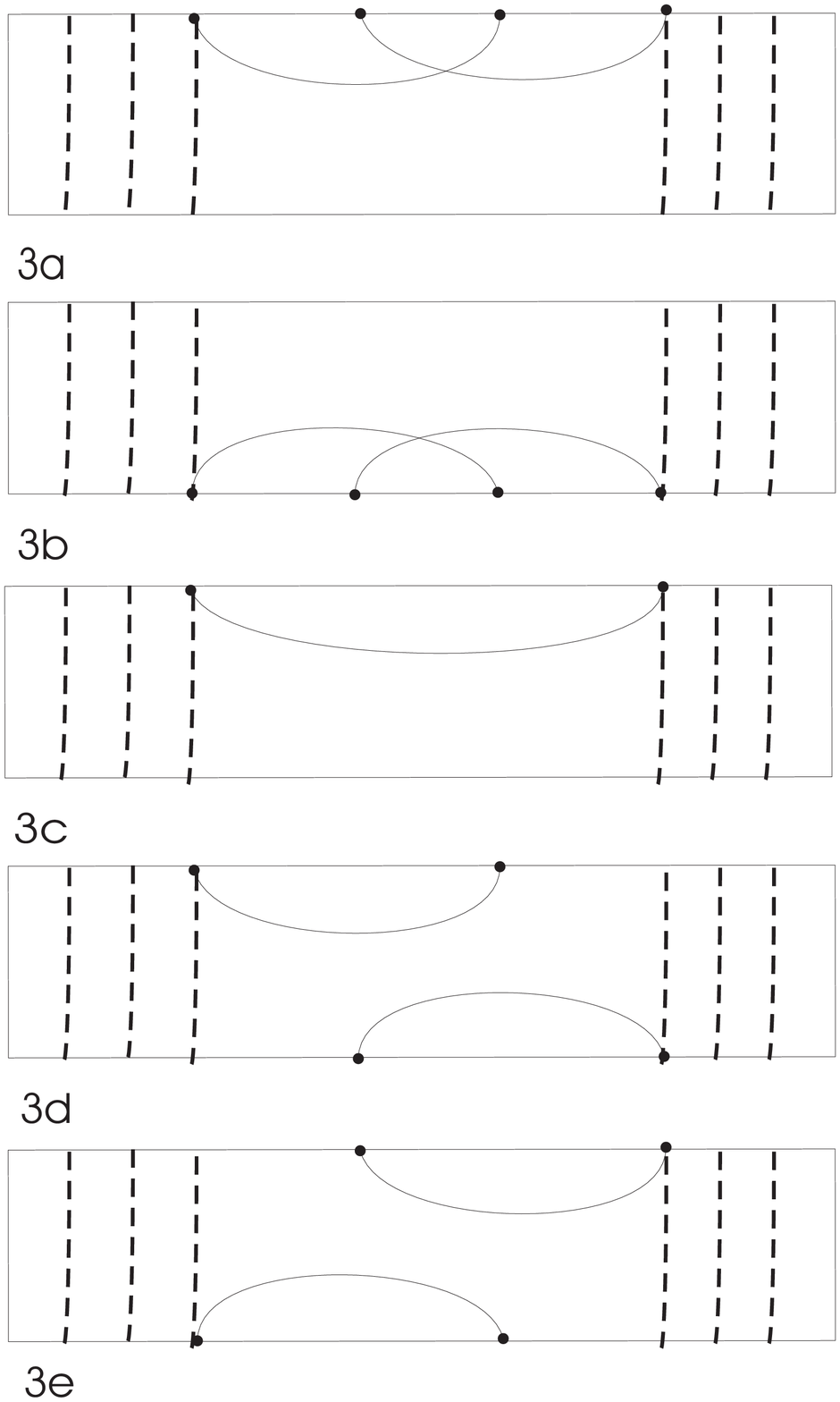,width=9.5cm}
}
\caption[]   

\end{figure}

\begin{figure}
\vspace*{3mm}
\centering{
\epsfig{file=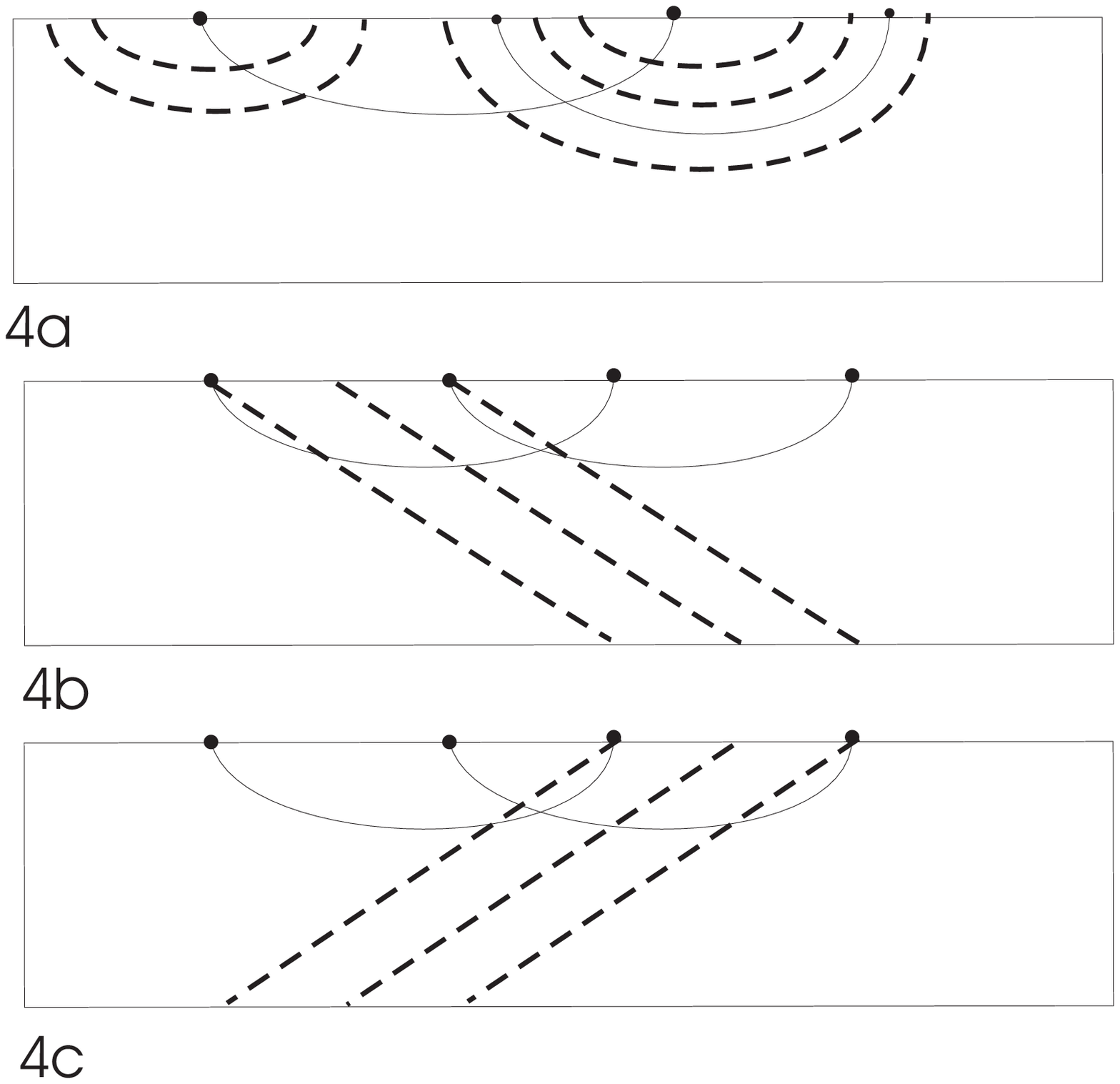,width=9.5cm}
}
\caption[]   

\end{figure}

\begin{figure}
\vspace*{3mm}
\centering{
\epsfig{file=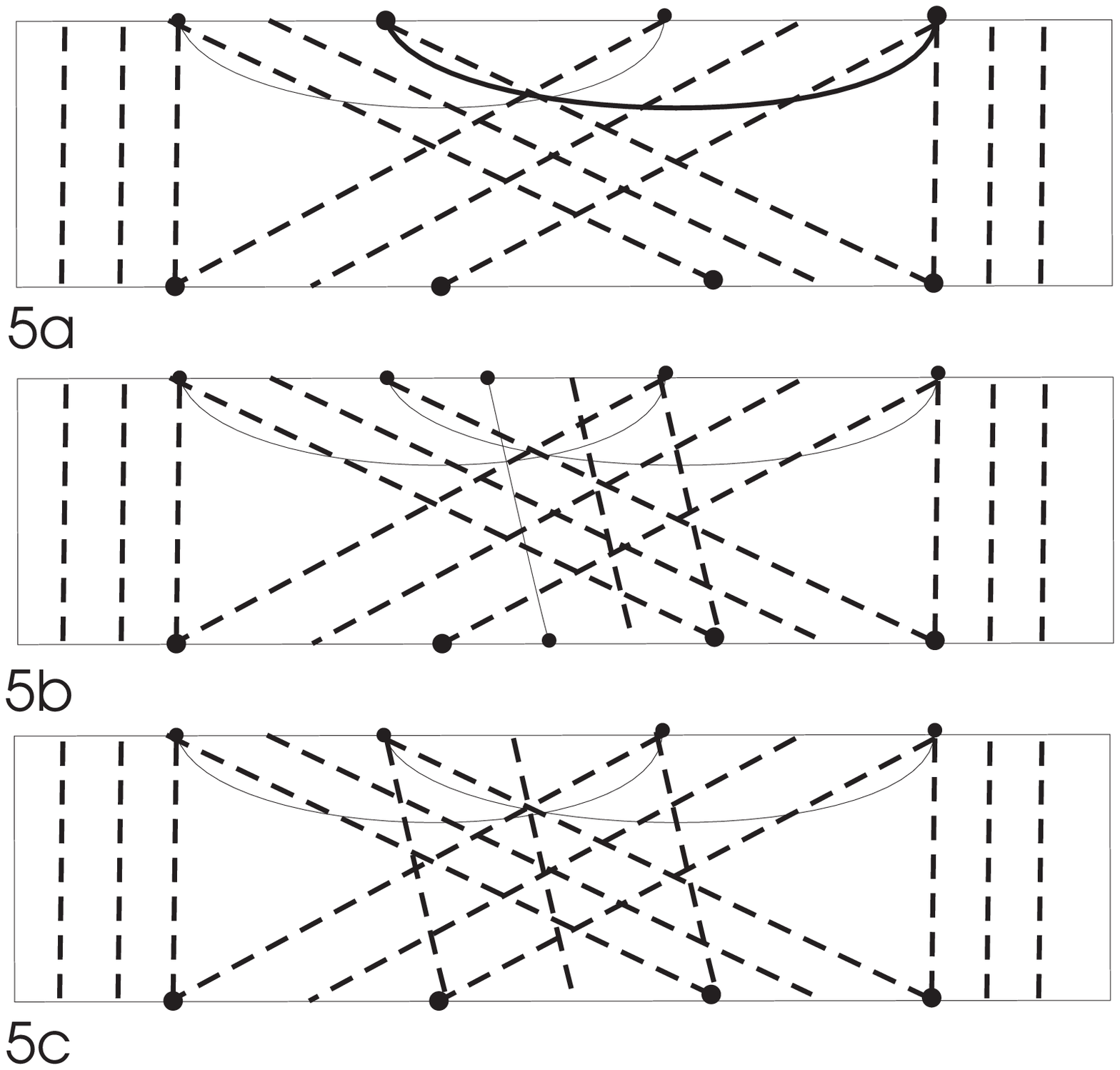,width=9.5cm}
}
\caption[]   

\end{figure}

\begin{figure}
\vspace*{3mm}
\centering{
\epsfig{file=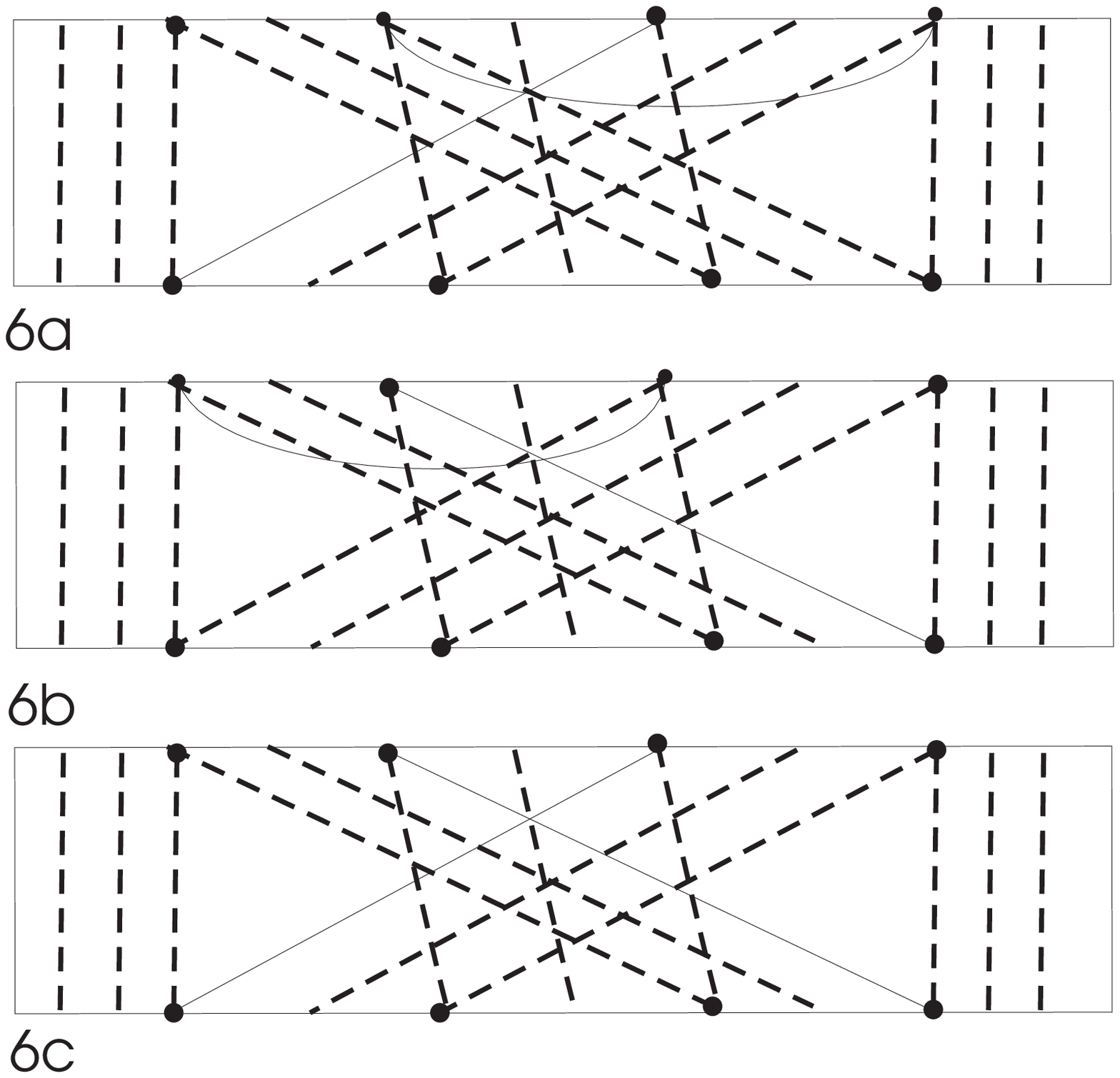,width=9.5cm}
}
\caption[]   

\end{figure}

\begin{figure}
\centering{
\epsfig{file=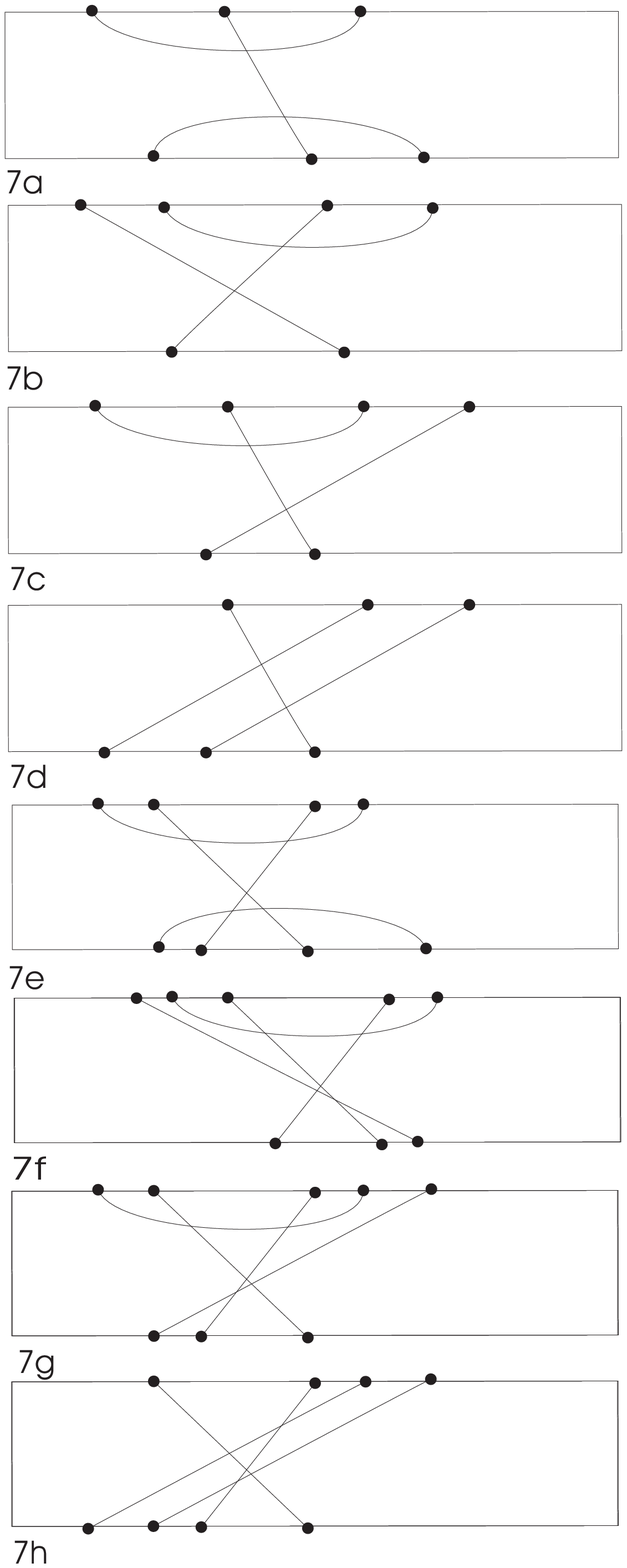,width=9.5cm}
}
\caption[]   

\end{figure}

\begin{figure}
\vspace*{3mm}
\centering{
\epsfig{file=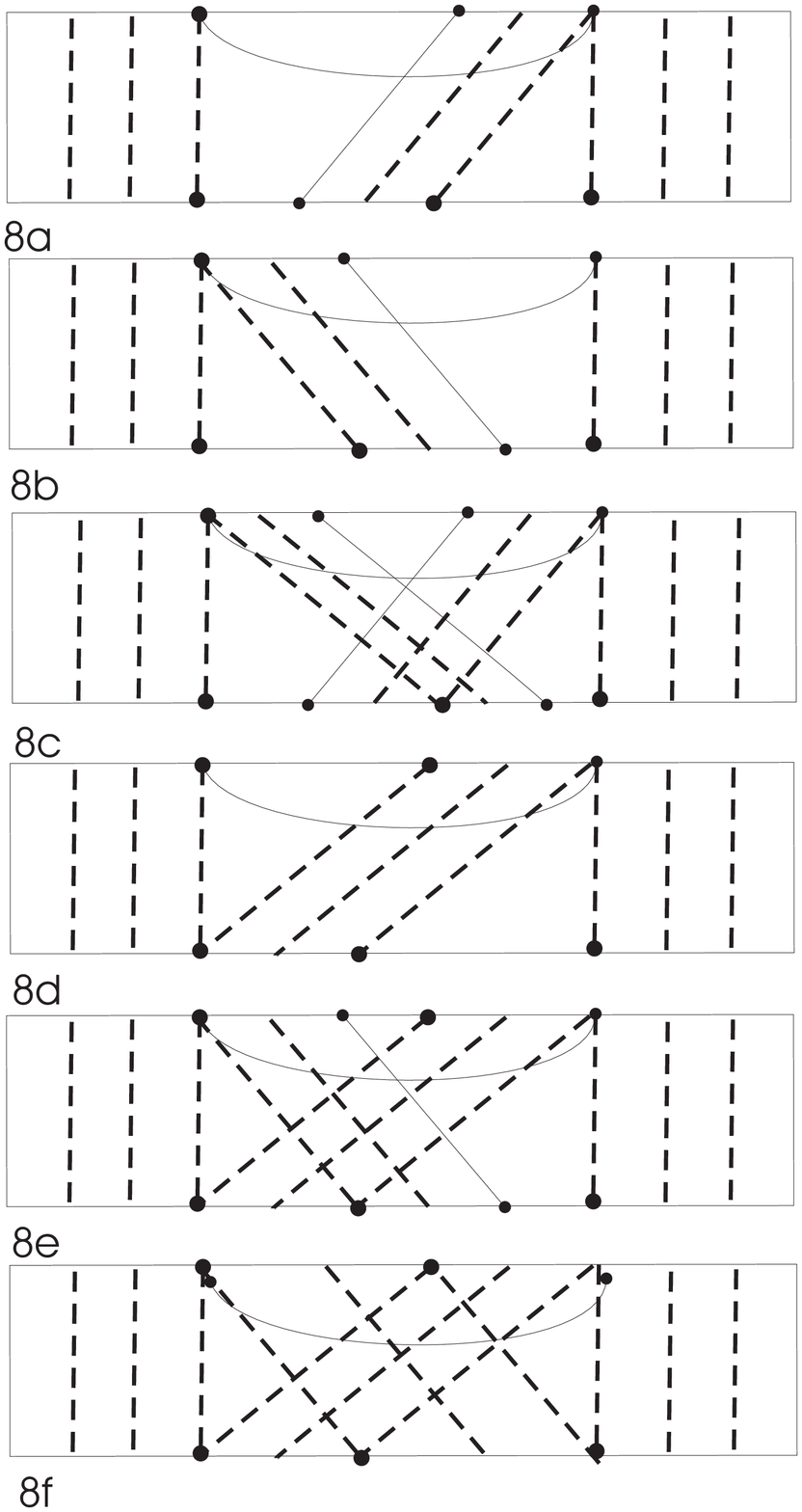,width=9.5cm}
}
\caption[]   

\end{figure}

\begin{figure}
\centering{
\epsfig{file=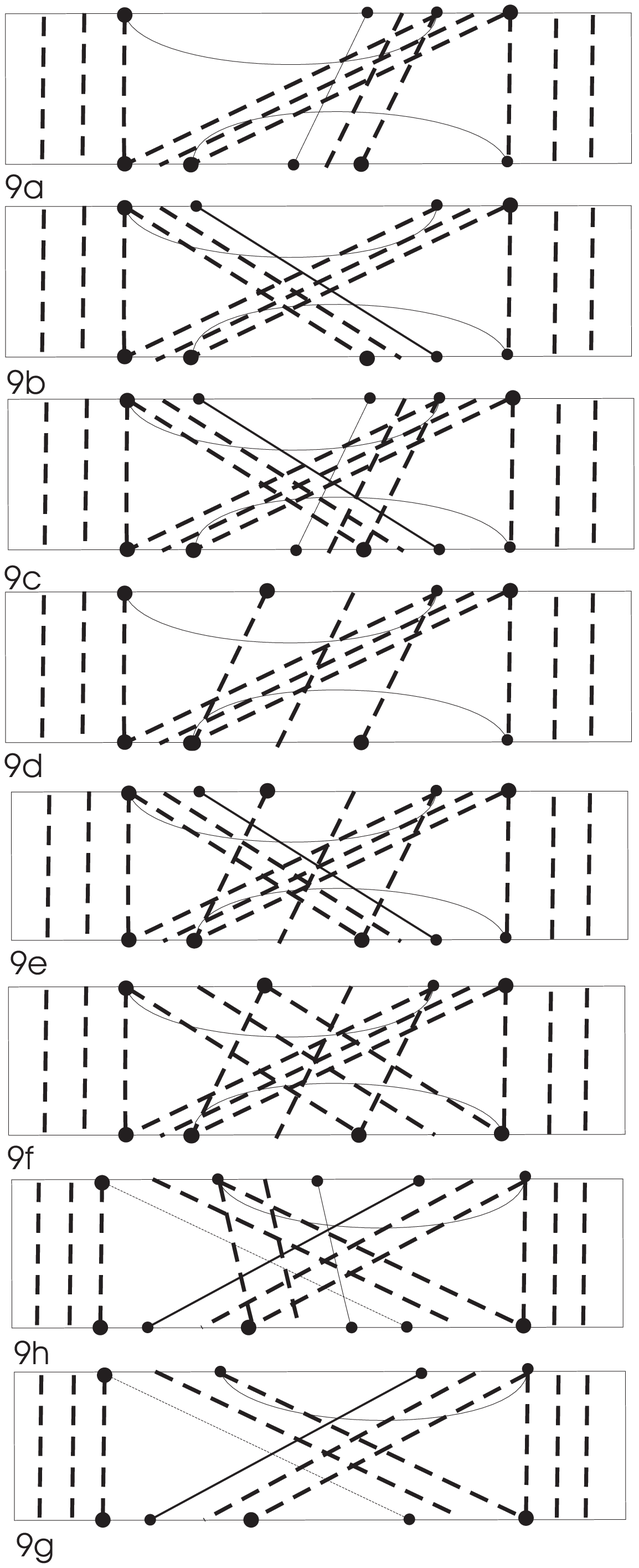,width=9.5cm}
}
\caption[]   

\end{figure}

\end{document}